\documentclass[a4paper,fleqn,usenatbib]{mnras}
\usepackage{newtxtext,newtxmath}
\usepackage[T1]{fontenc}
\usepackage{ae,aecompl}
\usepackage{graphicx}
\usepackage{float}
\usepackage{caption}
\usepackage{subcaption}
\usepackage{multicol}
\usepackage{multirow}
\usepackage{flushend}
\usepackage{comment}

\title[RQ quasars with VLBA]{The radio emission in radio-quiet quasars: the VLBA perspective}

\author[Sina Chen et al.]{
Sina Chen $^{1}$ \thanks{E-mail: sina.chen@campus.technion.ac.il},
Ari Laor $^{1}$,
Ehud Behar $^{1}$,
Ranieri D. Baldi $^{2}$,
and Joseph D. Gelfand $^{3}$.
\\
$^{1}$ Physics Department, Technion, Haifa 32000, Israel \\
$^{2}$ INAF - Istituto di Radioastronomia, via Gobetti 101, 40129 Bologna, Italy \\
$^{3}$ NYU Abu Dhabi, PO Box 129188, Abu Dhabi, UAE}

\date{Accepted XXX, Received YYY, in original form ZZZ.}

\pubyear{2023}

\begin{document}
\label{firstpage}
\pagerange{\pageref{firstpage}--\pageref{lastpage}}
\maketitle

\begin{abstract}

The origin of the radio emission in radio-quiet quasars (RQQ) is not established yet.
We present new VLBA observations at 1.6 and 4.9~GHz of ten RQQ (nine detected), which together with published earlier observations of eight RQQ (five detected), forms a representative sample of 18 RQQ drawn from the Palomar-Green sample of low $z$ ($< 0.5$) AGN.
The spectral slope of the integrated emission extends from very steep ($\alpha < -1.98$) to strongly inverted ($\alpha = +2.18$), and the slopes of nine of the 14 objects are flat ($\alpha > -0.5$).
Most objects have an unresolved flat-spectrum core, which coincides with the optical {\it Gaia} position.
The extended emission is generally steep-spectrum, has a low brightness temperature ($< 10^7$~K), and is displaced from the optical core (the {\it Gaia} position) by $\sim 5-100$~pc.
The VLBA core flux is tightly correlated with the X-ray flux, and follows a radio to X-ray luminosity relation of $\log L_{\rm R}/L_{\rm X} \simeq -6$, for all objects with a black hole mass $\log M_{\rm BH}/M_{\odot} < 8.5$.
The flatness of the core emission implies a compact source size ($\lesssim 0.1$~pc), which likely originates from the accretion disk corona.
The mas-scale extended emission is optically thin and of clumpy structure, and is likely produced by an outflow from the center. 
Radio observations at higher frequencies can further test the accretion disk coronal emission interpretation for the core emission in RQQ.

\end{abstract}


\section{Introduction}

Active galactic nuclei (AGN) can be divided into radio-loud (RL) and radio-quiet (RQ) based on their radio loudness $R$, which is the ratio of radio 5\,GHz to optical 4400\,$\text{\AA}$ flux density \citep{Kellermann1989}.
The majority ($\sim$90\%) of AGN are RQ ($R \le 10$), and only a fraction ($\sim$10\%) are RL.
The radio loudness can also be defined using the ratio of the 5\,GHz radio luminosity to the 2--10\,keV X-ray luminosity \citep{Terashima2003}, where RL objects have $\log R_{\rm X} > -3.5$ \citep{Panessa2007,Laor2008}.
The radio emission in RL quasars (RLQ) is generally dominated by a powerful relativistic jet \citep{Urry1995,Blandford2019}.
In RQ quasars (RQQ), the radio emission can be produced by a variety of emission mechanisms, from the host galaxy scales ($\sim$ kpc) down to the innermost accretion disk scales ($<$ 0.01\,pc) \citep{Panessa2019}.

We elaborate below briefly on possible radio emission mechanisms in RQQ from large to small scales, and how they can be tested based on the spectral slope $\alpha$ ($S_{\nu} \propto \nu^{\alpha}$) and the spatial extension of the radio emission.

{\em Host galaxy star formation (SF).}
AGN activity and enhanced SF are often related. The SF radio emission \citep{Bell2003} may be responsible for the bulk of the radio emission observed in RQQ \citep{Kimball2011,Condon2013,Zakamska2016}.
If so, the radio emission would be predominantly optically thin synchrotron radiation ($\alpha < -0.5$), and its radio morphology would be similar to other tracers of SF (e.g.\ H$\alpha$ line and infrared emission) in the host galaxy.

{\em An AGN-driven wind interacting with the host interstellar medium (ISM).}
Wide angle winds are a common feature of AGN, and are believed to be the dominant AGN feedback mechanism on the host SF, and may eventually lead to the black hole (BH) mass versus the bulge mass relation \citep{Magorrian1998,Laor1998,Ho1999,Gebhardt2000,Gultekin2009,Kormendy2013}.
The interaction between the fast ($\sim$ thousands of km\,s$^{-1}$) wind and the surrounding ISM would generate shocks that accelerate particles to relativistic energies \citep[e.g.][]{Jiang2010}.
Relativistic particles in the shocked gas magnetic field results in a steep spectrum ($\alpha < -0.5$) optically thin synchrotron emission \citep{Zakamska2016}, likely at the host galaxy inner regions.

{\em Thermal free-free emission.}
The intense ultra-violet radiation of AGN photoionizes large volumes of ambient gas, as indicated by the strength of the narrow and broad line emission in AGN. Photoionization calculations indicate that the free-free emission of this gas may be detectable in millimeter waveband, with the characteristic flat spectral slope of $\alpha = -0.1$ \citep{Baskin2021}.

{\em A low-power jet.}
The jet mechanism is identical to the one in RL AGN, but the luminosity scales down by a factor of typically $10^3$ \citep[e.g.][]{Falcke1995}.
In this case, RQQ may show similar phenomenology to RLQ, steep spectra in lobe dominated objects, and flat spectra in core dominated objects. The jet emission may produce a linear structure, which may extend only on sub-arcsec scales, and is possibly resolved only on mas scales.

{\em Coronal emission.}
RQQ follow tightly the radio versus X-ray luminosity relation of coronally active stars \citep{Guedel1993} $L_{\rm R} / L_{\rm X} \simeq 10^{-5}$ \citep{Laor2008}.
Since the AGN X-ray emission most likely originates from hot coronal gas at the inner accretion disk, it is possible that the radio emission of RQQ is also of coronal origin \citep{Gallimore1997}. As the emission originates on sub-pc scales, it will remain unresolved with the VLBA, and will be characterized by an optically thick flat/inverted spectrum.
Furthermore, the jet and outflow base may physically coincide with the corona where the relativistic particles are being accelerated \citep{Blundell1998,Merloni2002,King2017}.

The relatively few available studies of RQQ \citep[e.g.][]{Kellermann1989,Kellermann1994,Barvainis1996,Kukula1998,Ulvestad2005,Leipski2006,Padovani2011,Doi2011,Zakamska2016,Silpa2020,Panessa2022b,Baldi2022,Chen2022a} generally lead to mixed results indicating that the origin of the radio emission in RQQ is still an open question. If a number of different mechanisms are indeed involved, then radio observations can be used as a powerful tool to probe the variety of emission mechanisms that are likely to occur in RQQ, unlike the situation in RLQ where the radio emission is dominated by a relativistic jet \citep{Panessa2019}.

Interesting hints are provided in \citet{Laor2019}, based on a compilation of various archival VLA observations of 25 RQ Palomar-Green (PG) quasars, mostly with the highest resolution A configuration.
This study found strong relations between the derived 5--8.5\,GHz spectral slope $\alpha_{5-8.5}$ of the core emission ($<$ 0.3\,arcsec) and a number of properties: the full width at half maximum (FWHM) of the broad H$\beta$ line, the Eddington ratio $L/L_{\rm Edd}$, and the Fe\,II/H$\beta$ line ratio.
All $L/L_{\rm Edd} > 0.3$ objects have steep spectra, i.e.\,$\alpha_{5-8.5} < -0.5$, and nearly all objects with $L/L_{\rm Edd} < 0.3$ have flat spectra, i.e.\,$\alpha_{5-8.5} > -0.5$.
In addition, the slopes of all objects with Fe\,II/H$\beta > 0.5$ are steep, and all flat slope objects have Fe\,II/H$\beta < 0.4$.
A possible scenario, suggested by \citet{Laor2019}, for the $\alpha_{5-8.5}$ and the $L/L_{\rm Edd}$ correlation is that the extended optically thin emission, which dominates in high $L/L_{\rm Edd}$ objects, is associated with an outflow, which is more prominent at high $L/L_{\rm Edd}$, such as radiation pressure driven winds.
In low $L/L_{\rm Edd}$ objects, the extended component does not dominate the total radio emission, and only a compact optically thick component, possibly associated with coronal emission, is present.

A striking result was found in a recent follow up exploratory VLBA study \citep{Alhosani2022}, where eight of the above 25 RQ PG quasars were observed with the VLBA in the L and C bands, of which four were selected to have a very steep spectrum, $\alpha_{5-8.5} < -1$, and four were selected to have an inverted spectrum, $\alpha_{5-8.5} > 0$.
They found that in three of the four flat-spectrum sources, the VLA core flux remains unresolved with the VLBA, while in three of the four steep-spectrum sources, no flux was detected ($<$ 5--10\% of the VLA flux).
This suggests a dichotomy in terms of the dominant radio emission mechanism in RQQ: extended emission in high $L/L_{\rm Edd}$ RQ AGN versus only a very compact core in low $L/L_{\rm Edd}$ RQ AGN.
This dichotomy also supports the $\alpha_{5-8.5}$ versus $L/L_{\rm Edd}$ correlation found in \citet{Laor2019}.
At low $L/L_{\rm Edd}$, the compact core (i.e.\ the disk corona) is primarily responsible for the total radio emission, which remains unresolved on the VLBA scales.
At high $L/L_{\rm Edd}$, the extended emission (i.e.\ outflows) dominates, and it may be resolved out on the VLBA scales.

In this work, we report new VLBA observations of a representative sample of 10 bright RQQ from the PG catalogue in the L (18\,cm) and C (6\,cm) bands. These observations complement the exploratory VLBA study \citep{Alhosani2022}, which focused on the eight more extreme RQ PG quasars, four with the steepest and four with the flattest $\alpha_{5-8.5}$ slopes.
Our sample is selected to cover a wide luminosity range ($-27 < M_{\rm V} < -21$) and a broad range of H$\beta$ FWHM values ($\sim$ 1000--10,000\,km\,s$^{-1}$).
In combination with the earlier exploratory VLBA observations, we aim to explore systematically the mas scale radio emission of RQQ, and address the following questions.
Do objects with steep VLA slope core emission generally lack emission on the VLBA scales?
In contrast, does the core emission in flat VLA slope objects remains unresolved on the VLBA scales?
Do the bimodal relations between $\alpha_{5-8.5}$ and $L/L_{\rm Edd}$ found in the VLA data also hold on the VLBA scales?
Does the radio emission switch from being compact optically thick at low $L/L_{\rm Edd}$, to extended optically thin at high $L/L_{\rm Edd}$?
In this work, we will focus on the origins of the AGN-driven wind, the low-power jet, and the coronal emission as possible emission mechanisms.
The host galaxy SF emission is generally diffuse and resolved out at the VLBA resolution.
Free-free emission is not expected to dominate in centimeter waveband, and its expected brightness temperature is too low to be detectable with the VLBA \citep{Baskin2021}.

The structure of this paper is as follows.
Section~\ref{sample_section} describes the sample selection, Section~\ref{reduce_section} presents the data reduction, Section~\ref{analyze_section} describes the data analysis, Section~\ref{result_section} presents the results, and we discuss them in Section~\ref{discussion_section}. The summary is given in Section~\ref{summary_section}.

\section{Sample selection} \label{sample_section}

The sample is selected from the 71 $z < 0.5$ PG RQQ \citep{Boroson1992}, which is the most extensively studied sample of Type 1 AGN: including e.g.\ a complete coverage of the overall SED \citep{Neugebauer1987,Sanders1989}, radio cm-band continuum and imaging \citep{Kellermann1989,Kellermann1994,Miller1993}, optical spectroscopy \citep{Boroson1992}, optical polarimetry \citep{Berriman1990}, UV spectroscopy \citep{Baskin2005}, and X-ray spectroscopy \citep{Brandt2000}. This sample revealed various interesting trends, such as the Eigenvector 1 set of emission line correlations \citep{Boroson1992}, the BH-bulge mass relation in AGN \citep{Laor1998}, and the BH mass and radio loudness relation \citep{Laor2000}.

A critical property of the sample is that it is optically selected, and is therefore not subject to radio bias. The sample covers AGN over a wide range of intrinsic properties, from Seyfert to quasar luminosities ($L_{\rm bol} \sim 10^{43} - 10^{46}$\,erg\,s$^{-1}$), and with a BH mass in the range of $M_{\rm BH} \sim 10^6 - 10^9\,M_{\odot}$. The wealth of additional information available for this sample in other bands will allow to explore possible relations of the nuclear spectral slope derived here with other emission and absorption properties \citep[e.g.][]{Boroson1992}.

\begin{figure}
\includegraphics[width=.5\textwidth, trim={7cm, 0cm, 8cm, 1cm}, clip]{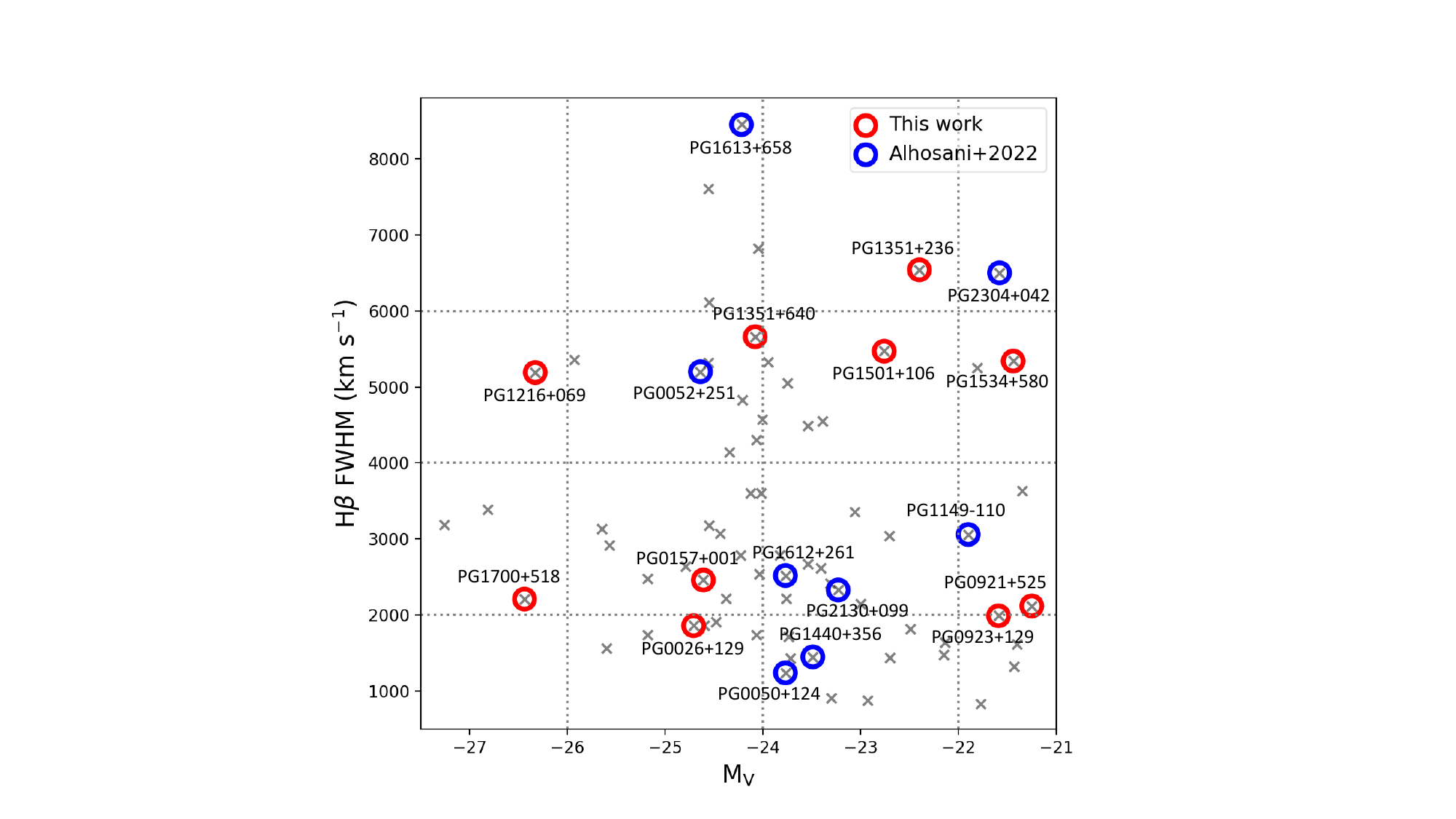}
\caption{The distribution of the 71 PG RQQ in the H$\beta$ FWHM versus optical luminosity $M_{\rm V}$ plane. We divide this distribution into four bins along each axis (dotted lines), which forms 16 regions, of which two are empty (the upper and lower left most bins). We select the object with a relatively high 5\,GHz flux in each region. This representative sample includes 14 objects. We also include in this paper the eight RQQ already observed in \citet{Alhosani2022} (marked in blue circles), of which four overlap with our sample. The two samples include a total of 18 objects, of which ten were observed in our new VLBA program (marked in red circles). The merged sample spreads over the observable ranges of $M_{\rm V}$ and H$\beta$ FWHM, and is likely representative of optically selected RQQ.}
\label{sample}
\end{figure}

Figure~\ref{sample} shows the distribution of the 71 PG RQQ in the H$\beta$ FWHM versus absolute magnitude $M_{\rm V}$ plane. These two parameters can be used to determine $M_{\rm BH}$ and $L/L_{\rm Edd}$. We divide the plane of Figure~\ref{sample} into four bins in $M_{\rm V}$ and four bins in H$\beta$ FWHM. This forms 16 regions, of which two are not populated.
In each of the 14 populated bins, we select the brightest object at 5\,GHz based on the VLA D configuration flux \citep{Kellermann1989}. 
This representative sample thus includes 14 objects, which have a wide spread in luminosity ($-27 < M_{\rm V} < -21$) and H$\beta$ FWHM ($\sim$ 1000--10,000\,km\,s$^{-1}$).
Despite the selection criterion of a relatively high radio flux in each bin, the distribution of $R$ values of this sample is consistent with that of the parent sample \citep{Kellermann1989}.

We merge the representative sample with the eight objects already observed with the VLBA in \citet{Alhosani2022}, of which four overlap our sample.
The merged sample therefore includes 18 objects.
Here we report the VLBA observations and the data reduction of the ten newly observed RQQ.
The data analysis and the discussion include all 18 objects, of which ten are from our new VLBA program and eight are from \citet{Alhosani2022}.

\section{Data reduction} \label{reduce_section}

The observations were carried out with the VLBA in the L and C bands using the 8--10 main VLBA stations (see Table~\ref{position}).
The angular resolution can reach 4.3\,mas in the 18\,cm band and 1.4\,mas in the 6\,cm band.
We used the digital down conversion observing system with a 2-bit sampling at a data rate of 4 Gbps and four intermediate frequency (IF) bands with dual polarization and 128 MHz bandwidth.
The IFs are centered at 1376, 1504, 1632, and 1760 MHz in the L band and at 4612, 4740, 4868, and 4996 MHz in the C band.
We performed phase-referencing continuum observations.
Each observation was about three hours long with scans at different frequencies interleaved, which can yield a better UV coverage.
A four minutes nodding cycle was used with two minutes on a target and one minute on a phase calibrator before and after the target.
A ``Fringe Finder'' (3C84, 3C345, 3C454.3, or 4C39.25) was observed twice when all the antennas were up.
This strategy yields an integration time of about one hour on the target at each frequency.

The data was calibrated using the VLBA data calibration pipeline procedure VLBARUN \footnote{\url{http://www.aips.nrao.edu/vlbarun.shtml}} in the Astronomical Image Processing System \citep[AIPS;][]{Greisen2003}.
A step-by-step usage of VLBARUN is described in Appendix C of the AIPS Cookbook \footnote{\url{http://www.aips.nrao.edu/cook.html}}.
The standard steps include first correcting the visibility for ionospheric delays and Earth Orientation Parameters, then applying amplitude corrections from digital sampling, finding the phase delays using the fringe finder and applying phase solutions.
An inspection on all baselines (pairs of antennas) for radio frequency interference (RFI) was performed. Data suffered from RFI was removed using the AIPS task UVFLG.
A self-calibration was applied on the calibrators.
Once the calibration procedure completed, we applied the final calibration table on the target and split the data of the target using the AIPS task SPLIT.

We imaged the visibility data of the target using the AIPS task IMAGR.
This task uses the CLEAN algorithm to deconvolve the ``dirty image'' with the point spread function (PSF), to obtain the residual and the ``clean image''.
We chose natural weighting (i.e.\ robust=5) which maximizes sensitivity at the expense of angular resolution.
The source images were not self-calibrated, as an additional self-calibration does not improve the images if the signal-to-noise ratio is not too high.
The final images were inspected using the AIPS task IMEAN to obtain the background noise in a source-free region.
The synthesized beam size can be seen in the header of the images.
The AIPS task IMFIT or JMFIT was used to model the source with a 2D Gaussian profile, to obtain the peak intensity, the integrated flux density, the source position, and the source sizes before and after deconvolution.
We leave the centroid location, peak intensity, major and minor axes, and position angle, as free parameters in the Gaussian model.
If the source in the C band shows two components, which are unresolved in the L band, we modelled the source with two Gaussian components in the L band.

In order to measure the spectral slope, which is less biased by the resolution in the different bands, we created tapered images via setting a UV range of 3000--50000\,k$\lambda$, corresponding to an angular resolution $\sim$ 5\,mas, in both L and C bands.
The lower limit is equivalent to the minimum UV range in the C band, to recover emission on similar scales in both bands, and the upper limit is equivalent to the maximum UV range in the L band, to obtain comparable resolutions in both bands.

\section{Data analysis} \label{analyze_section}

\begin{figure*}

\begin{subfigure}{.75\textwidth}
\centering
\includegraphics[width=.48\textwidth, trim={8cm, 0cm, 8cm, 0cm}, clip]{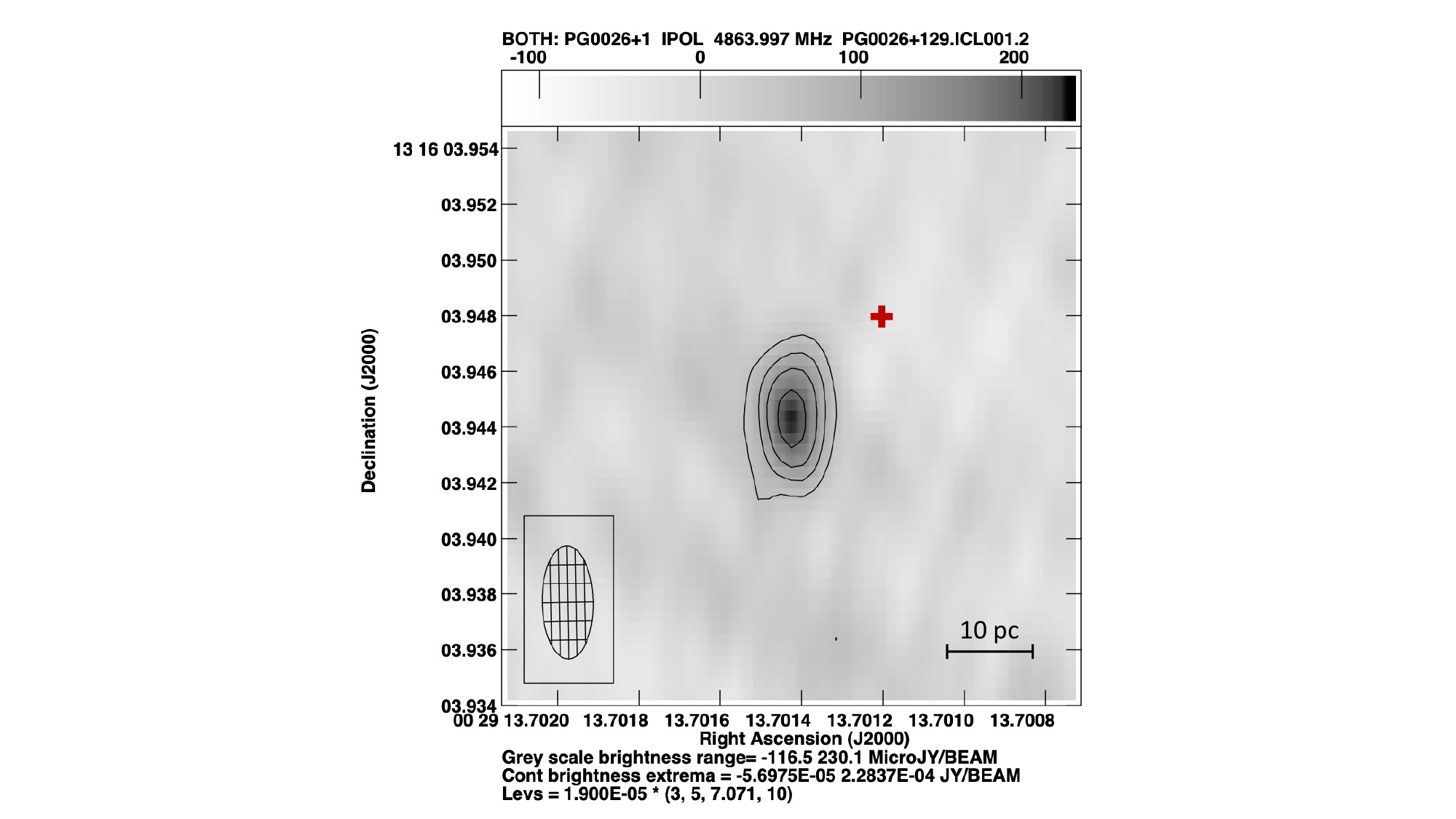}
\includegraphics[width=.48\textwidth, trim={8cm, 0cm, 8cm, 0cm}, clip]{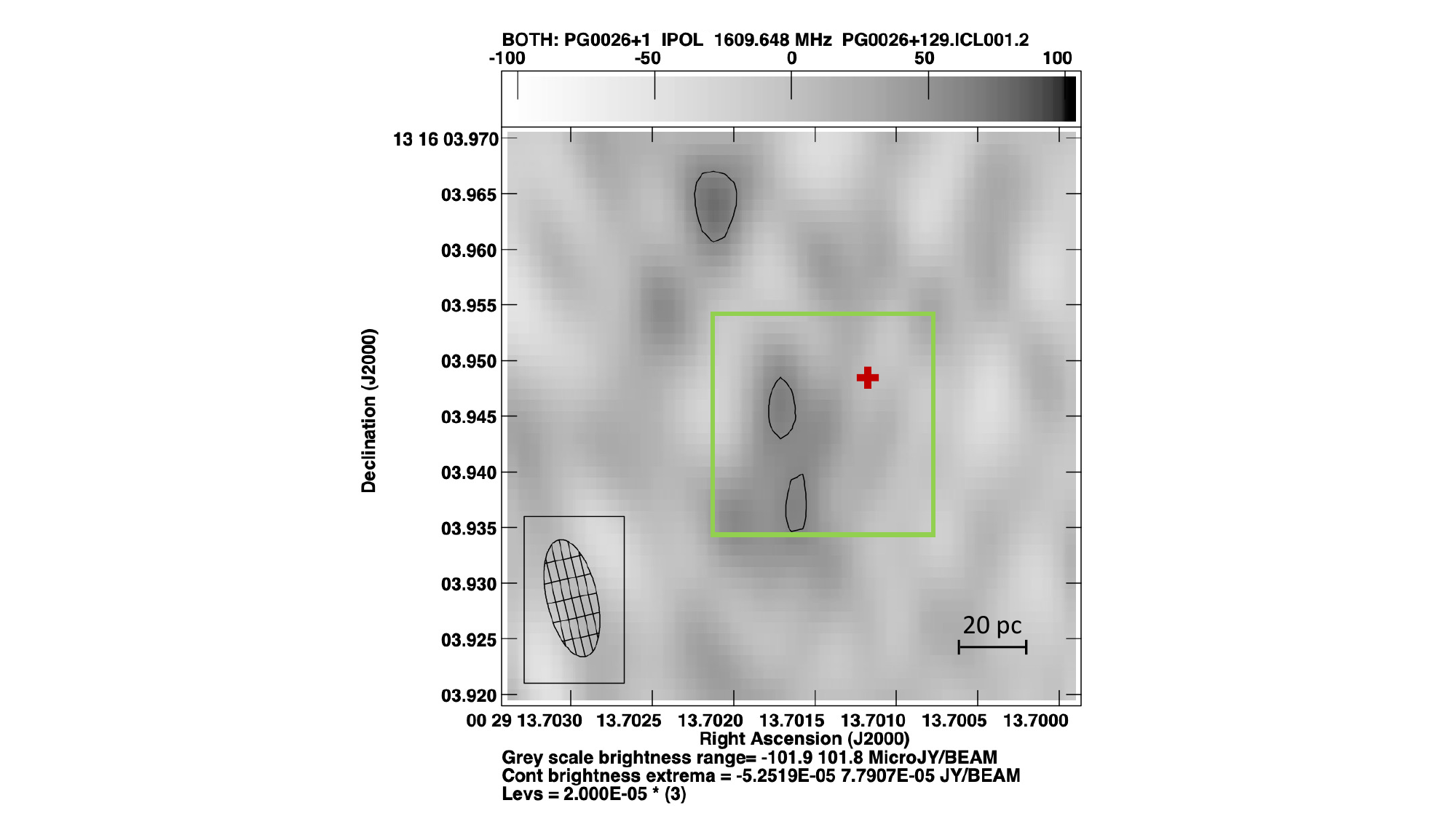}
\caption{PG0026+129}
\label{PG0026+129}
\end{subfigure}

\begin{subfigure}{.75\textwidth}
\centering
\includegraphics[width=.48\textwidth, trim={8cm, 0cm, 8cm, 0cm}, clip]{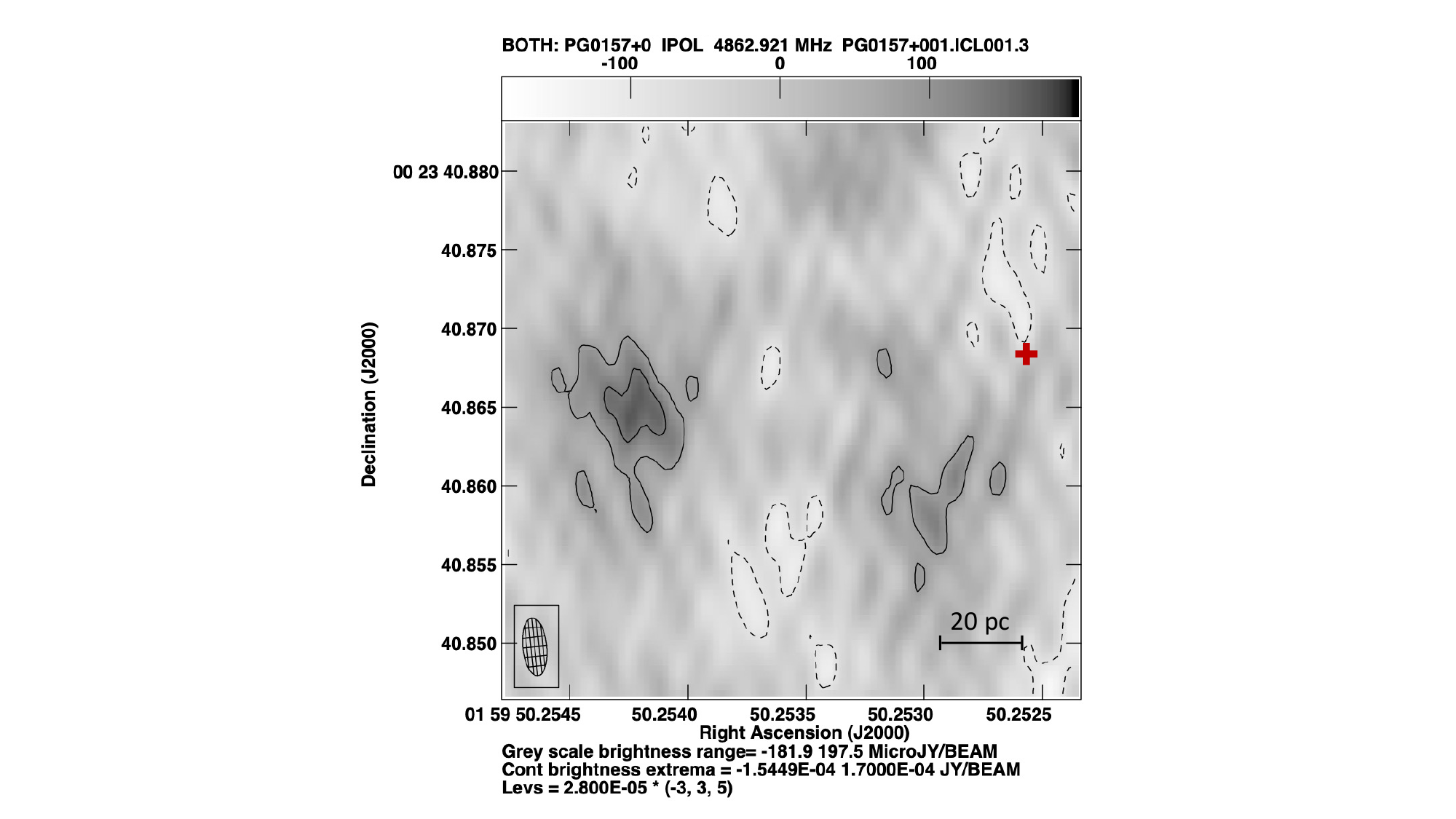}
\includegraphics[width=.48\textwidth, trim={8cm, 0cm, 8cm, 0cm}, clip]{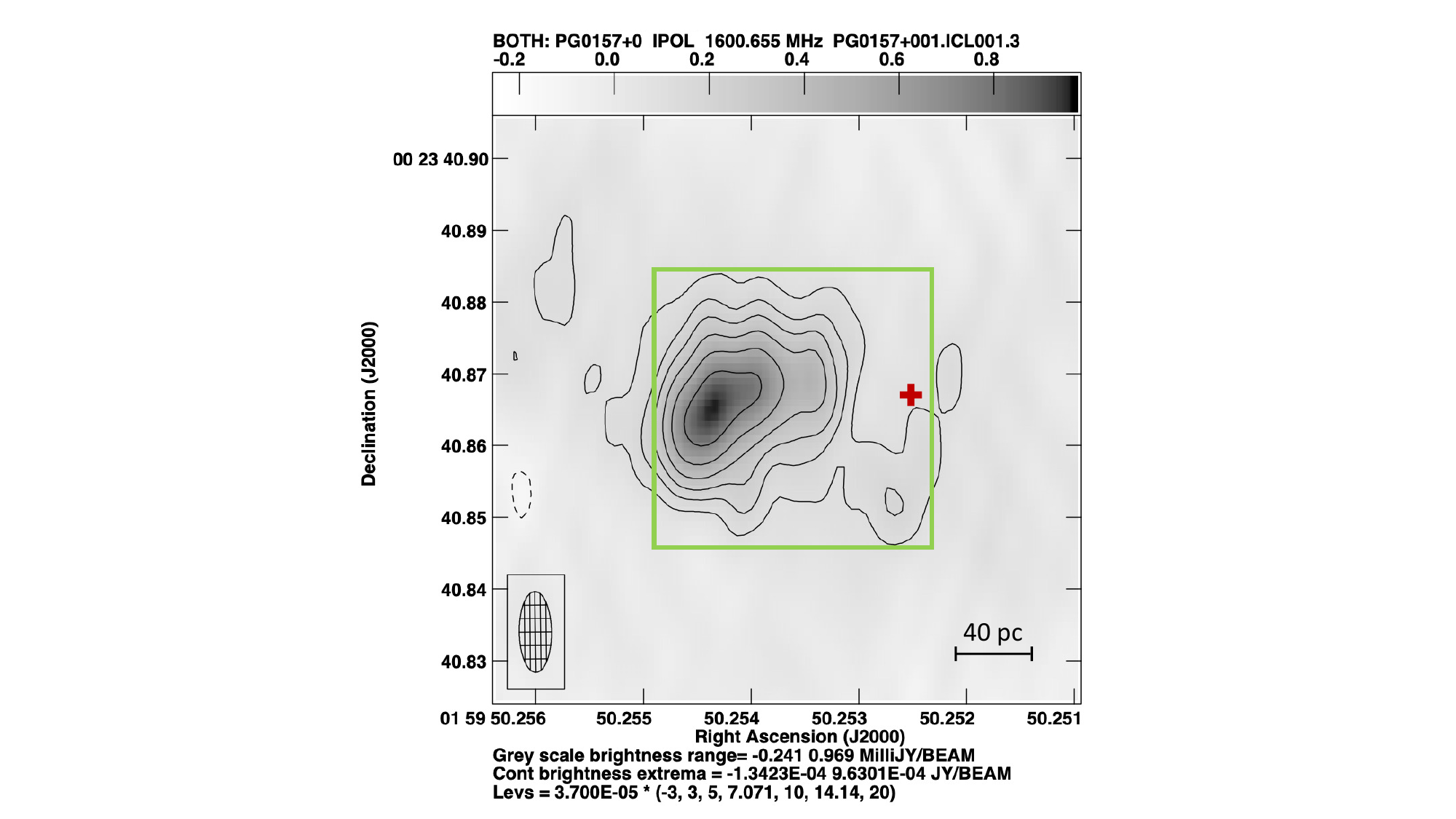}
\caption{PG0157+001}
\label{PG0157+001}
\end{subfigure}

\begin{subfigure}{.75\textwidth}
\centering
\includegraphics[width=.48\textwidth, trim={8cm, 0cm, 8cm, 0cm}, clip]{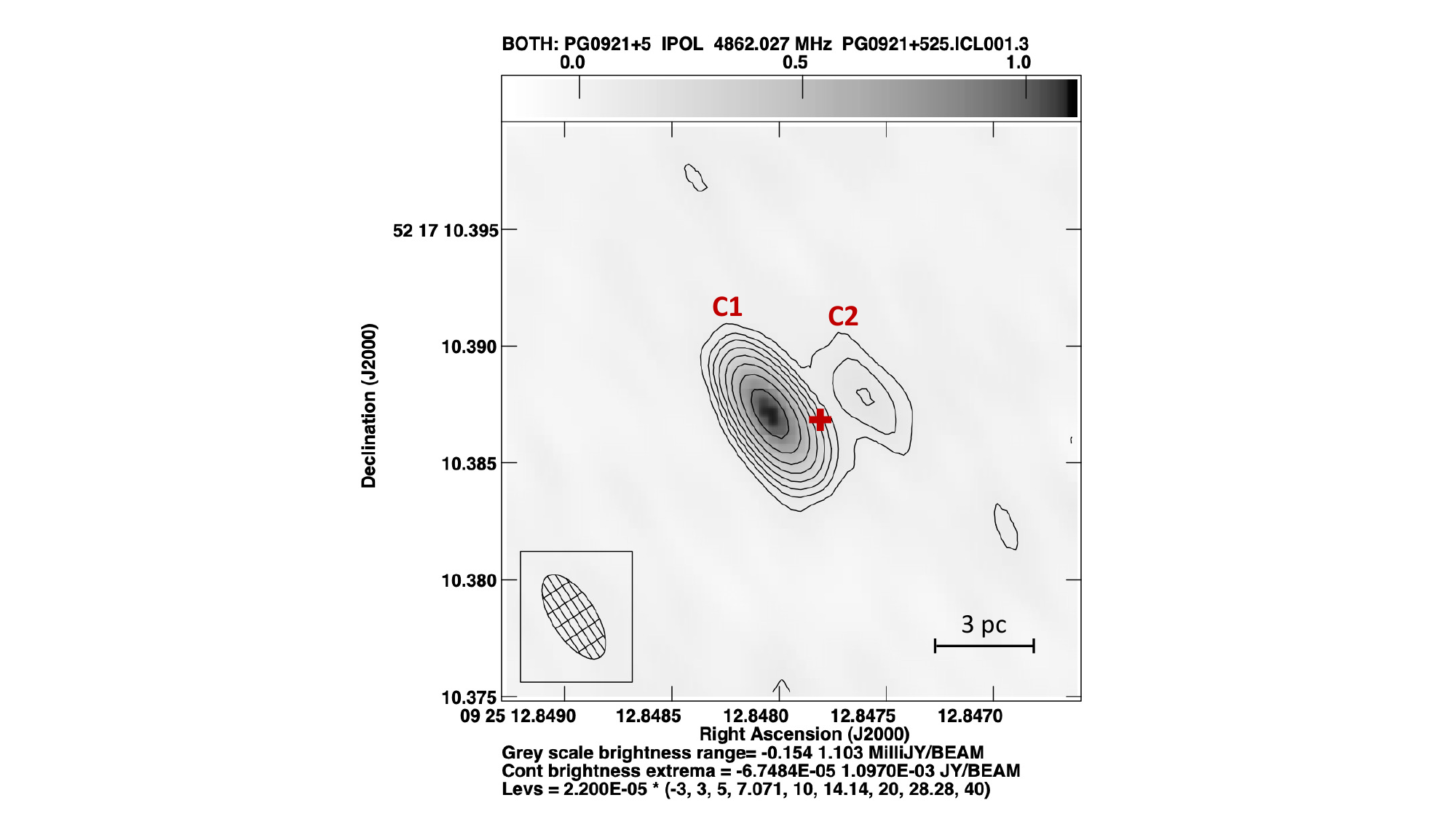}
\includegraphics[width=.48\textwidth, trim={8cm, 0cm, 8cm, 0cm}, clip]{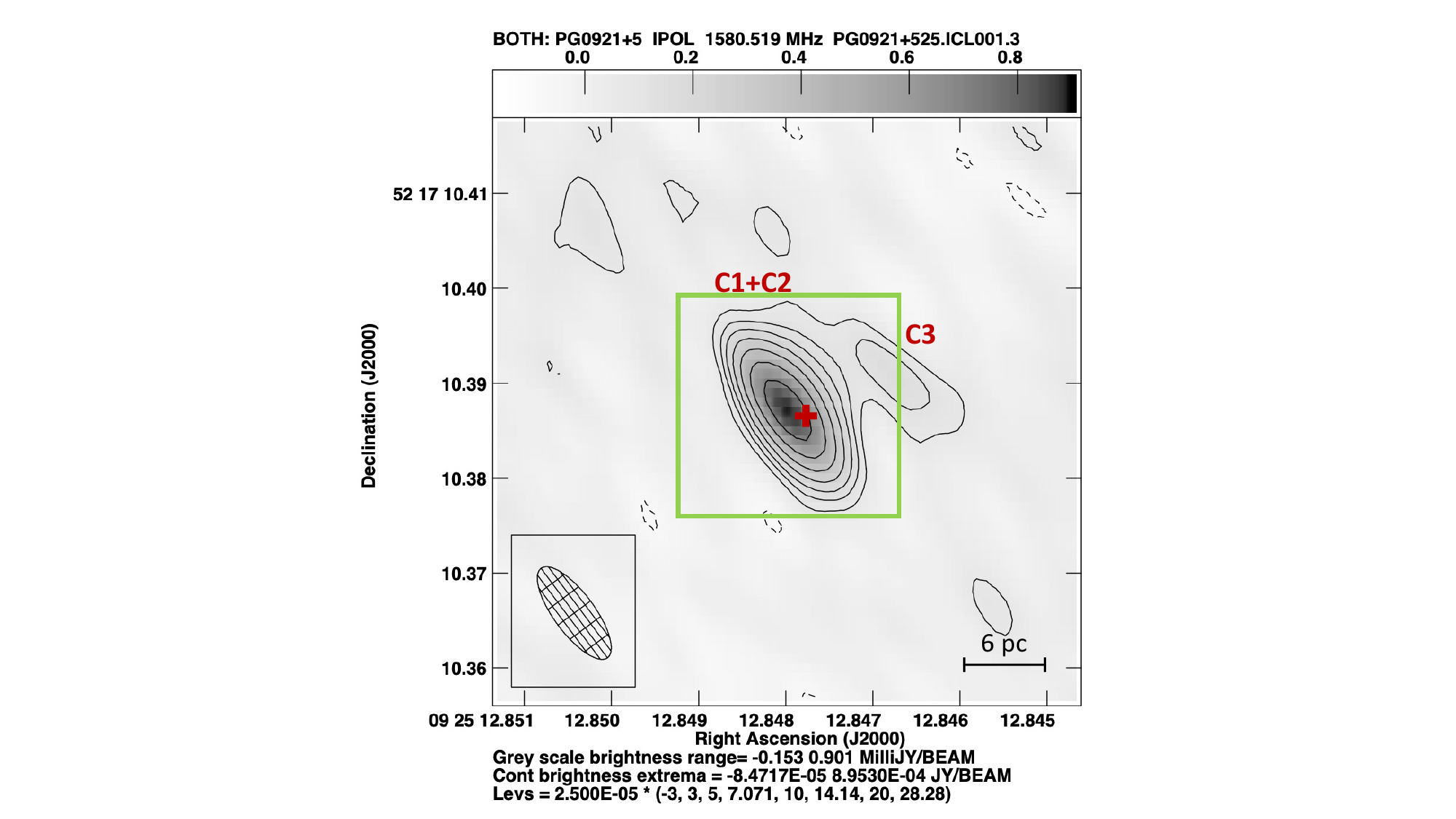}
\caption{PG0921+525}
\label{PG0921+525}
\end{subfigure}

\caption{The radio maps at 4.9 and 1.6\,GHz. The background noise RMS is reported in Table~\ref{flux}. The contour levels are at $-3, 3, 5 \times \sqrt{2}^n, n \in [0,15]$. The size and orientation of the synthesized beam is shown in the lower-left corner. The grey scale indicates the intensity in unit of Jy\,beam$^{-1}$ in a linear scale. The red crosses mark the {\it Gaia} position. The green squares at 1.6\,GHz mark the scale at 4.9\,GHz. If the object has more than one component, different components are marked.}
\label{maps}
\end{figure*}

\begin{figure*}
\ContinuedFloat

\begin{subfigure}{.75\textwidth}
\centering
\includegraphics[width=.48\textwidth, trim={8cm, 0cm, 8cm, 0cm}, clip]{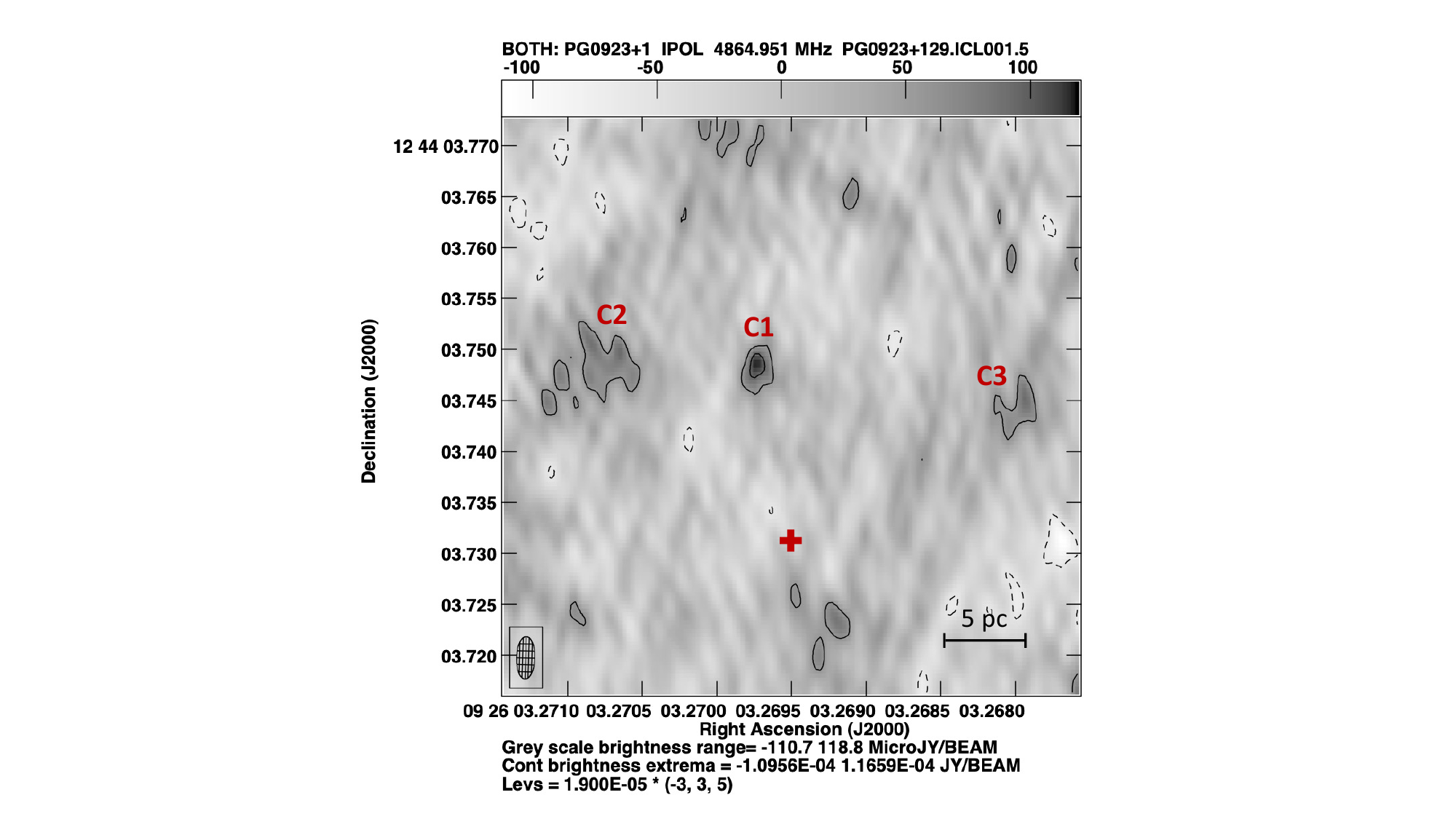}
\includegraphics[width=.48\textwidth, trim={8cm, 0cm, 8cm, 0cm}, clip]{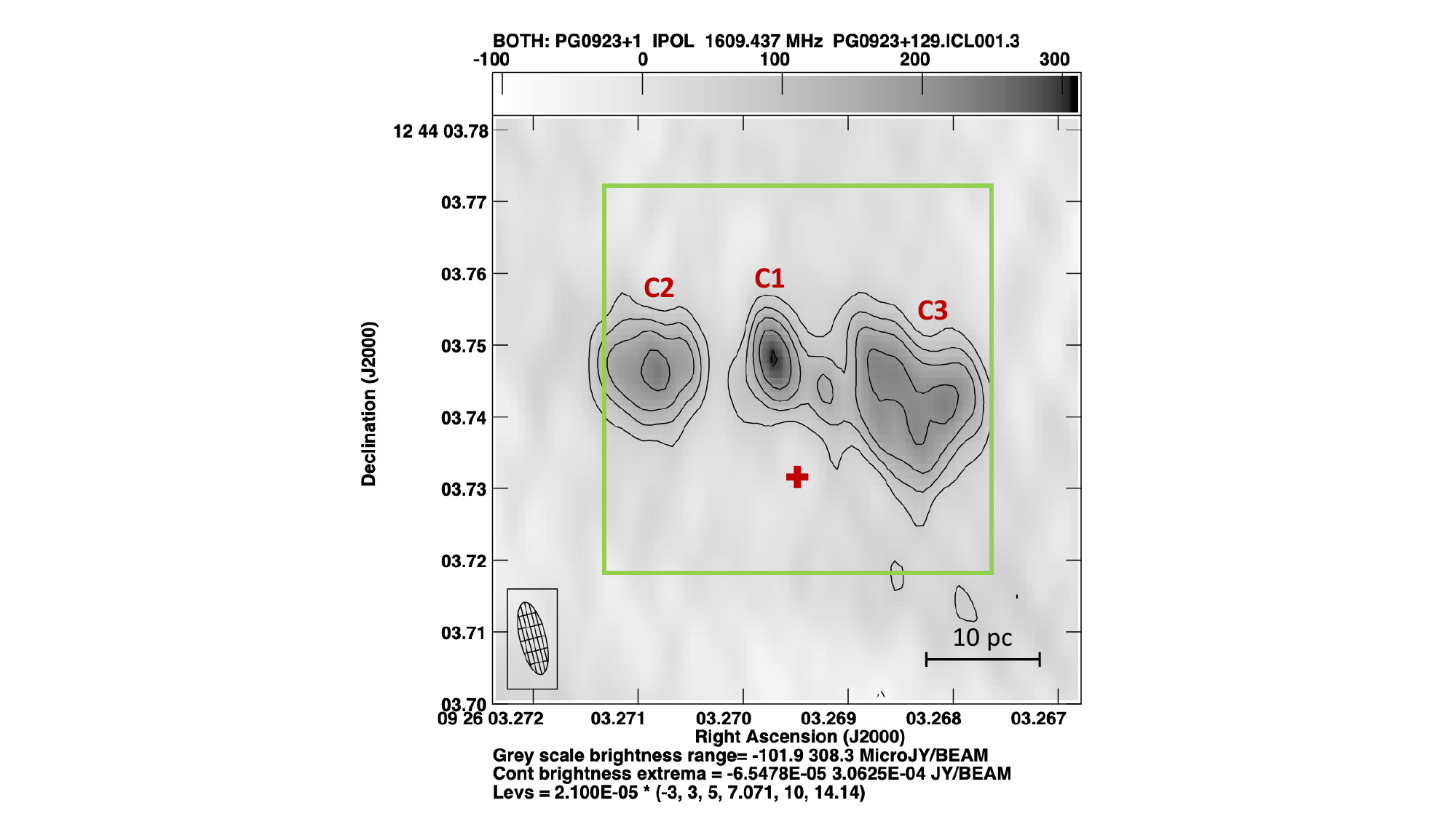}
\caption{PG0923+129}
\label{PG0923+129}
\end{subfigure}

\begin{subfigure}{.75\textwidth}
\centering
\includegraphics[width=.48\textwidth, trim={8cm, 0cm, 8cm, 0cm}, clip]{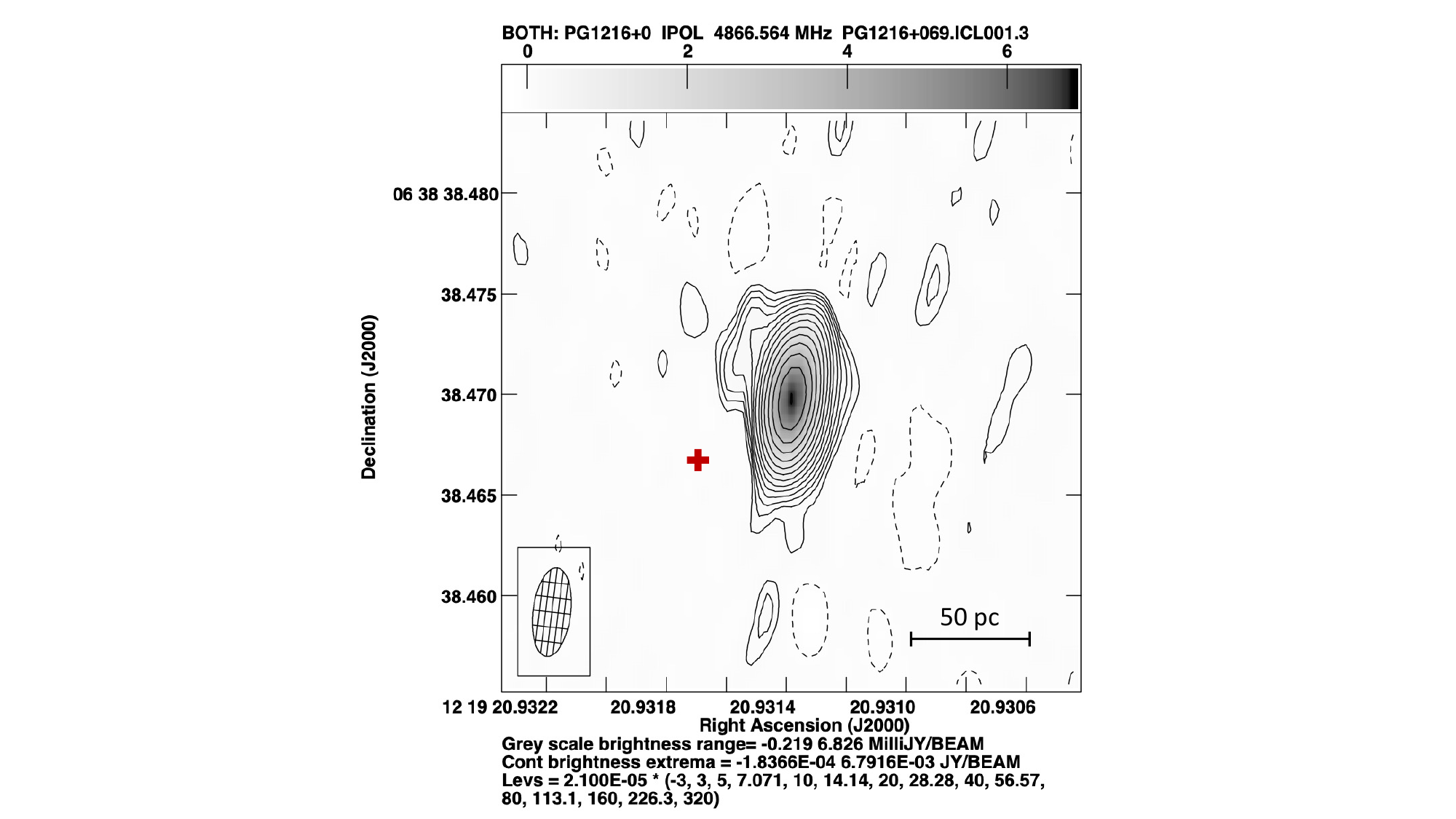}
\includegraphics[width=.48\textwidth, trim={8cm, 0cm, 8cm, 0cm}, clip]{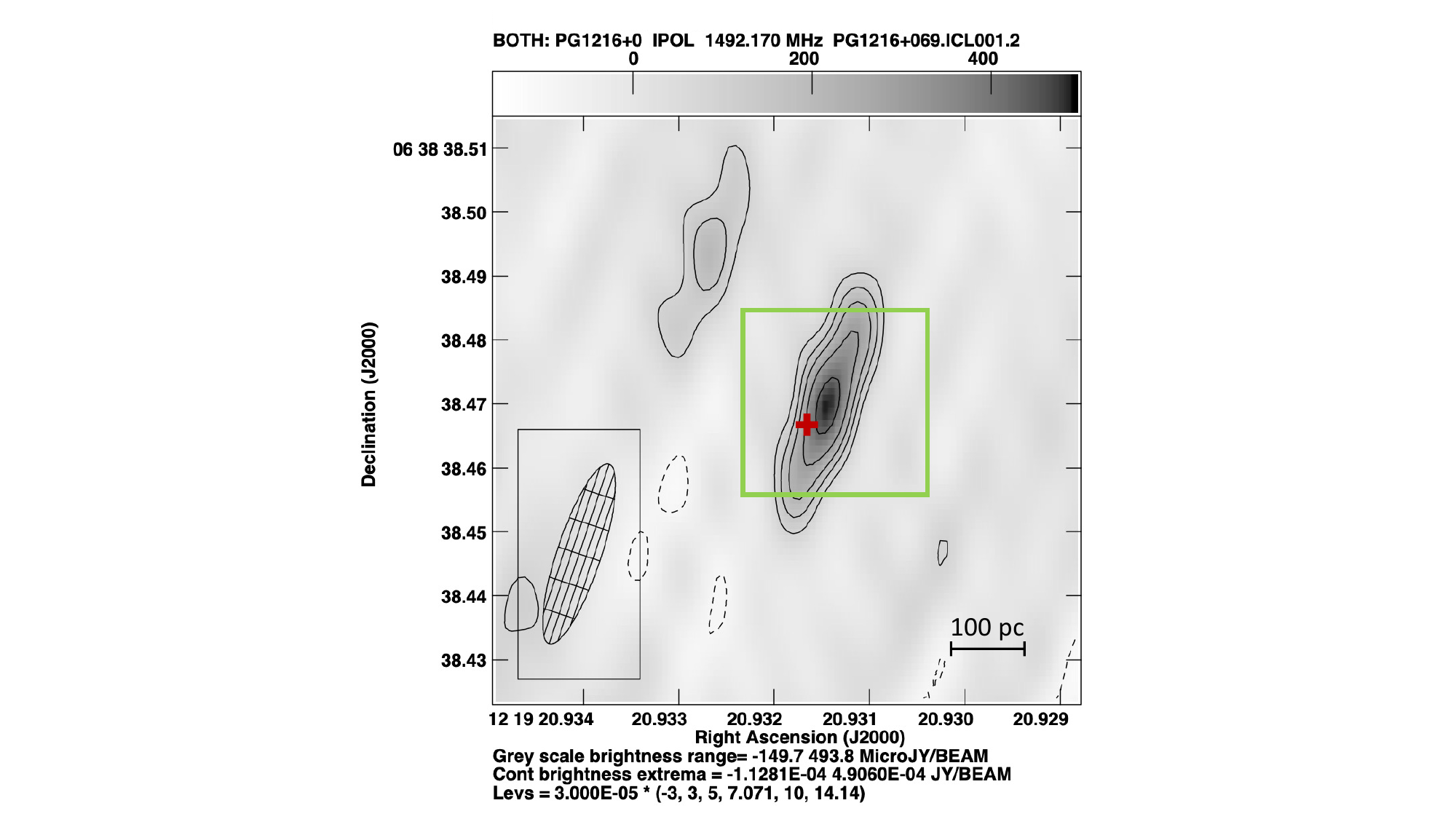}
\caption{PG1216+069}
\label{PG1216+069}
\end{subfigure}

\begin{subfigure}{.75\textwidth}
\centering
\includegraphics[width=.48\textwidth, trim={8cm, 0cm, 8cm, 0cm}, clip]{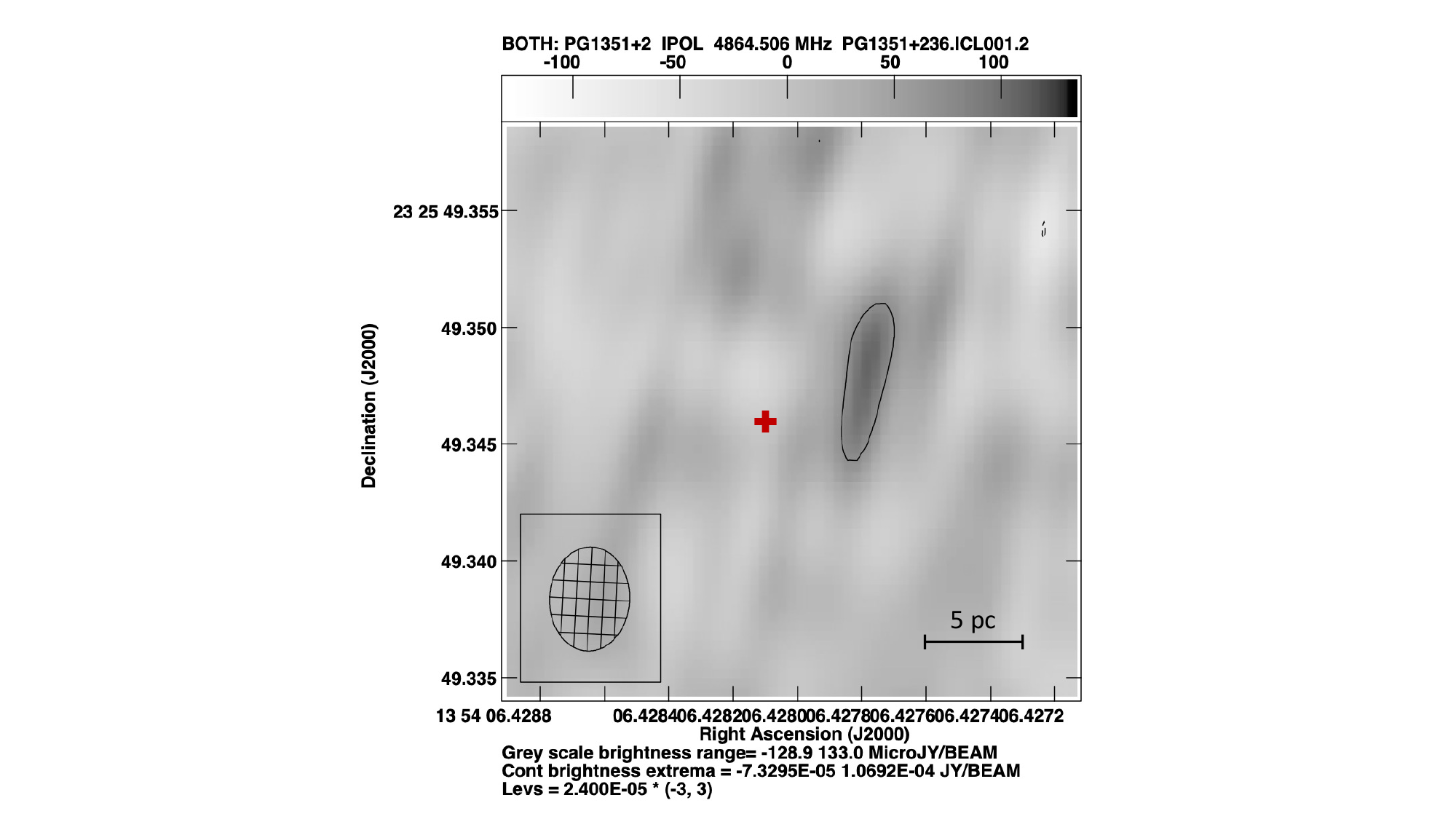}
\includegraphics[width=.48\textwidth, trim={8cm, 0cm, 8cm, 0cm}, clip]{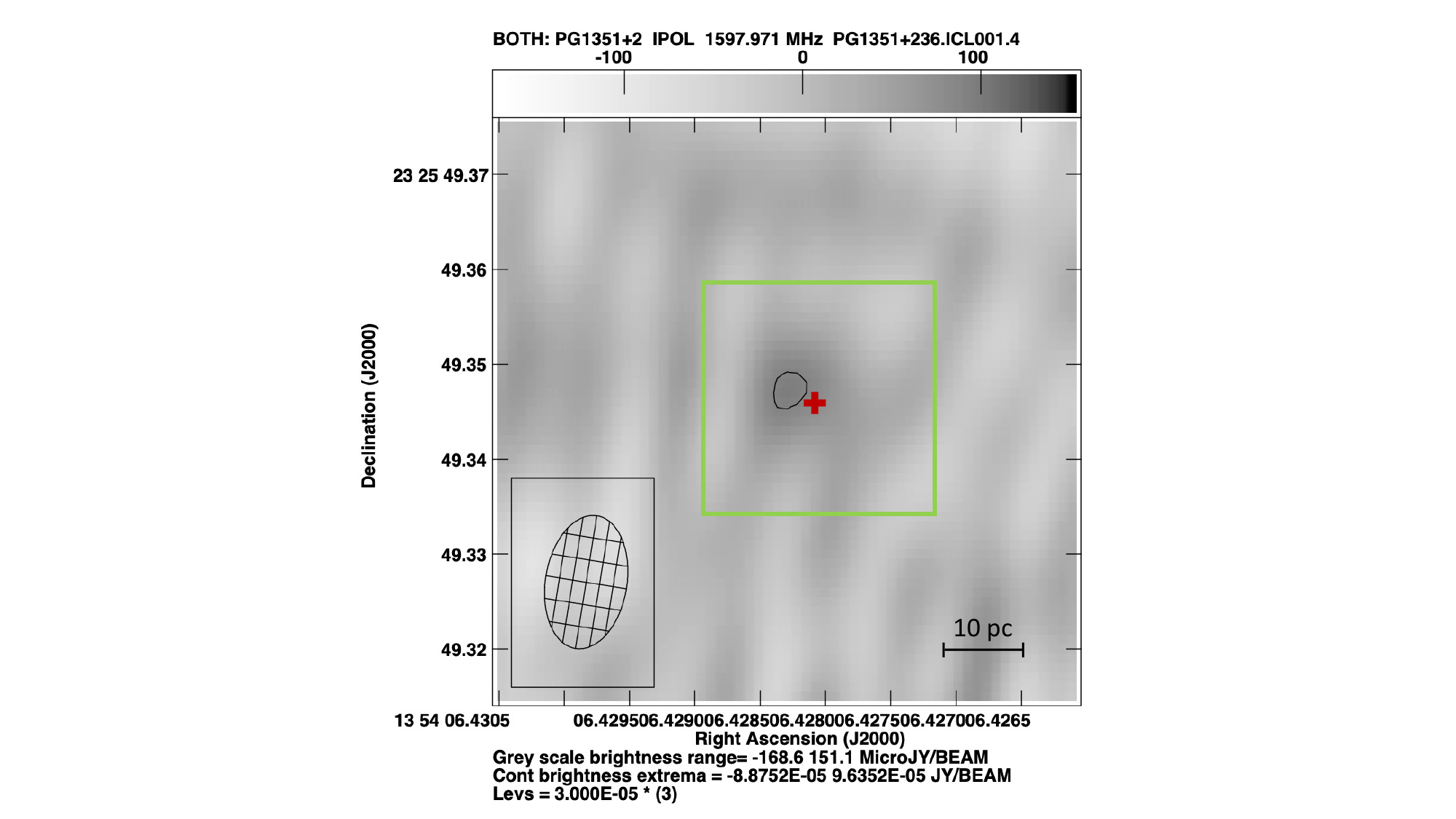}
\caption{PG1351+236}
\label{PG1351+236}
\end{subfigure}

\caption{Continued.}
\end{figure*}

\begin{figure*}
\ContinuedFloat

\begin{subfigure}{.75\textwidth}
\centering
\includegraphics[width=.48\textwidth, trim={8cm, 0cm, 8cm, 0cm}, clip]{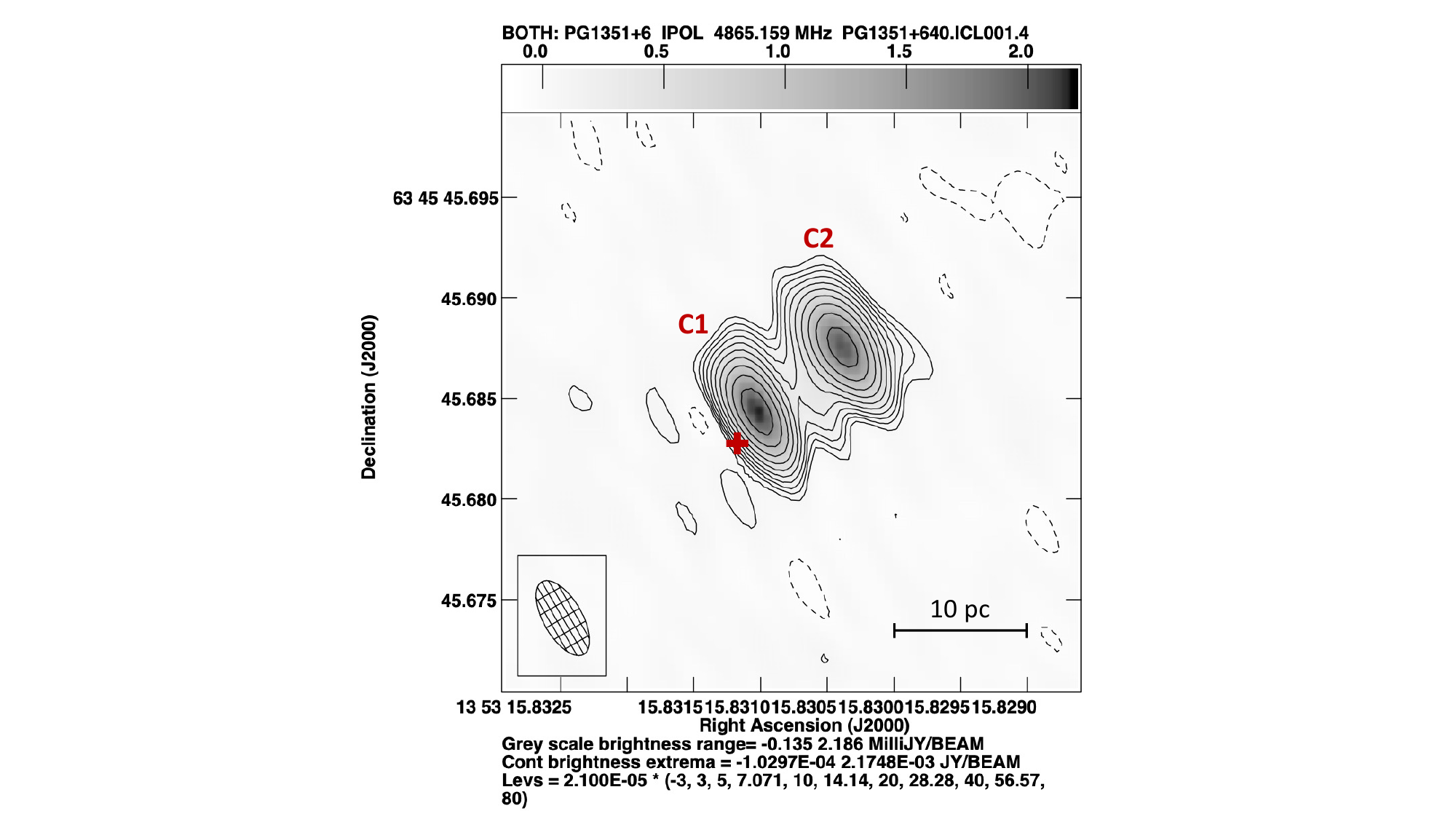}
\includegraphics[width=.48\textwidth, trim={8cm, 0cm, 8cm, 0cm}, clip]{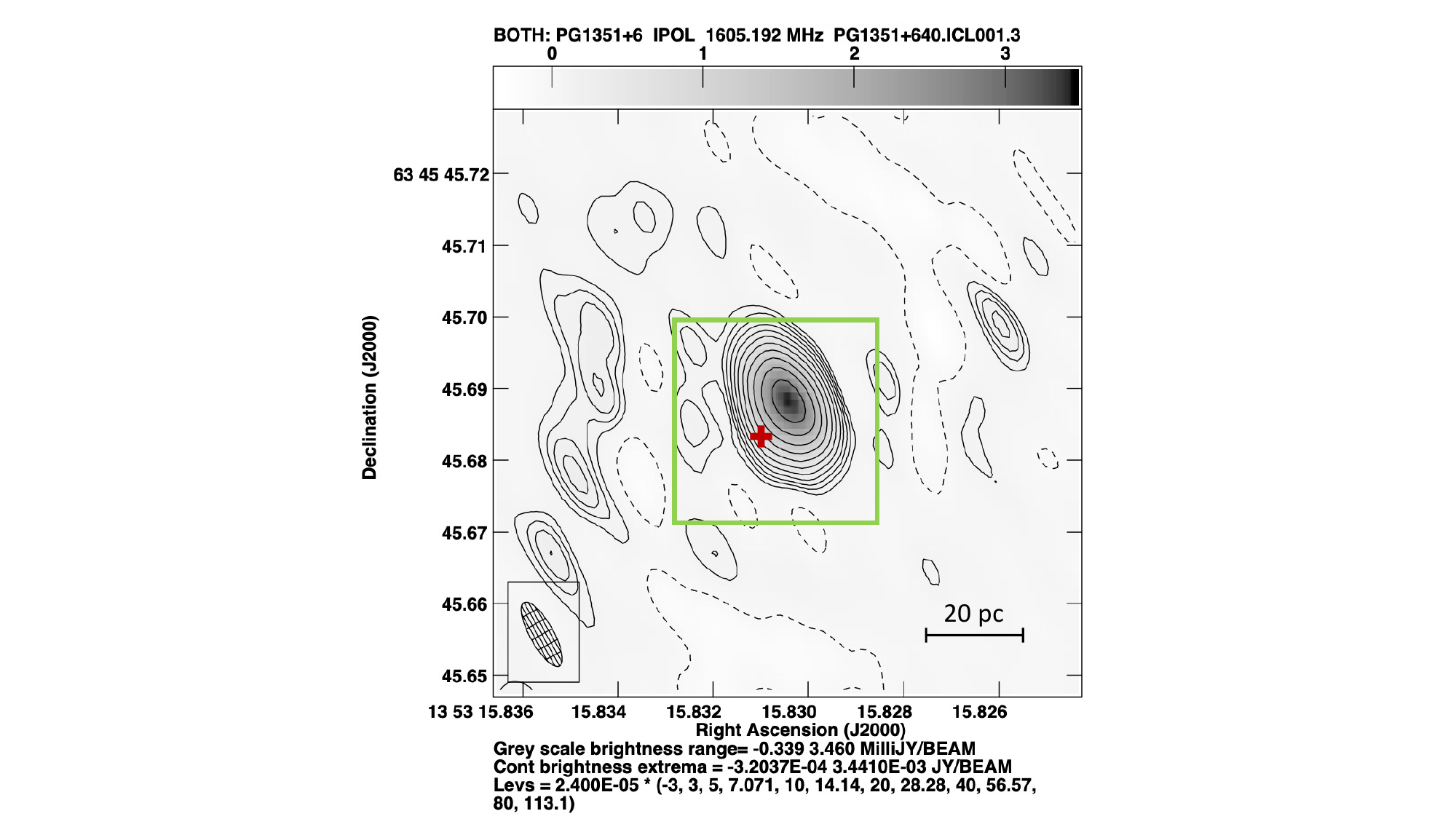}
\caption{PG1351+640}
\label{PG1351+640}
\end{subfigure}

\begin{subfigure}{.75\textwidth}
\centering
\includegraphics[width=.48\textwidth, trim={8cm, 0cm, 8cm, 0cm}, clip]{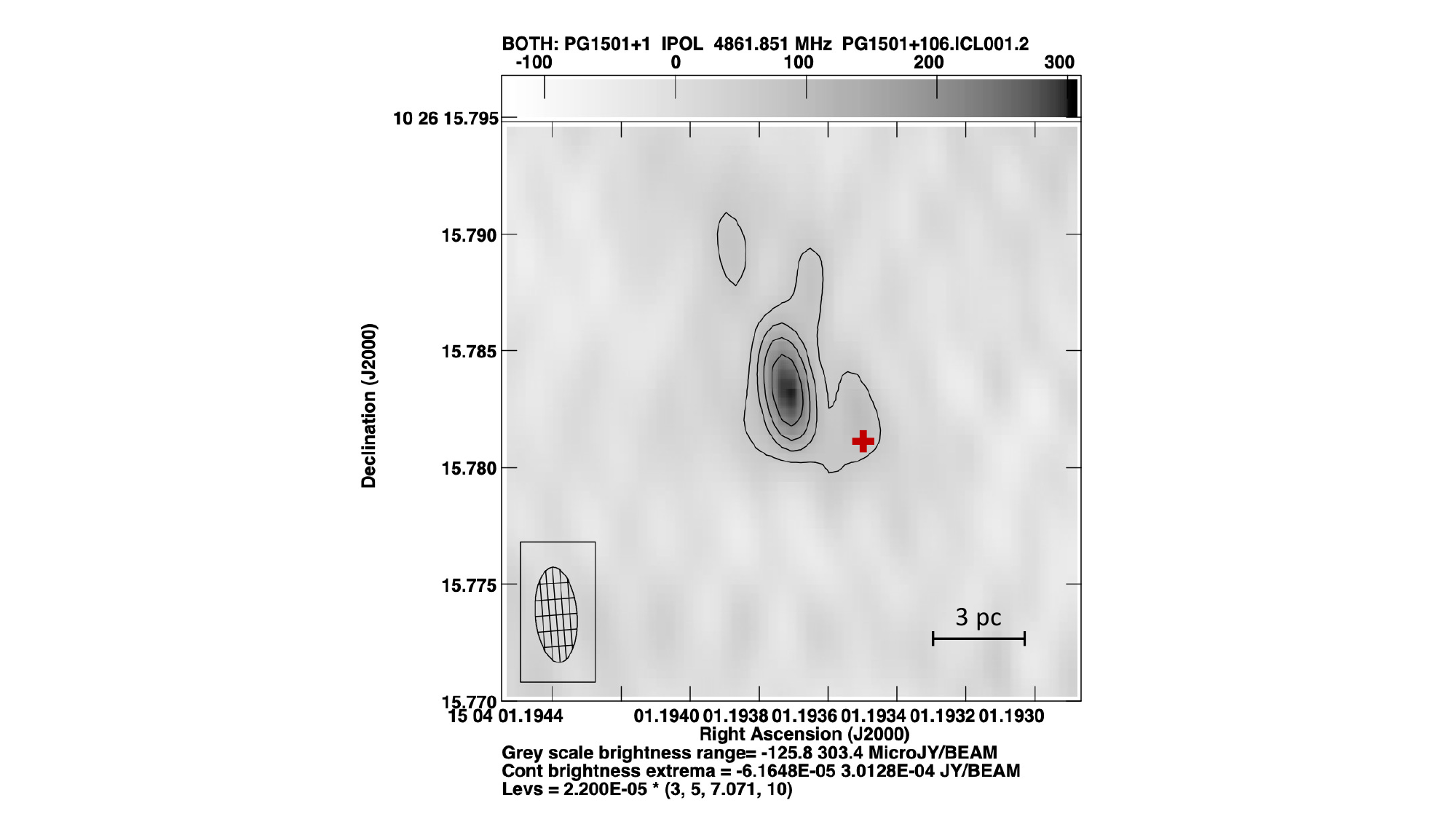}
\includegraphics[width=.48\textwidth, trim={8cm, 0cm, 8cm, 0cm}, clip]{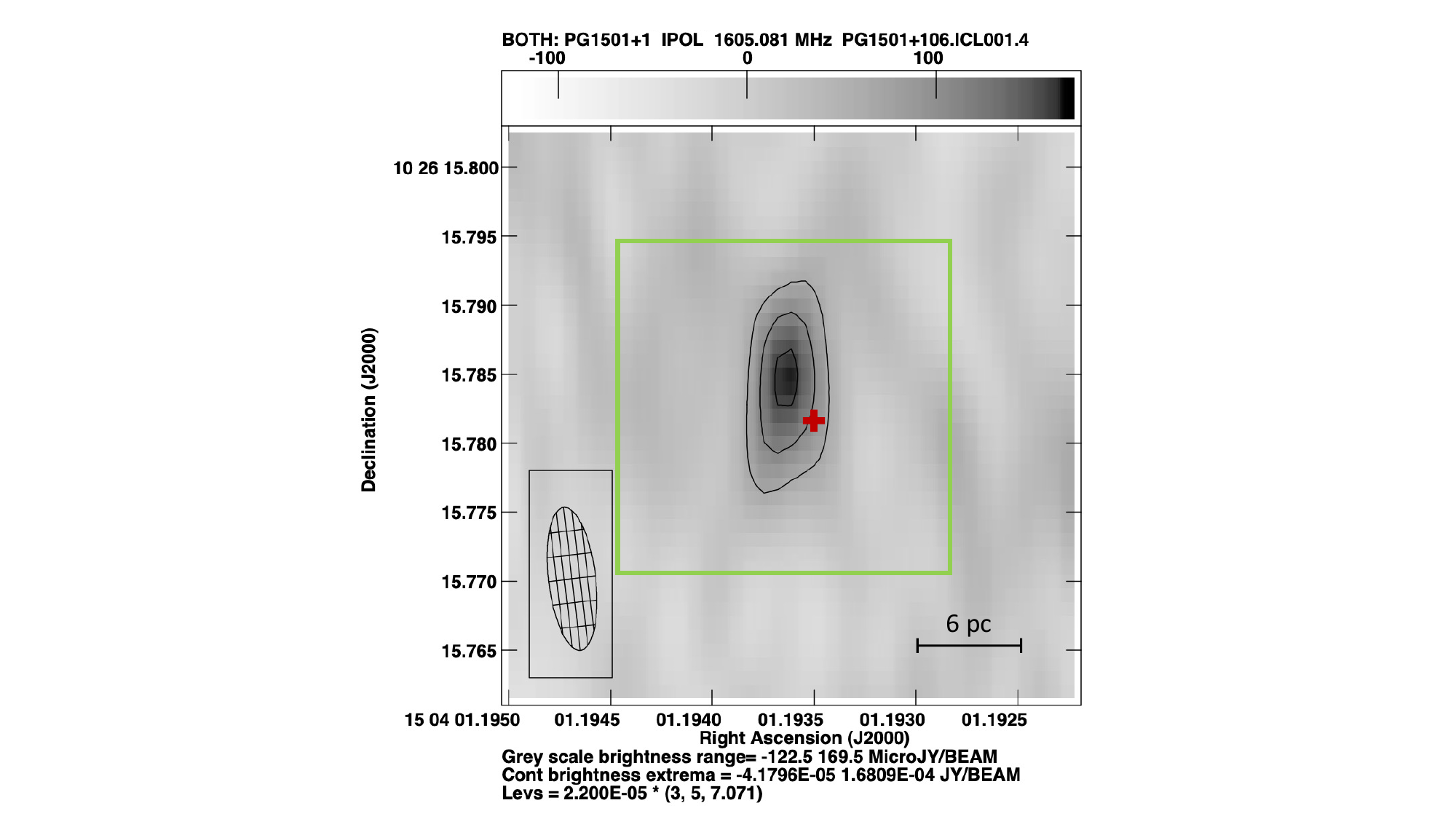}
\caption{PG1501+106}
\label{PG1501+106}
\end{subfigure}

\begin{subfigure}{.75\textwidth}
\centering
\includegraphics[width=.48\textwidth, trim={8cm, 0cm, 8cm, 0cm}, clip]{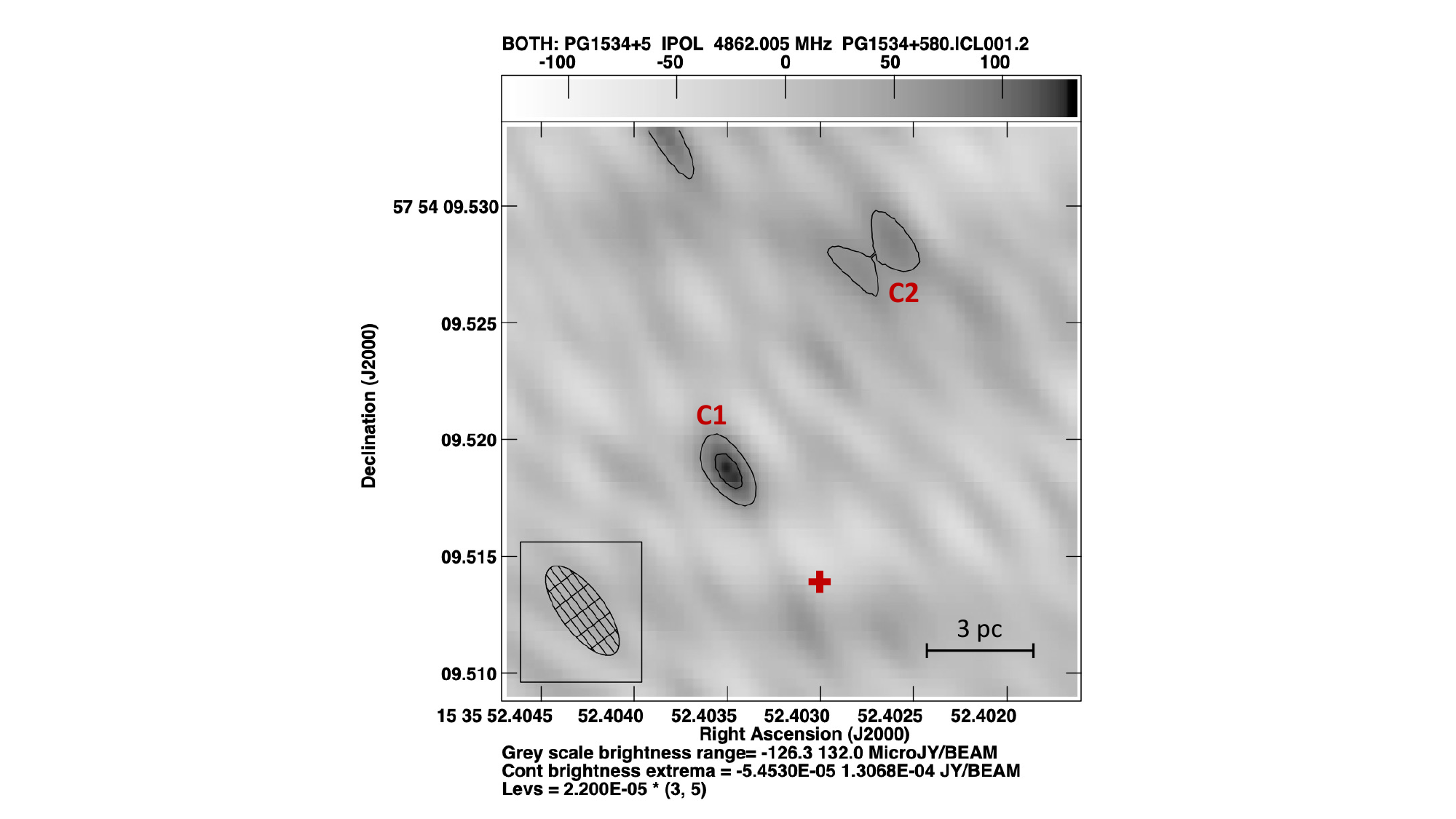}
\includegraphics[width=.48\textwidth, trim={8cm, 0cm, 8cm, 0cm}, clip]{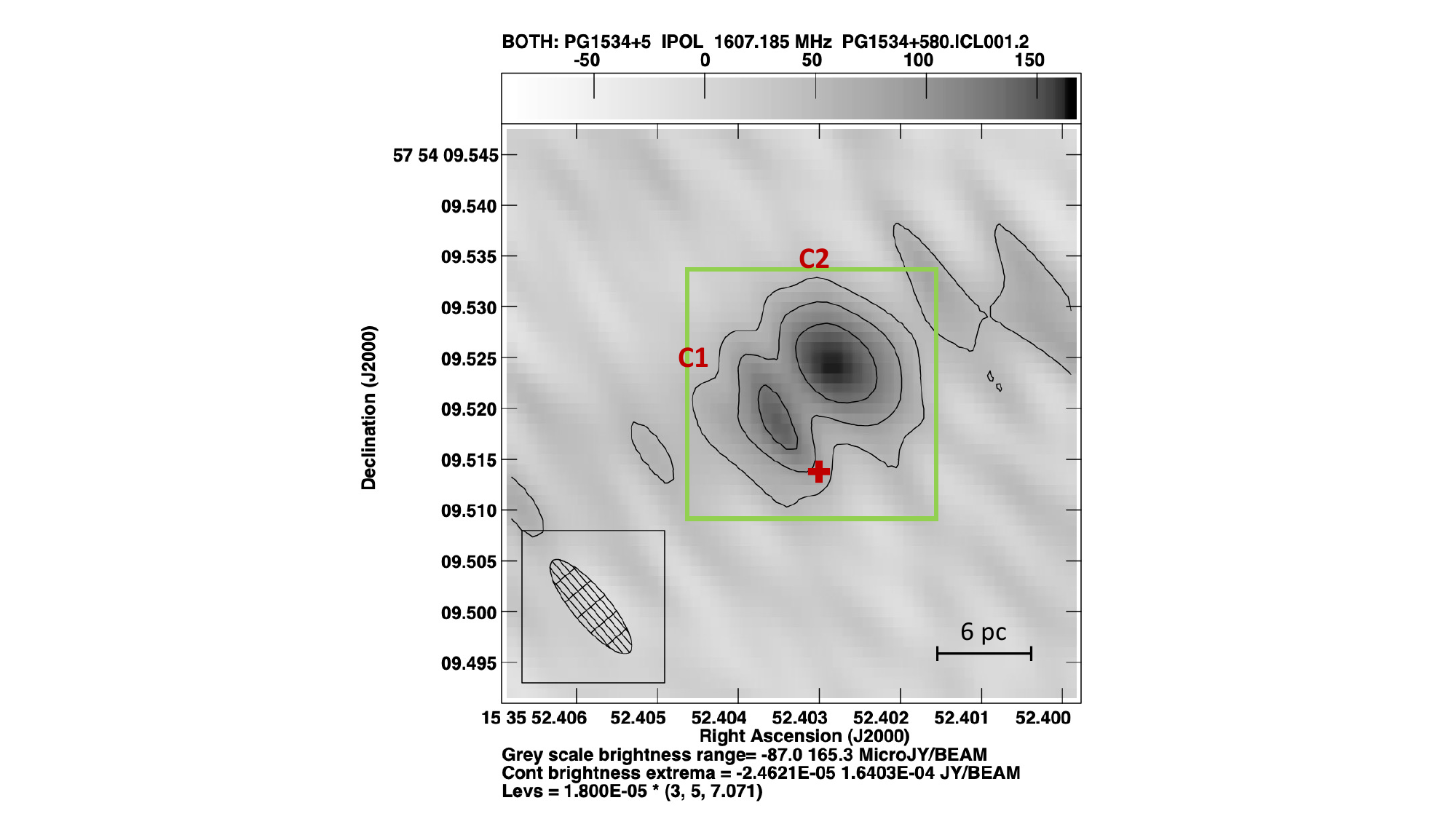}
\caption{PG1534+580}
\label{PG1534+580}
\end{subfigure}

\caption{Continued.}
\end{figure*}

\begin{figure*}
\ContinuedFloat

\begin{subfigure}{.75\textwidth}
\centering
\includegraphics[width=.48\textwidth, trim={8cm, 0cm, 8cm, 0cm}, clip]{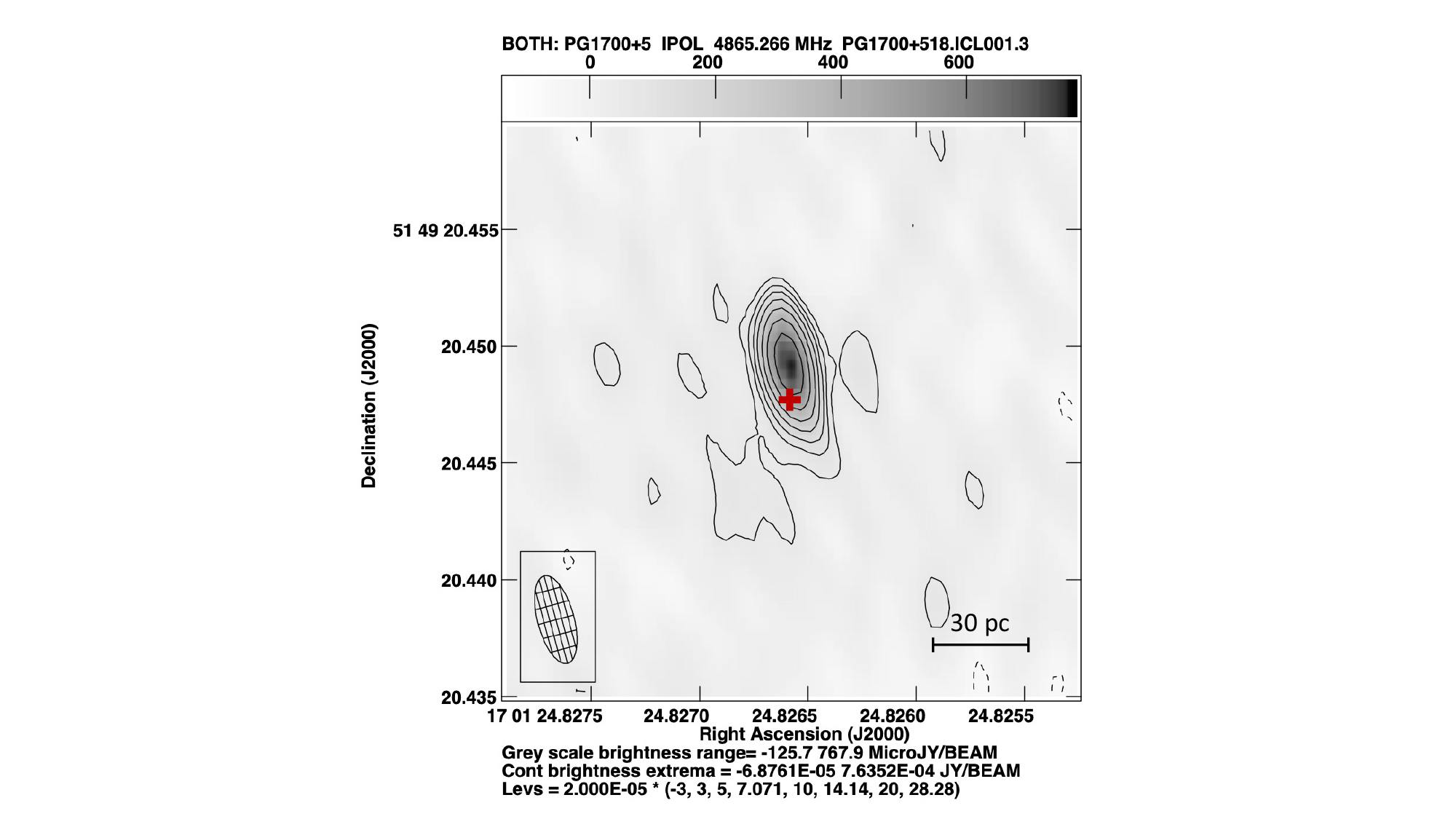}
\includegraphics[width=.48\textwidth, trim={8cm, 0cm, 8cm, 0cm}, clip]{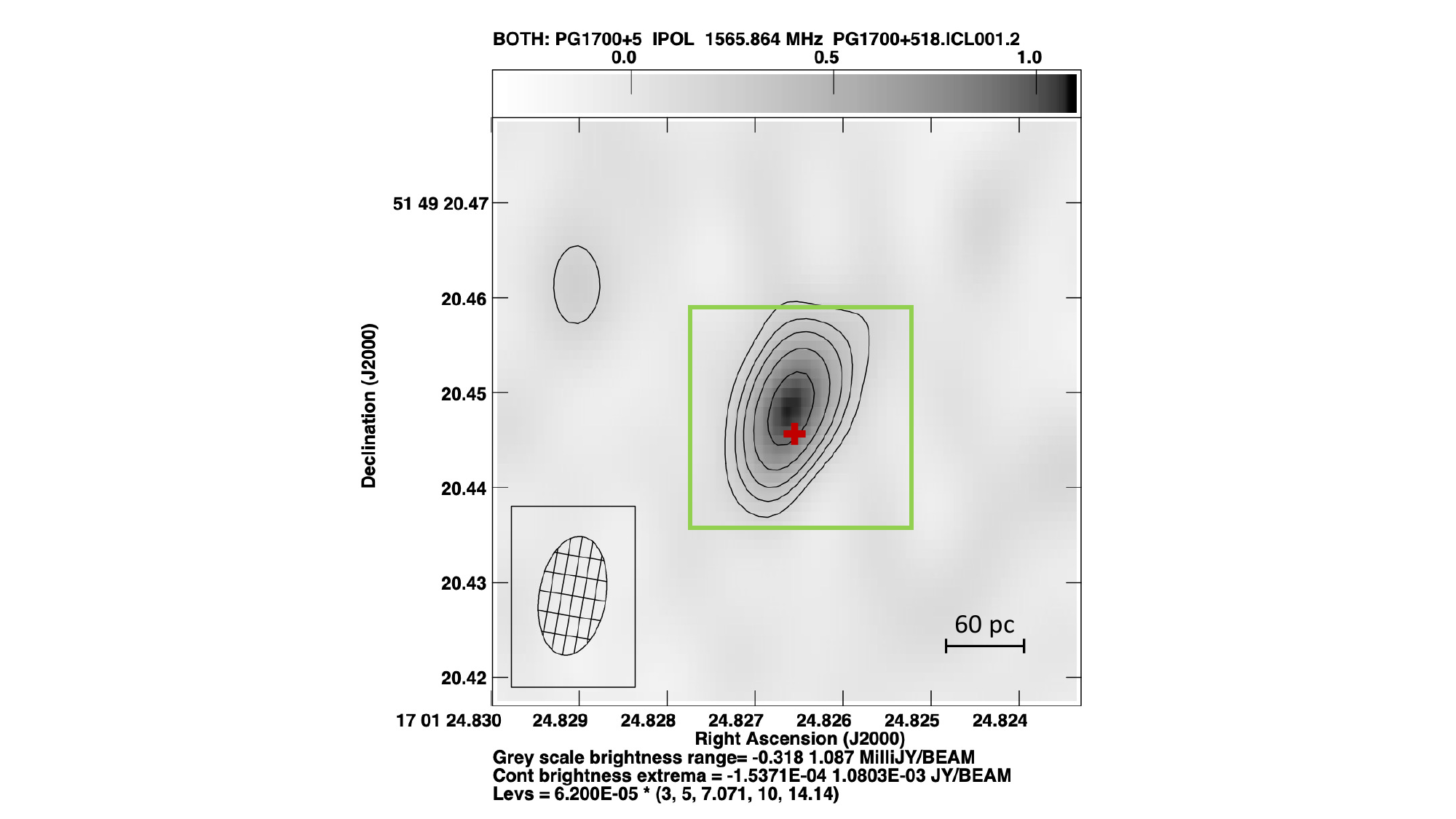}
\caption{PG1700+518}
\label{PG1700+518}
\end{subfigure}

\caption{Continued.}
\end{figure*}

\begin{figure*}

\begin{subfigure}{.45\textwidth}
\centering
\includegraphics[width=\textwidth, trim={8cm, 0cm, 8cm, 0cm}, clip]{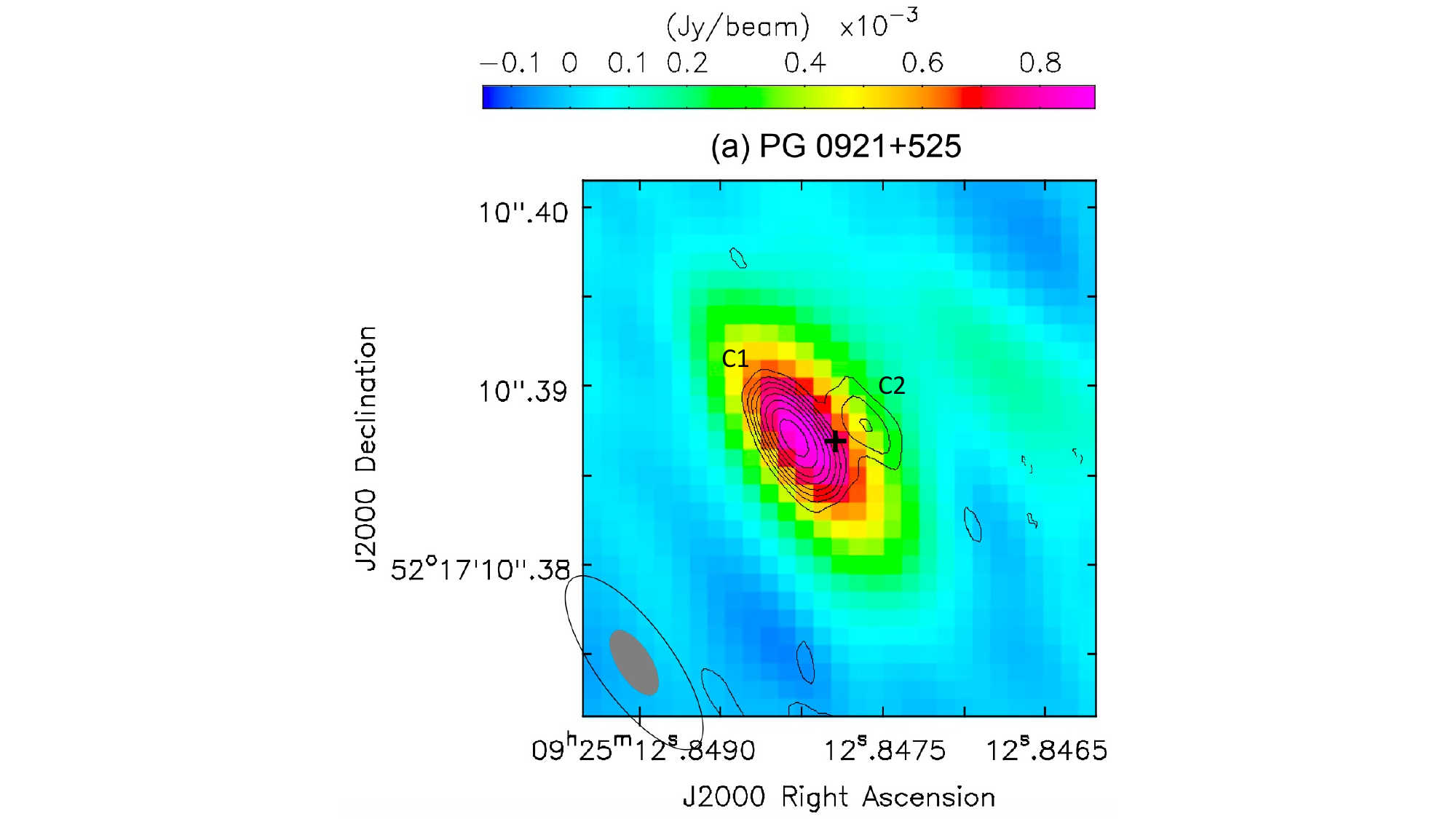} 
\label{PG0921_overlap}
\end{subfigure}
\begin{subfigure}{.45\textwidth}
\centering
\includegraphics[width=\textwidth, trim={8cm, 0cm, 8cm, 0cm}, clip]{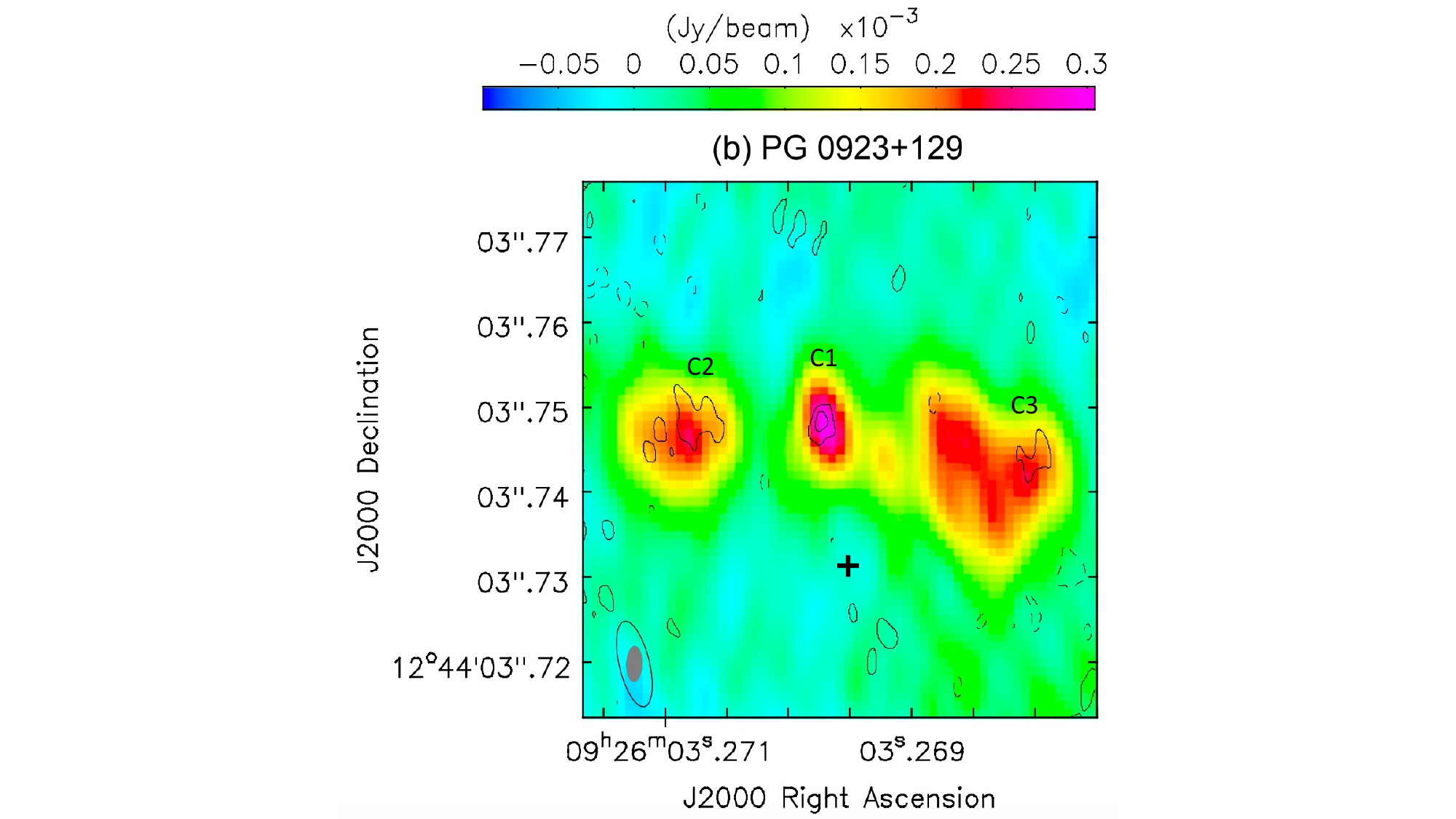} 
\label{PG0923_overlap}
\end{subfigure}

\begin{subfigure}{.45\textwidth}
\centering
\includegraphics[width=\textwidth, trim={8cm, 0cm, 8cm, 0cm}, clip]{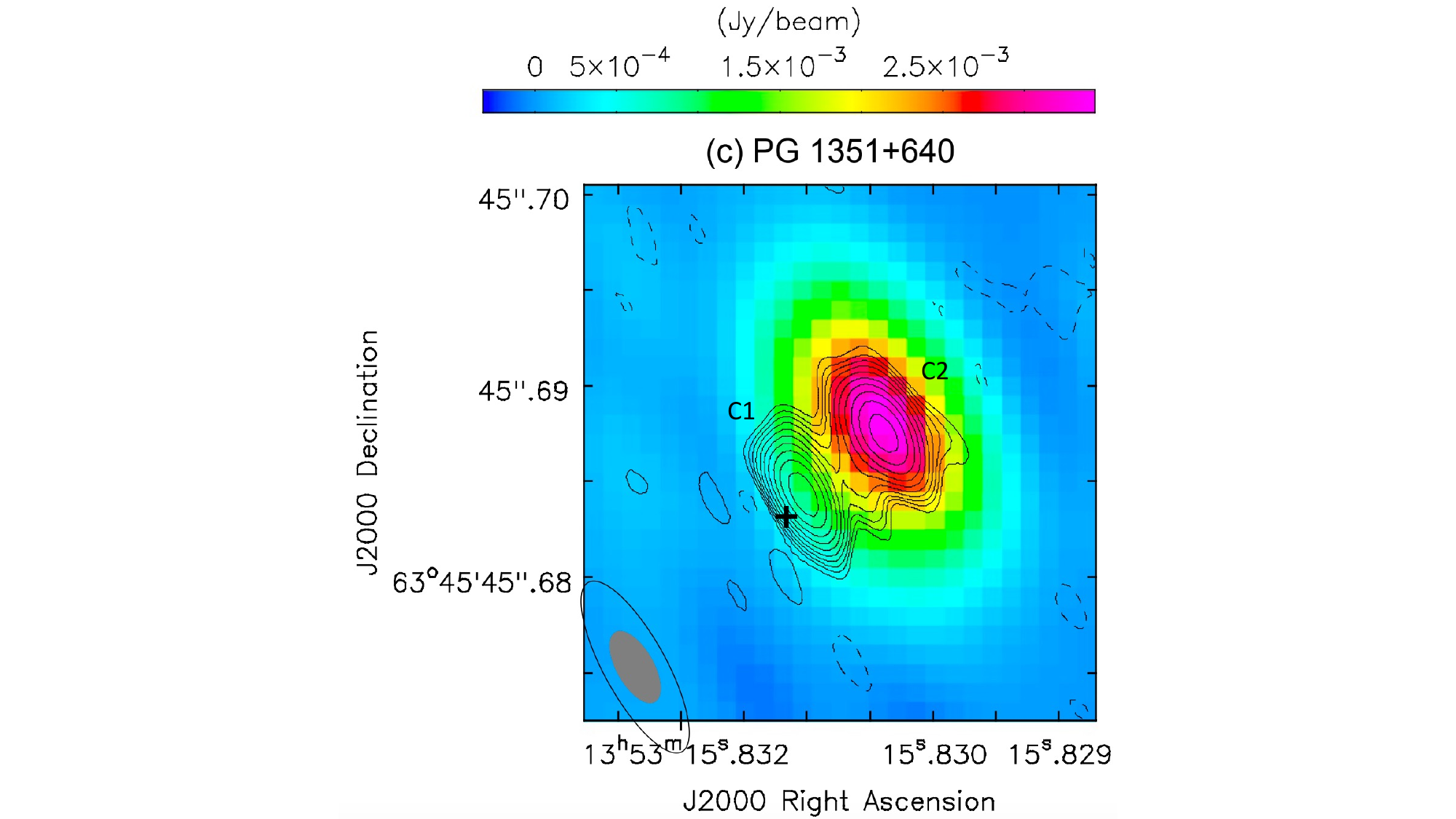} 
\label{PG1351_overlap}
\end{subfigure}
\begin{subfigure}{.45\textwidth}
\centering
\includegraphics[width=\textwidth, trim={8cm, 0cm, 8cm, 0cm}, clip]{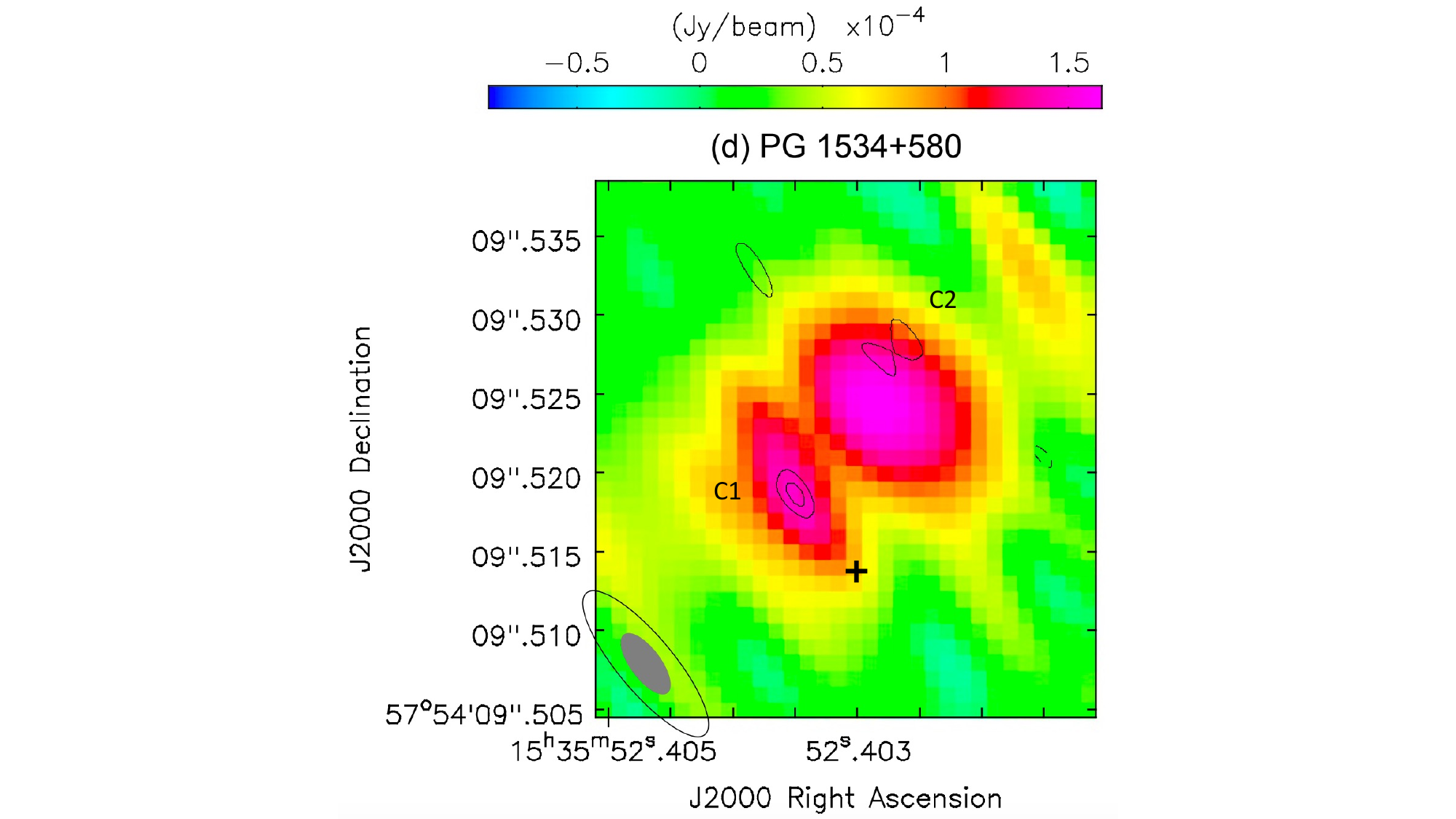} 
\label{PG1534_overlap}
\end{subfigure}

\caption{An overlap of the radio maps at 1.6 and 4.9\,GHz for the objects with multiple components. The colour scales show the 1.6\,GHz images. The contour levels, which are the same as Figure~\ref{maps}, represent the 4.9\,GHz images. The size and orientation of the synthesized beam in both bands is shown in the lower-left corner. The black crosses mark the {\it Gaia} position. The components are also labelled.}
\label{overlap}
\end{figure*}

We consider a 5$\sigma$ level as the detection criterion, where $\sigma$ is the background noise, and a 5$\sigma$ upper limit on the flux density if the source is not detected.
Figure~\ref{maps} presents the radio maps of the 10 RQ PG quasars observed with the VLBA at 1.6 and 4.9\,GHz.
The objects are generally detected in both bands, excluding PG0026+129 which is detected only at 4.9\,GHz, and PG1351+236 which is not detected in either bands.
Four objects (PG0921+525, PG0923+129, PG1351+640, and PG1534+580) exhibit two or three components, which are marked in Figure~\ref{maps}.
Figure~\ref{overlap} shows an overlap of the 1.6 and 4.9\,GHz images of these four objects, in order to see the alignment of different components in the two bands.
Table~\ref{position} lists the VLBA positions of each of the components.
The C band coordinates are listed if the source is detected in this band. Otherwise, the L band coordinates are listed.
The distances between the VLBA and the {\it Gaia} positions \citep{Gaia2016,Gaia2022} are also reported in Table~\ref{position}.
The astrometric uncertainty of the {\it Gaia} DR3 with respect to the optical spectroscopy is about 1--10\,mas \citep{Khamitov2023}.
For objects with more than one component, we identify the closest component to the {\it Gaia} position as the core of the AGN (labelled as C1), where the supermassive BH resides.

Table~\ref{size} lists the sizes of the synthesized beam and the source before and after deconvolution of the full resolution maps.
If the deconvolved source size is smaller than half of the beam size, we consider the source as unresolved.
Table~\ref{flux} reports the flux densities of the full resolution maps and the tapered maps in the L and C bands.
For sources with only one component, we use the peak intensity, which is the unresolved flux density in a single beam, as the core flux density, and the total flux density is the greater value between the peak intensity and the integrated flux density.
For sources with more than one component, we use the peak intensity of the core component as the core flux density, and the total flux density is the sum of the total flux density of all components.
If an object or a component is not detected, we use the 5$\sigma$ upper limit on the flux density.
The extended flux density of each source is defined as the total flux density minus the core flux density.
The extended flux density is used only if it is larger than a 3$\sigma_{\rm err}$ level, where $\sigma_{\rm err}$ is the uncertainty of the extended flux density given in the error propagation.
Otherwise, a 3$\sigma_{\rm err}$ upper limit on the extended flux density is used.

Table~\ref{literature} lists the BH mass, $M_{\rm BH}$, and the bolometric luminosity, $L_{\rm bol}$, of the objects \citep{Davis2011,Laor2019}.
The Eddington ratio, $L/L_{\rm Edd}$, is calculated using $M_{\rm BH}$ and $L_{\rm bol}$, where $L_{\rm Edd} = 1.3 \times 10^{38} (M_{\rm BH}/M_{\odot})$.
Table~\ref{literature} also reports the radio to X-ray luminosity ratio, $L_{\rm R}/L_{\rm X}$, using both the VLBA core flux and the VLA A configuration core flux at 5\,GHz. The X-ray flux at 0.2--12.0\,keV is from the {\it XMM-Newton} DR12 catalog \citep{Webb2020}.
If the object is not detected with the VLBA, we set an upper limit on the VLBA $L_{\rm R}/L_{\rm X}$ ratio.
PG1351+236 is not detected in either radio or X-ray, so its flux ratio is unknown.

Table~\ref{slope} lists the VLBA spectral slopes between 1.6 and 4.9\,GHz of the core flux density $\alpha_{\rm core}$, the total flux density $\alpha_{\rm total}$, and the extended flux density $\alpha_{\rm extended}$, for each object.
Table~\ref{slope} also includes $\alpha_{\rm total}$ of the individual components in four of the nine objects where more than one component is detected.
The spectral slopes are measured based on the flux densities in the tapered maps, which have comparable resolutions and cover emission on similar scales at both 1.6 and 4.9\,GHz.
PG1351+236 is not detected in either bands, and is therefore not listed.
The additional extended components are often detected in only one band.
Specifically, in PG0921+525, C3 is not detected at 4.9\,GHz, and C1 and C2 are detected in the 4.9\,GHz full resolution map and unresolved in the 4.9\,GHz tapered map and the 1.6\,GHz maps.
In PG0923+129, all three components are detected at 1.6\,GHz, but only C1 is detected at 4.9\,GHz.
PG1351+640 exhibits two components at 4.9\,GHz, but only a single component at 1.6\,GHz, most likely because the separation between C1 and C2 of $\sim 5$\,mas is too small to be resolved at the 1.6\,GHz resolution.
We identify C1 as the core because it is closer to the {\it Gaia} position (Table~\ref{position}).
In addition, the 1.6\,GHz flux of C2 is one order of magnitude higher than that of C1, while their 4.9\,GHz fluxes are comparable (Table~\ref{flux}), which suggests that the slope of C1 is flatter than that of C2 (Table~\ref{slope}).
In PG1534+580, the two components are detected at 1.6\,GHz, but only C1 is detected at 4.9\,GHz.

Table~\ref{slope} reports the brightness temperature of all components
\begin{equation}
T_{\rm B} = 1.8 \times 10^{9} (1+z) \frac{S_\nu}{\nu^2 \theta_{\rm max} \theta_{\rm min}}
\end{equation}
\citep[e.g.][]{Ulvestad2005}, where $S_\nu$ is the flux density in mJy, $\nu$ is the observing frequency in GHz, and $\theta_{\rm max}$ and $\theta_{\rm min}$ are the major and minor axes of the source size in mas.
We compute the brightness temperature at 4.9\,GHz, when available, which provides a higher resolution.
Otherwise, we calculate the brightness temperature at 1.6\,GHz.
In objects where $\theta_{\rm max}$ and/or $\theta_{\rm min}$ are smaller than half of the beam size, the measured $T_{\rm B}$ is a lower limit.

Table~\ref{slope} lists the VLBA core to total flux ratio, and the VLBA total flux to the VLA A configuration core flux ratio, both at 5\,GHz.
These ratios provide a measure of the compactness of the mas and arcsec scale emission.
PG0921+525, PG0923+129, and PG1534+580 have an upper limit on the VLBA total flux density given that part of the extended emission is undetected at 5\,GHz.
This yields a lower limit on the VLBA core/total flux ratio and an upper limit on the VLBA/VLA flux ratio.

Five of the eight objects observed by \citet{Alhosani2022} were detected, of which four (PG0052+251, PG1149$-$110, PG1612+261, and PG2304+042) have one component, and one (PG0050+124) has two components, where C1 and C2 are the extended and the core components (reverse of the convention here).
PG1612+261 was only detected in the L band, and the other four objects were detected in both L and C bands.
A taper was applied by \citet{Alhosani2022} in the L band with a limit on the minimum baseline, to recover a comparable scale of emission in both bands, but it was not applied in the C band with a limit on the maximum baseline, to make comparable resolution images in both bands.
We use the spectral slopes in \citet{Alhosani2022}, which were computed using the total flux density of each component, as their $\alpha_{\rm total}$.
The $\alpha_{\rm total}$ of all components in PG0050+124 is calculated using the sum of the total flux density of the two components.
We do not use the peak intensity in \citet{Alhosani2022} to estimate the spectral slopes, $\alpha_{\rm core}$ and $\alpha_{\rm extended}$, as they are biased due to the different resolutions.
The VLBA core/total flux ratio at 5\,GHz in PG0052+251, PG1149$-$110, and PG2304+042, is estimated using the peak intensity and the total flux density.
In PG0050+124, the VLBA core/total flux ratio is the core flux (C2) to the total flux of the two components (C1+C2).
PG1612+261 was not detected in the C band, its VLBA core/total flux ratio is thus not available.
The other parameters, including the distance between the VLBA and the {\it Gaia} positions (Table~\ref{position}), the VLBA and the VLA radio to X-ray luminosity ratios (Table~\ref{literature}), the brightness temperature, and the VLBA/VLA flux ratio at 5\,GHz (Table~\ref{slope}) are calculated as described above.

The sample extends over a wide range of BH mass $6.8 \lesssim \log M_{\rm BH}/M_{\odot} \lesssim 9.1$, luminosity $44.47 \lesssim \log L_{\rm bol} \lesssim 46.61$, and Eddington ratio $-1.64 \lesssim \log L/L_{\rm Edd} \lesssim +0.43$.
The VLBA flux densities $S_{\rm total}$ are in the range of 0.1--6.9\,mJy at 4.9\,GHz and 0.36--8.72\,mJy at 1.6\,GHz.
The ranges of the spectral slope and the brightness temperature are $-1.98 \lesssim \alpha_{\rm total} \lesssim +2.18$ and $6.0 \lesssim \log T_{\rm B} \lesssim 8.5$.
The $\log L_{\rm R}/L_{\rm X}$ is distributed from $-6.6$ to $-2.7$ using the VLBA flux and from $-6.1$ to $-2.3$ using the VLA flux.
The span of the VLBA core/total flux ratio is 0.1--1.0 and that of the VLBA/VLA flux ratio is 0--1.6.

\section{Results} \label{result_section}

\subsection{The $L_{\rm R}/L_{\rm X}$ relation}

\begin{figure*}
\centering
\includegraphics[width=.48\textwidth, trim={5cm, 0cm, 6cm, 2cm}, clip]{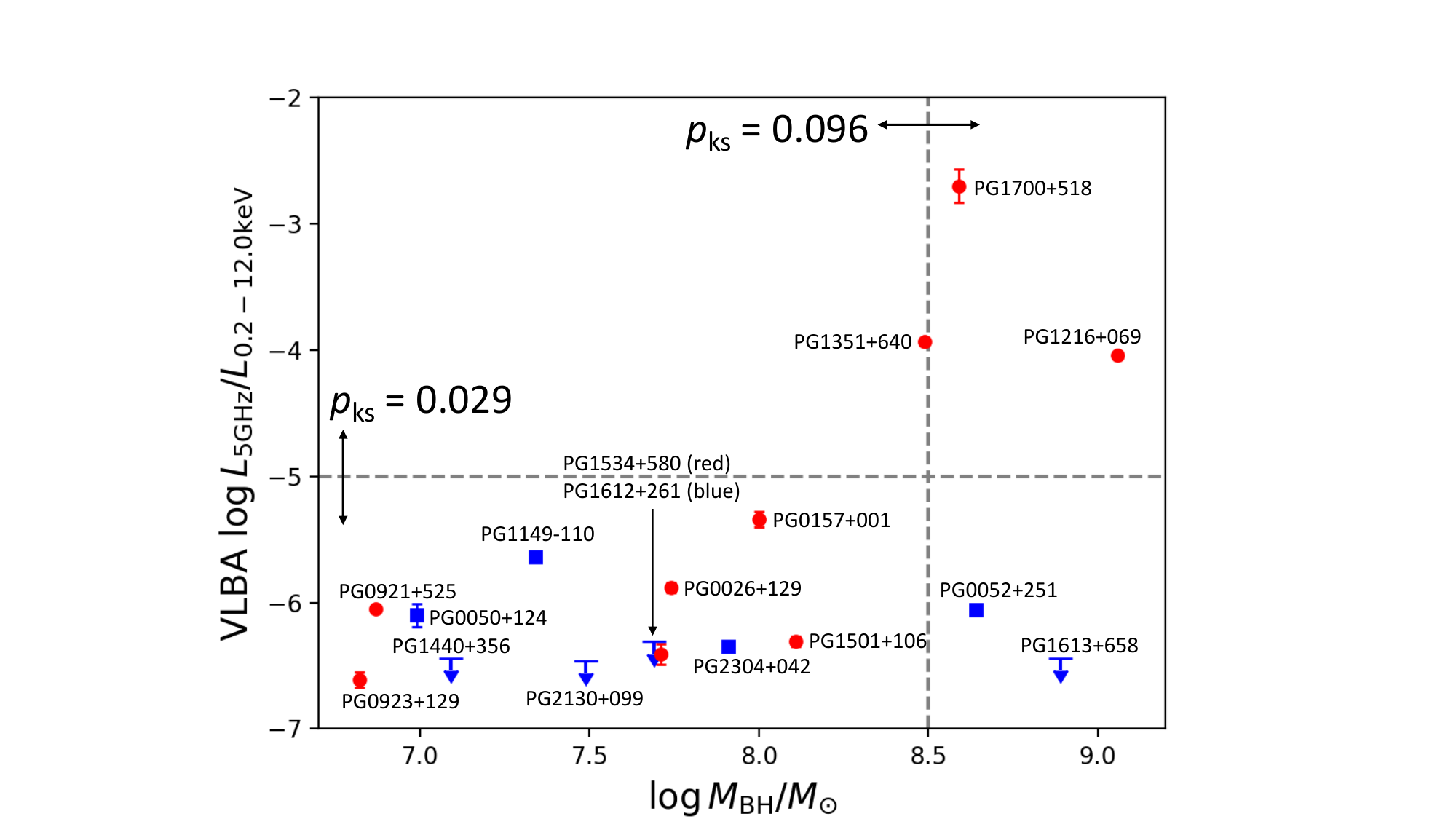}
\includegraphics[width=.48\textwidth, trim={5cm, 0cm, 6cm, 2cm}, clip]{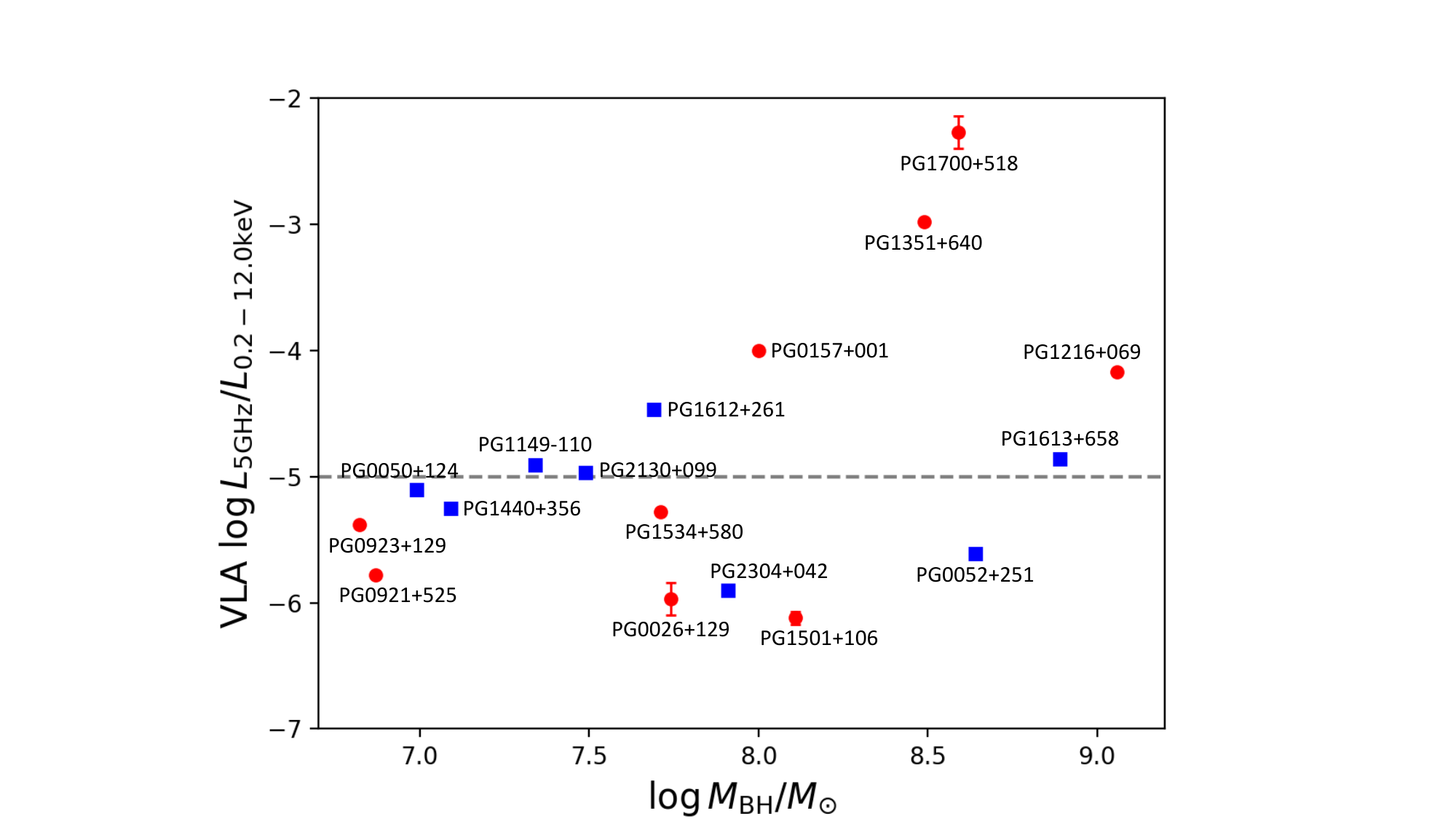}
\caption{The radio 5\,GHz to X-ray 0.2--12.0\,keV luminosity ratio as a function of the BH mass. The VLBA core flux on $\sim 3$\,mas scale is shown in the \textit{left panel}, and the VLA unresolved flux with the A configuration on $\sim 300$\,mas is shown in the \textit{right panel}. The red circles and blue squares represent the nine objects in our new VLBA observations and the eight objects from \citet{Alhosani2022}, respectively.
The VLBA scale emission is characterized by $\log L_{\rm R}/L_{\rm X} \simeq -6$ with three outliers at $\log L_{\rm R}/L_{\rm X} \gtrsim -4$ and $\log M_{\rm BH}/M_{\odot} \gtrsim 8.5$.
The $p$-values of the KS test on $L_{\rm R}/L_{\rm X}$ with $\log M_{\rm BH}/M_{\odot} \gtrsim 8.5$ or $\lesssim 8.5$ and on $M_{\rm BH}$ with $\log L_{\rm R}/L_{\rm X} > -5$ or $< -5$ are labelled in the \textit{left panel}.
In contrast, the VLA scale emission shows $\log L_{\rm R}/L_{\rm X} \simeq -5$, with a rising trend which starts already at $\log M_{\rm BH}/M_{\odot} \sim 7$. The mas scale $L_{\rm R}$ shows a tighter relation with $L_{\rm X}$ than the sub-arcsec scale $L_{\rm R}$, which may indicate a common origin of the mas scale radio and X-ray sources, as expected for coronal emission.}
\label{Lrx+Mbh}
\end{figure*}

Figure~\ref{Lrx+Mbh} compares the distribution of $L_{\rm R}/L_{\rm X}$ as a function of $M_{\rm BH}$ using the VLBA (left panel) and the VLA (right panel) core flux both at 5\,GHz, including nine objects in our new VLBA observations and eight objects from \citet{Alhosani2022}.
The VLA flux is measured with the A configuration, where the resolution beam is $\sim 0.3$\,arcsec, and the VLBA resolution beam is typically $\sim 3$\,mas, which is 100 times smaller.

The VLA scale emission follows $\log L_{\rm R}/L_{\rm X} \simeq -5$ (a mean value of $-5.3$ and a scatter of 0.6 if the three outliers mentioned below are excluded), as found for the complete sample of 71 RQQ in the PG sample \citep{Laor2008}.
There is a trend where $L_{\rm R}/L_{\rm X}$ increases gradually with $M_{\rm BH}$ starting at $\log M_{\rm BH}/M_{\odot} \sim 7$, which is consistent with the dependence of the optically defined radio loudness $R$ on $M_{\rm BH}$ \citep{Laor2000}.
The VLBA scale emission follows $\log L_{\rm R}/L_{\rm X} \simeq -6$ with a smaller scatter (a mean value of $-6.2$ and a scatter of 0.3 if the three outliers mentioned below are excluded).
At $\log M_{\rm BH}/M_{\odot} \gtrsim 8.5$, three of the five objects become more RL with $\log L_{\rm R}/L_{\rm X} \gtrsim -4$.
The Kolmogorov-Smirnov (KS) test on $L_{\rm R}/L_{\rm X}$ suggests that the objects with $\log M_{\rm BH}/M_{\odot} \gtrsim 8.5$ or $\lesssim 8.5$ come from different populations at a confidence level of 90\% ($p = 0.096$).
The lower $L_{\rm R}/L_{\rm X}$ of the mas emission compared to the sub-arcsec emission is expected, as the VLBA/VLA flux ratio is generally below unity.
The mas scale $L_{\rm R}$ appears to show a tighter relation with $L_{\rm X}$ than the sub-arcsec scale $L_{\rm R}$, with a smaller scatter in the mas scale $L_{\rm R}/L_{\rm X}$ (0.3) than the sub-arcsec scale $L_{\rm R}/L_{\rm X}$ (0.6), which may reflect the fact that the VLBA core emission is highly compact, and originates close to the X-ray source.
In contrast, the VLA unresolved emission of $\sim 300$\,mas, corresponding to $\sim 200-2000$\,pc, may be associated with additional extended emission and may also depend on the host galaxy properties.
A larger sample is required to establish the tightness of the VLBA based $L_{\rm R}/L_{\rm X}$ relation.

The three outliers at $\log L_{\rm R}/L_{\rm X} \gtrsim -4$ are all at $\log M_{\rm BH}/M_{\odot} \gtrsim 8.5$.
The KS test on the $M_{\rm BH}$ distribution of the objects with $\log L_{\rm R}/L_{\rm X} > -5$ or $< -5$ shows that a similar $M_{\rm BH}$ distribution can be excluded at a confidence level of 97\% ($p = 0.029$).
The three outliers are,
PG1216+069, which is likely very compact as it shows significant variability and appears to be synchrotron self-absorbed (see Section~\ref{core_discussion}),
PG1351+640, which may have a mildly relativistic jet on a scale of 10\,pc \citep{Wang2023b} and appears to be intermediate \citep[3<R<200;][]{Kellermann1989,Falcke1996} between RL and RQ AGN (see Section~\ref{jet_discussion}),
and PG1700+518, which is a broad absorption line (BAL) quasar with a very low X-ray flux \citep{Pettini1985,Ballo2011,Luo2013} that leads to the high $L_{\rm R}/L_{\rm X}$ ratio, and it also shows a fast outflow (see Section~\ref{core_discussion}).

\subsection{The spectral slope}

The spectral slope of the 5--8.5\,GHz VLA unresolved emission of a sample of 25 RQQ shows a trend with $L/L_{\rm Edd}$ \citep{Laor2019}.
Objects with $L/L_{\rm Edd} \lesssim 0.3$ have flat slopes, while higher $L/L_{\rm Edd}$ objects generally show steep slopes.
This suggests that compact optically thick radio emission at low $L/L_{\rm Edd}$ switches to more extended optically thin radio emission at higher $L/L_{\rm Edd}$, which is possibly a switch from the compact coronal emission to a larger scale wind emission.
The follow-up VLBA study in \citet{Alhosani2022} of the four flattest and four steepest objects in \citet{Laor2019}, confirms this interpretation, as the radio emission of three of the four flat spectrum objects remains unresolved on mas scales, while the radio emission of three of the four steep spectrum objects is resolved out with the VLBA and must be produced on larger scales.

\begin{figure*}
\centering
\includegraphics[width=.48\textwidth, trim={5cm, 0cm, 6cm, 2cm}, clip]{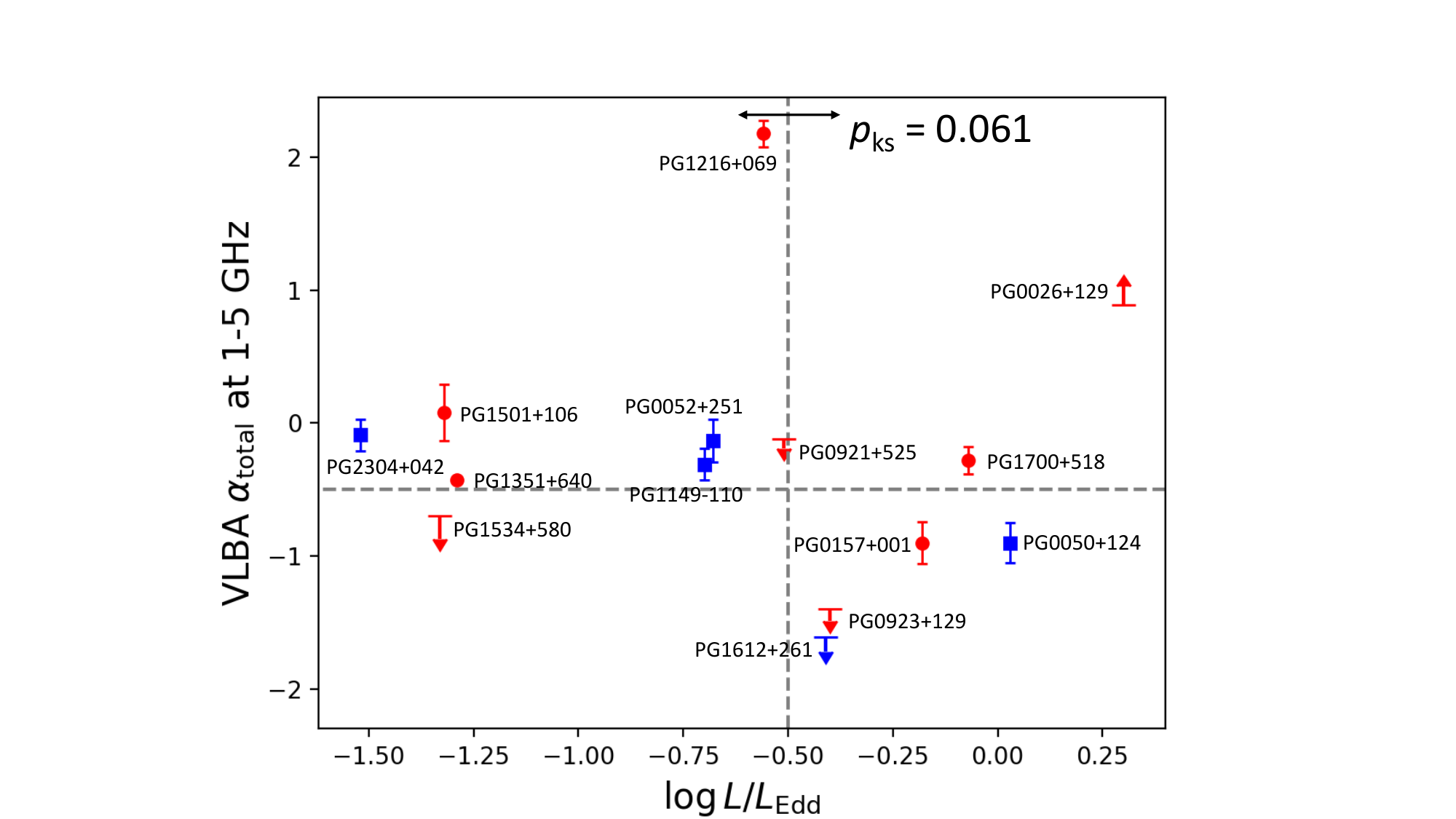}
\includegraphics[width=.48\textwidth, trim={5cm, 0cm, 6cm, 2cm}, clip]{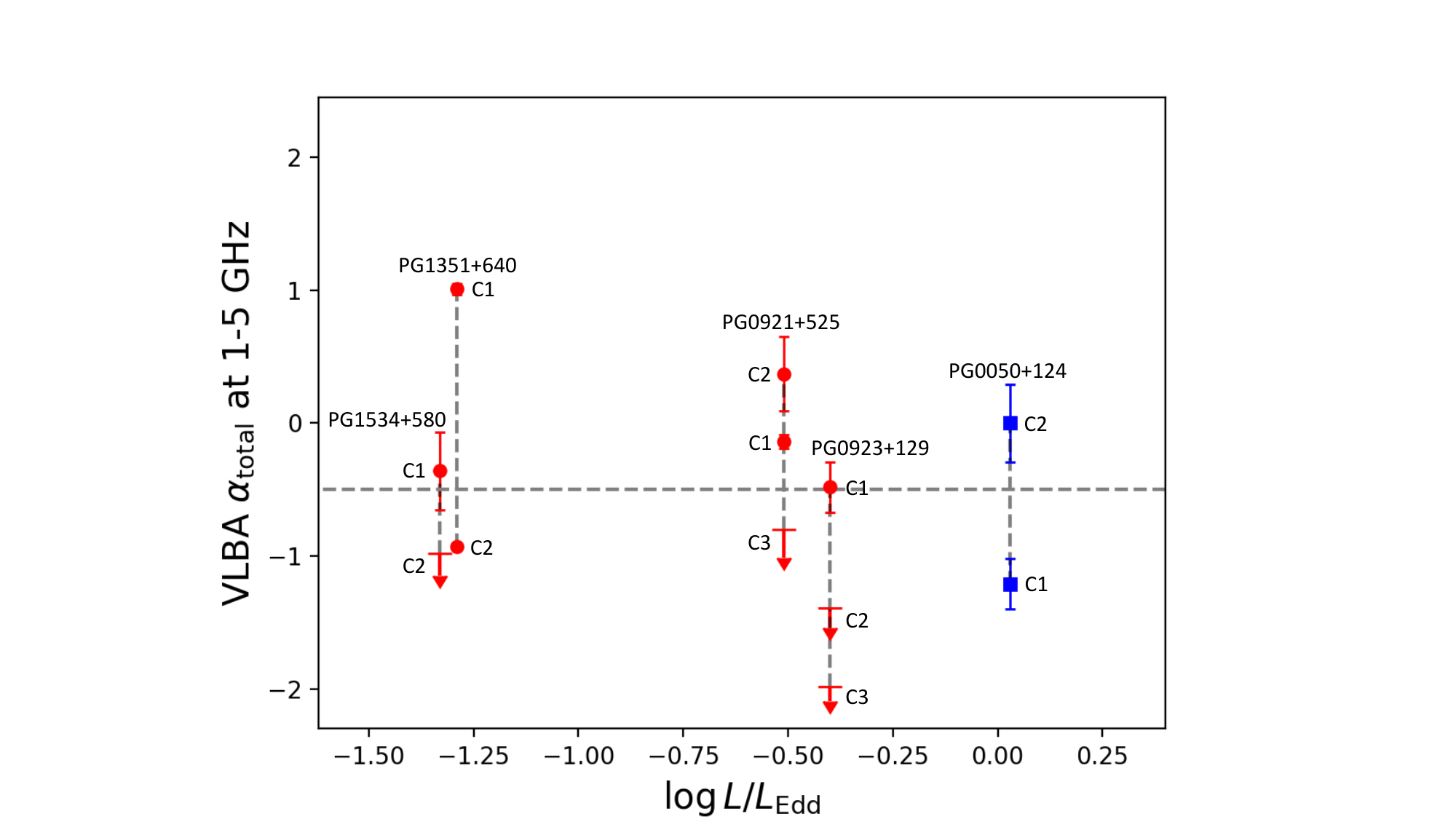}
\caption{\textit{Left panel:} The VLBA spectral slope of the total flux density $\alpha_{\rm total}$ at 1.6--4.9\,GHz as a function of the Eddington ratio $L/L_{\rm Edd}$.
The red circles and blue squares represent the nine objects detected in our new VLBA observations and the five objects detected in \citet{Alhosani2022}, respectively.
Although no significant correlation is found between $\alpha_{\rm total}$ and $L/L_{\rm Edd}$, there is a trend: The radio emission in only one of the eight objects at $\log L/L_{\rm Edd} < -0.5$ is steep, and in only two of the six objects at $\log L/L_{\rm Edd} > -0.5$ are flat.
The $p$-value of the KS test on $\alpha_{\rm total}$ with $\log L/L_{\rm Edd} > -0.5$ or $< -0.5$ are labelled.
\textit{Right panel:} The $\alpha_{\rm total}$ of individual components in the five objects with more than one component.
Each object is comprised of a flat compact component plus steep extended components.}
\label{total}
\end{figure*}

Figure~\ref{total} (left panel) shows an extension study of the correlation between $\alpha_{\rm total}$ and $L/L_{\rm Edd}$ in a larger and more representative sample with a total of 14 RQ PG quasars, including nine objects detected in our new VLBA observations and five objects detected in \citet{Alhosani2022}.
Although there is only a weak trend of $\alpha_{\rm total}$ with $L/L_{\rm Edd}$, the slopes of seven of the eight objects with $\log L/L_{\rm Edd} < -0.5$ are flat ($\alpha_{\rm total} > -0.5$), only one is steep ($\alpha_{\rm total} < -0.5$).
In contrast, of the six objects with $\log L/L_{\rm Edd} > -0.5$, only two are flat.
The KS test on the slopes suggests that the objects with $\log L/L_{\rm Edd} > -0.5$ or $< -0.5$ come from different populations at a confidence level of 94\% ($p = 0.061$).
Two objects in Figure~\ref{total} (left panel) are outliers, which have strongly inverted spectra ($\alpha_{\rm total} \sim 1-2$), PG1216+069, which is consistent with the flat spectra in the low $L/L_{\rm Edd}$ regime, and PG0026+129, which stands in contrast with most other high $L/L_{\rm Edd}$ objects.

The correlation between $\alpha_{5-8.5}$ and $L/L_{\rm Edd}$ is tighter for the VLA scale emission \citep{Laor2019}, which may indicate that the correlation is driven by the larger scale radio emission, which becomes more prominent at higher $L/L_{\rm Edd}$ objects.
On the VLBA scale, most of the larger scale radio emission is resolved out, and the emission is dominated by the flat compact core emission.
This suggests that the flat compact core emission is present in most of the RQQ regardless of the Eddington ratio, and its contribution dilutes the trend of $\alpha_{\rm total}$ with $L/L_{\rm Edd}$.
Figure~\ref{total} (right panel) further confirms the presence of the core emission in five objects with multiple components, of which four are from our VLBA observations and one is from \citet{Alhosani2022}.
Each of the five objects shows a flat compact component, and one or more steep extended components.

\begin{figure*}
\centering
\includegraphics[width=.48\textwidth, trim={5cm, 0cm, 6cm, 2cm}, clip]{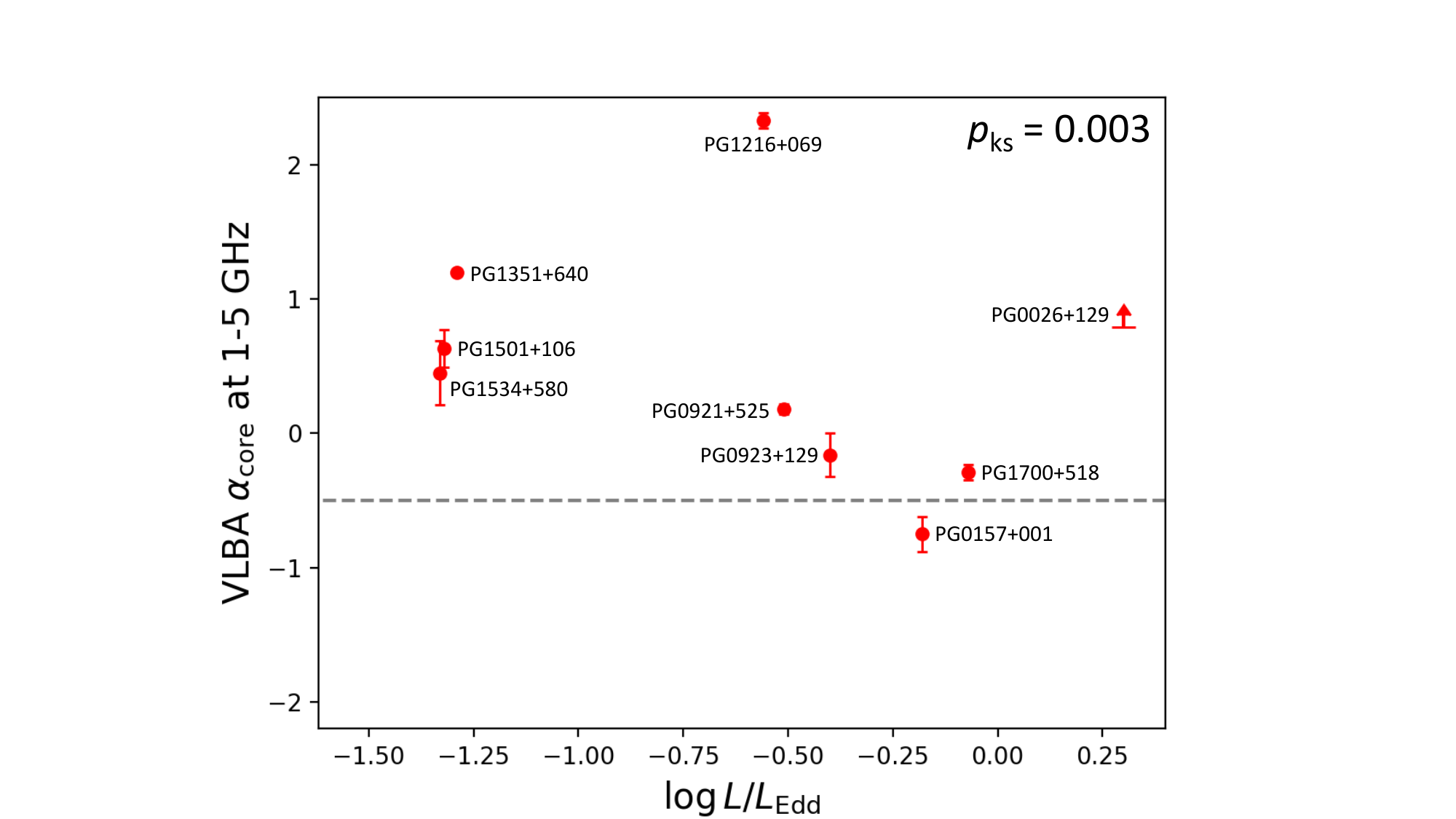}
\includegraphics[width=.48\textwidth, trim={5cm, 0cm, 6cm, 2cm}, clip]{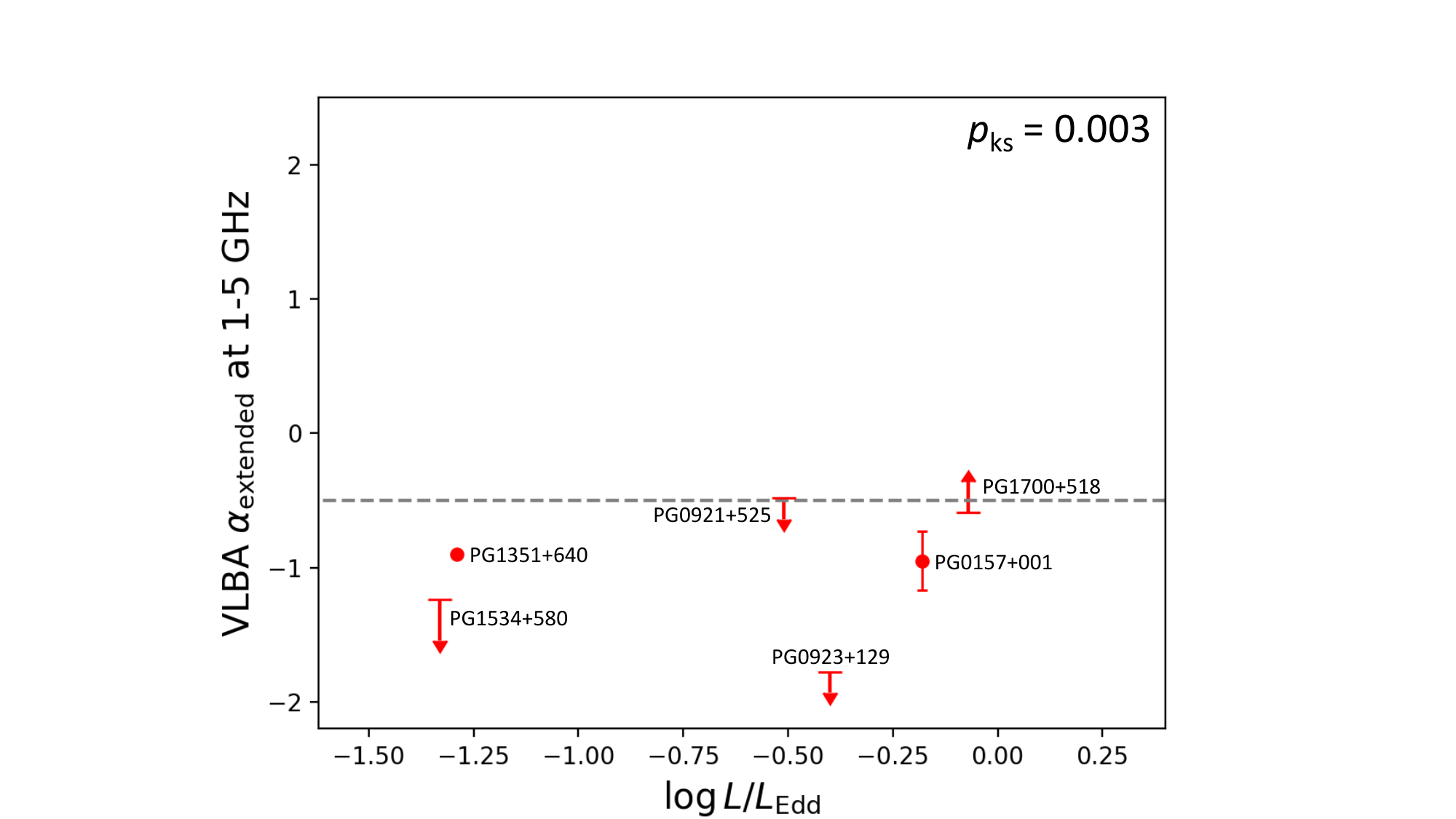}
\caption{The VLBA spectral slope of the core flux density $\alpha_{\rm core}$ (\textit{left panel}) and extended flux density $\alpha_{\rm extended}$ (\textit{right panel}) at 1.6--4.9\,GHz as a function of the Eddington ratio $L/L_{\rm Edd}$ for the nine objects detected in our new VLBA observations.
The core emission is generally flat, while the extended emission is generally steep. There is a possible trend of $\alpha_{\rm core}$ getting steeper as $L/L_{\rm Edd}$ increases, while $\alpha_{\rm extended}$ shows no trend.
The $p$-value of the KS test on the slopes of the core and extended components are labelled in both panels.}
\label{core}
\end{figure*}

\begin{figure*}
\centering
\includegraphics[width=.48\textwidth, trim={4.5cm, 0cm, 6cm, 2cm}, clip]{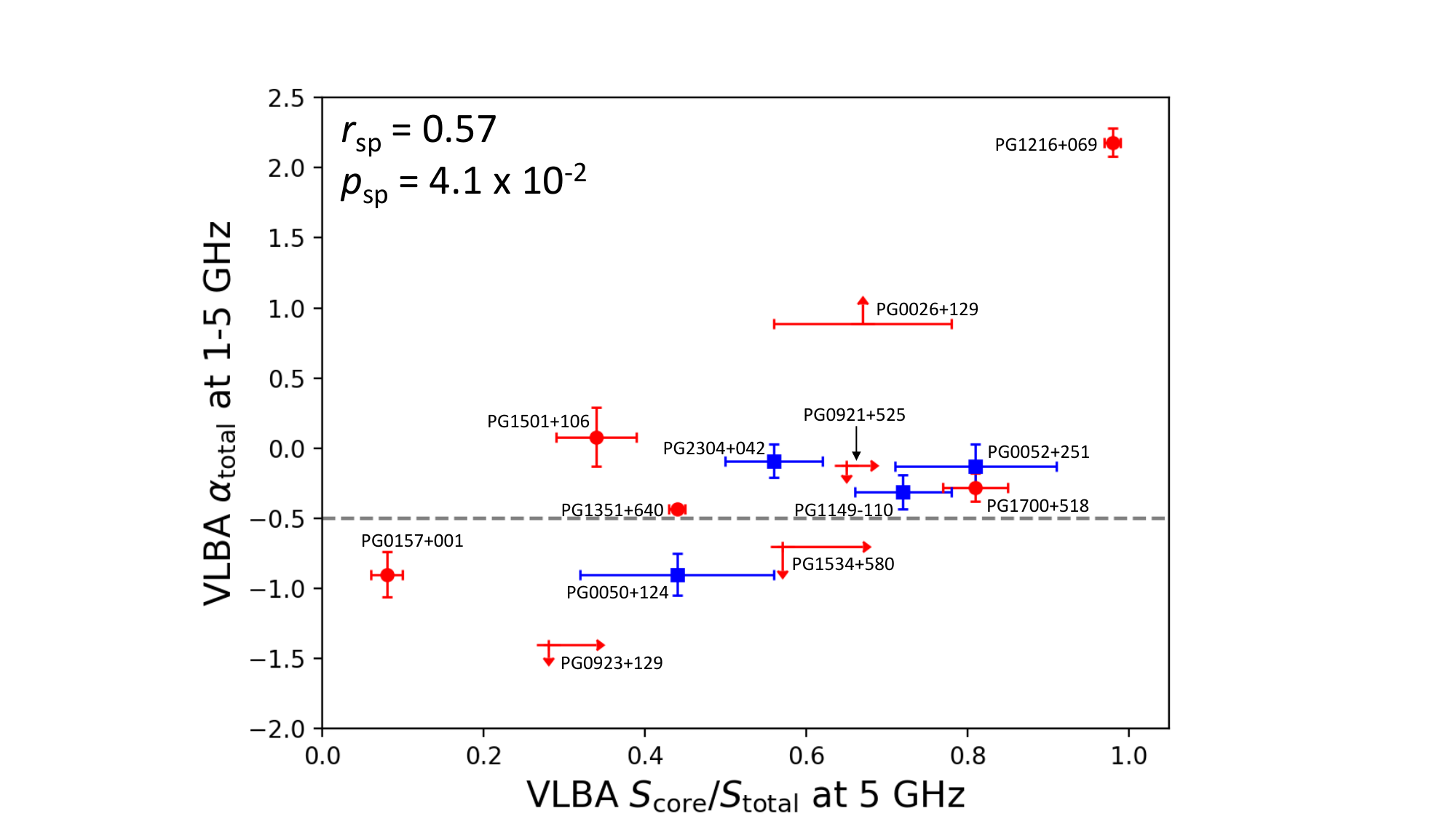}
\includegraphics[width=.48\textwidth, trim={4.5cm, 0cm, 6cm, 2cm}, clip]{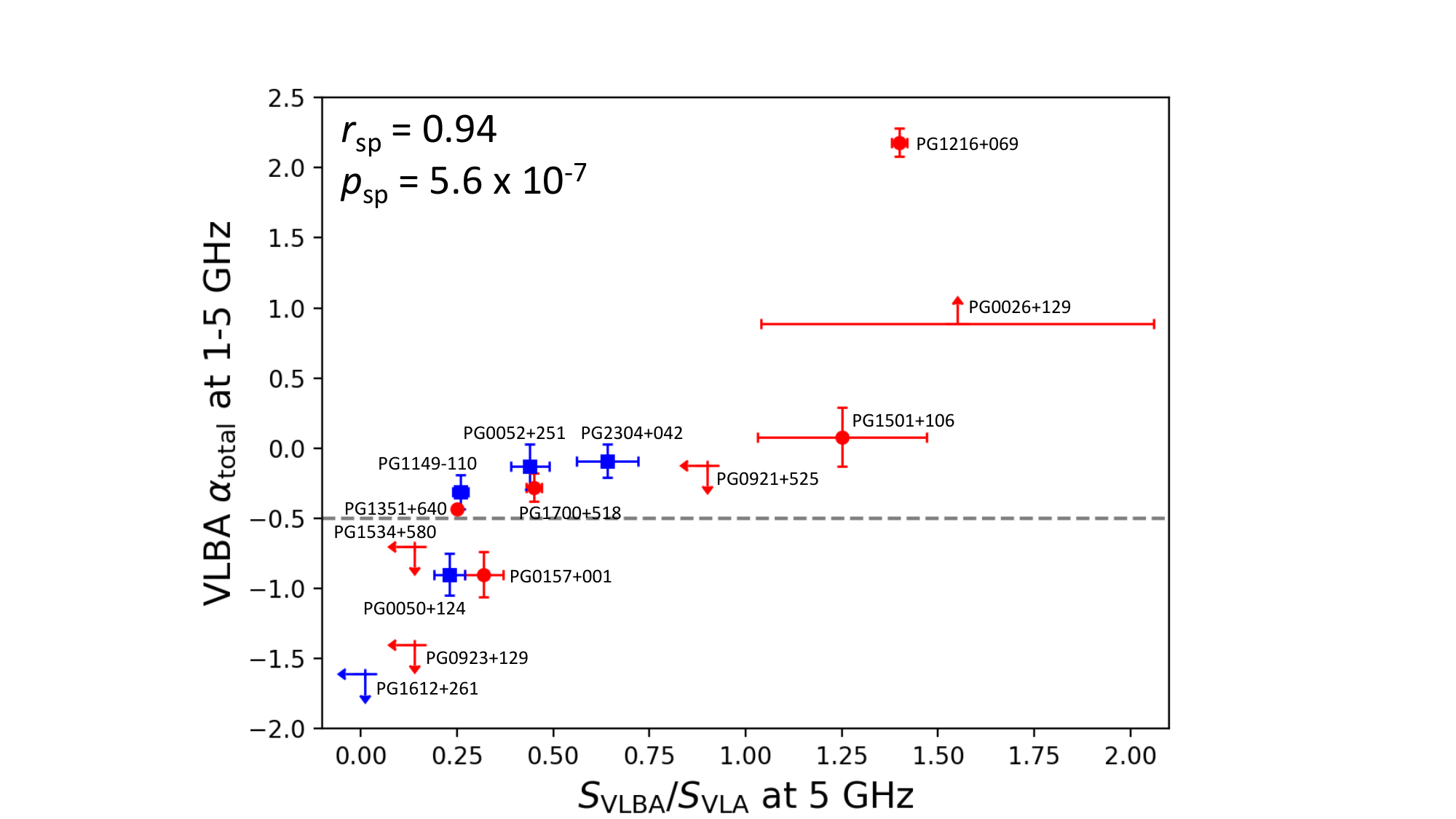}
\caption{The VLBA spectral slope of the total flux density $\alpha_{\rm total}$ at 1.6--4.9\,GHz as a function of the VLBA core/total flux ratio (\textit{left panel}) and the VLBA/VLA flux ratio (\textit{right panel}) for the 14 RQ PG quasars.
A similar trend of increasing flatness with increasing compactness is seen.
The VLBA slope is flat for $S_{\rm core}/S_{\rm total} > 0.5$, i.e.\ when the VLBA core emission dominates.
A flat $\alpha_{\rm total}$ also implies that the VLBA scale ($\sim 3$\,mas) emission dominates the VLA scale ($\sim 300$\,mas) emission.
The Spearman correlation, $r$ and $p$ values, are labelled in both panels.}
\label{fc}
\end{figure*}

Figure~\ref{core} presents the dependence of $\alpha_{\rm core}$ on $L/L_{\rm Edd}$ (left panel), and of $\alpha_{\rm extended}$ on $L/L_{\rm Edd}$ (right panel) for the nine objects detected in our new VLBA observations.
The distributions of $\alpha_{\rm core}$ and $\alpha_{\rm extended}$ are clearly different.
The KS test on the slopes suggests that the core and extended components are drawn from different distributions at a confidence level of 99.7\% ($p = 0.003$).
In eight of the nine objects, $\alpha_{\rm core}$ is flat.
In contrast, $\alpha_{\rm extended}$ is steep within 1$\sigma$ uncertainty in all six objects where it is detected.
In PG0157+001, both $\alpha_{\rm core}$ and $\alpha_{\rm extended}$ are steep.
The VLBA core/total flux ratio is $\sim 0.1$, which is exceptionally low as the median core/total flux ratio is $\sim 0.6$ (see Figure~\ref{fc}).
A compact optically thick flat source is not detected. The core which is detected resides far from the {\it Gaia} position (Table~\ref{position}), and is likely not the true core but rather optically thin emission away from the center.

Although neither $\alpha_{\rm core}$ or $\alpha_{\rm extended}$ are significantly correlated with $L/L_{\rm Edd}$, $\alpha_{\rm core}$ appears to show a trend of getting steeper with increasing $L/L_{\rm Edd}$.
To summarize, Figure~\ref{core} demonstrates that most objects have a compact core with a flat spectrum on mas scales, while the extended emission, when detected, has a steep spectrum.

Figure~\ref{fc} shows $\alpha_{\rm total}$ as a function of the VLBA core/total flux ratio (left panel), and the VLBA total flux to the VLA A configuration core flux ratio (right panel), both ratios measured at 5\,GHz, for the 14 RQ PG quasars.
There is a clear trend where $\alpha_{\rm total}$ increases with increasing flux ratios, that is the source becomes more compact as the emission becomes flatter.
The Spearman correlation gives $r = 0.57$ and $p = 4.1 \times 10^{-2}$ for the VLBA core/total flux ratio and $r = 0.94$ and $p = 5.6 \times 10^{-7}$ for the VLBA/VLA flux ratio.
The VLBA emission becomes generally flat ($\alpha_{\rm total} > -0.5$) in seven of the eight objects where the core emission dominates the total emission ($S_{\rm core}/S_{\rm total} > 0.5$).
This is consistent with the results above that the core emission is flat, while the extended emission is steep.
The correlation becomes significantly stronger when we take the VLBA to the VLA ($\sim 300$\,mas) flux ratio.
In five of the nine objects where $\alpha_{\rm total} > -0.5$, the VLBA emission also dominates the VLA emission on sub-arcsec scales ($S_{\rm VLBA}/S_{\rm VLA} > 0.5$).
In contrast, if the steep extended emission dominates on the VLBA scales, there is significantly more (a factor of 5--10) extended emission which contributes further out on the VLA scales.

In three of the 14 objects, $S_{\rm VLBA}/S_{\rm VLA} > 1$, which indicates the objects are variable.
Their variability implies that the source sizes are smaller than the light crossing times.
The earlier VLA observations in \citet{Kellermann1989} were carried out in 1983 or 40 years ago, which implies a size $\lesssim 10$\,pc.
Much tighter constraints on the size are given by the flat spectral slopes.
Since the spectra are optically thick, the source sizes at 5\,GHz are constrained to be $\sim 0.01-0.4$\,pc \citep{Laor2008}, which is consistent with the observed variability timescale.

\begin{figure*}
\centering
\includegraphics[width=.48\textwidth, trim={5cm, 0cm, 6cm, 2cm}, clip]{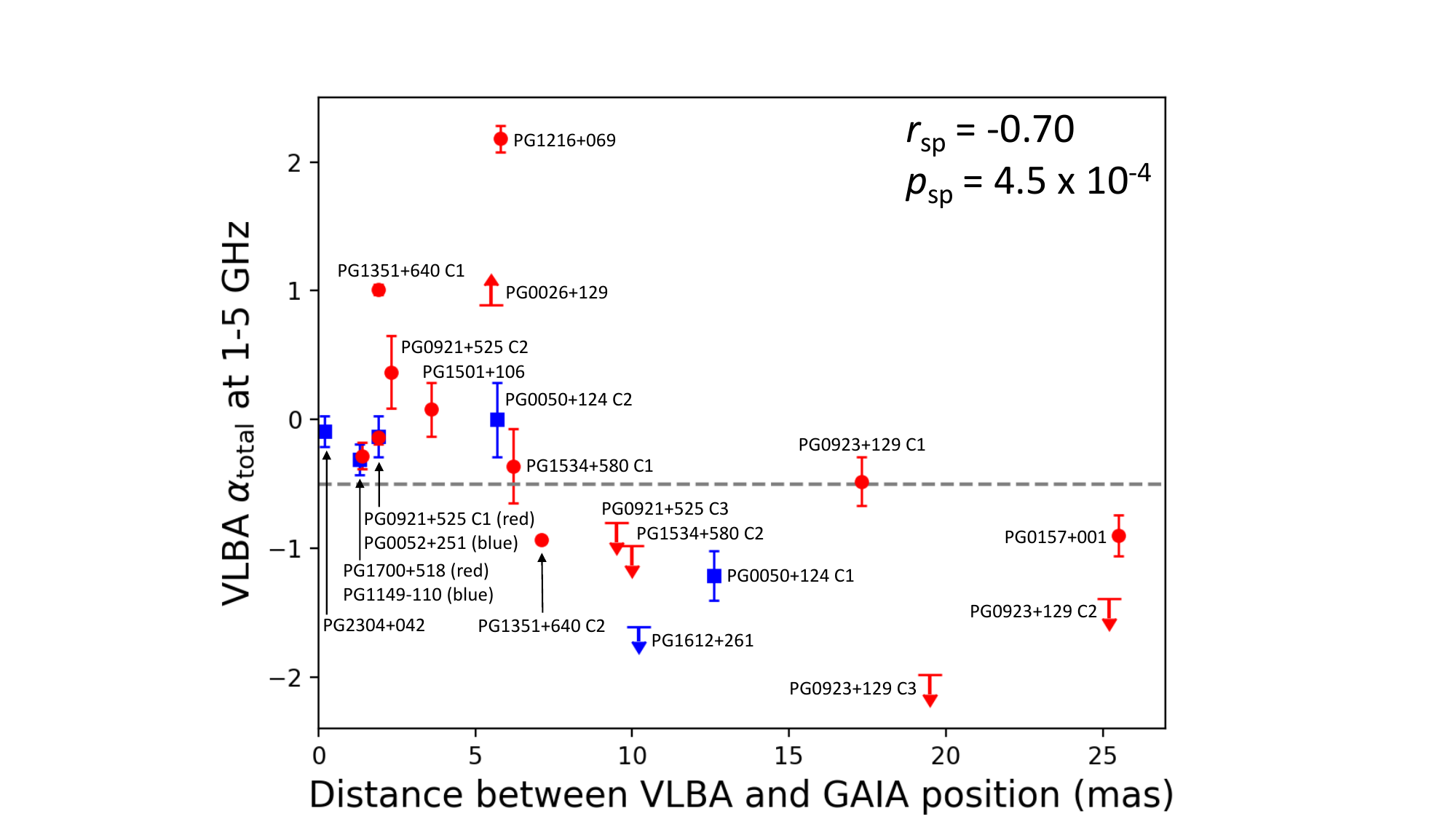}
\includegraphics[width=.48\textwidth, trim={5cm, 0cm, 6cm, 2cm}, clip]{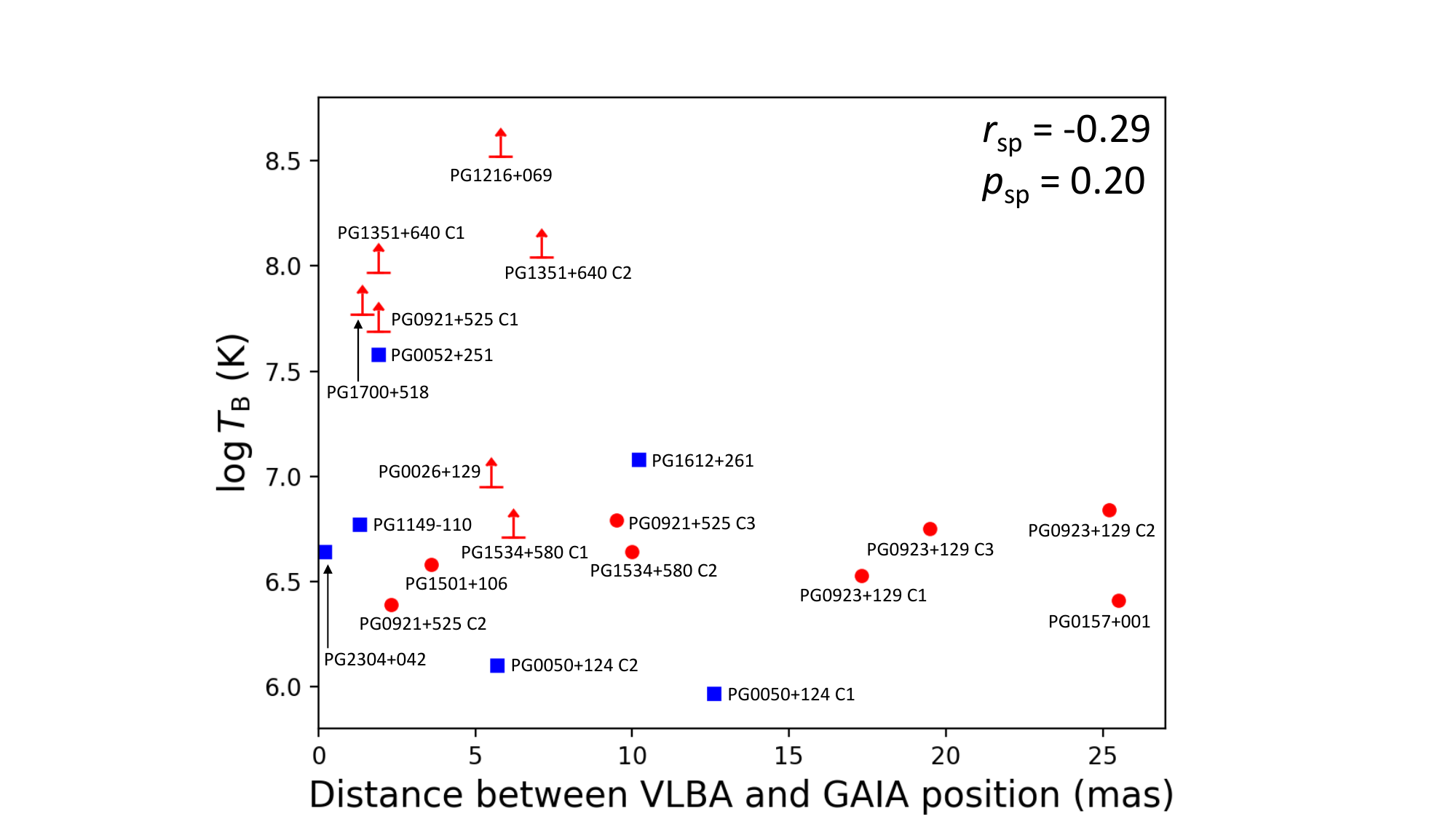}
\caption{The VLBA spectral slope of the total flux density $\alpha_{\rm total}$ at 1.6--4.9\,GHz (\textit{left panel}) and the brightness temperature $T_{\rm B}$ (\textit{right panel}) as a function of the distance between the VLBA and the {\it Gaia} positions in mas for individual components of the 14 RQ PG quasars.
The flat optically thick sources are all located within the astrometric position accuracy of the optical center, and are likely associated with the AGN core, while the steep optically thin sources are all physically offset from the core by $\gtrsim 10$\,mas, which typically corresponds to a few 10s of pc.
Similarly, the components with $T_{\rm B} \gtrsim 10^7$\,K are all located within the {\it Gaia} position uncertainty, while the components away from the center are all at $T_{\rm B} \lesssim 10^7$\,K.
The Spearman correlation, $r$ and $p$ values, are labelled in both panels.}
\label{distance}
\end{figure*}

Figure~\ref{distance} shows the VLBA slope $\alpha_{\rm total}$ (left panel) and the brightness temperature $T_{\rm B}$ (right panel) as a function of the distance between the VLBA and the {\it Gaia} positions for individual components of the 14 RQ PG quasars.
In general, the flat slope components are closer ($\lesssim$ 7\,mas) to the {\it Gaia} position, and their offsets are consistent with the {\it Gaia} astrometric uncertainty ($\sim$ 1--10\,mas) \citep{Khamitov2023}.
In contrast, the steep slope components are all located further out ($\gtrsim$ 10\,mas).
The Spearman correlation indeed shows that the slope is correlated with the distance from the {\it Gaia} position with $r = -0.70$ and $p = 4.5 \times 10^{-4}$.
No significant correlation is present between $T_{\rm B}$ and the distance from the {\it Gaia} position, where $r = -0.29$ and $p = 0.20$.
However, we do see a trend that all sources with $T_{\rm B} \sim 10^7 - 10^9$\,K are $\lesssim$ 7\,mas offset from the {\it Gaia} position, and all sources with an offset $\gtrsim$ 10\,mas have $T_{\rm B} \sim 10^6 - 10^7$\,K.
These trends suggest that the flat optically thick sources may be located at the physical center of the AGN, while the steep optically thin components are located $\sim$ a few tens of pc away from the central core (see the physical scale in unit of pc\,mas$^{-1}$ in Table~\ref{position}).

\section{Discussion} \label{discussion_section}

The main result of our study is that most of the RQQ have a compact core, which is still unresolved on pc scales, regardless of the Eddington ratio.
The compact core emission is flat at 1.5--5\,GHz, it overlaps with the optical core ({\it Gaia}) position, and can reach $T_{\rm B} \gtrsim 10^7$\,K.
The physical size of an optically thick synchrotron source can be estimated via \citep[Eq.22 in][]{Laor2008}
\begin{equation}
R_{\rm RS} = 0.47 L_{30}^{0.4} L_{46}^{0.1} \nu_{\rm p}^{-1}
\end{equation}
where $R_{\rm RS}$ is the radius of the radio sphere in pc, $\nu_{\rm p}$ is the turnover frequency in GHz, $L_{30}$ is the luminosity density at the turnover frequency in 10$^{30}$\,erg\,s$^{-1}$\,Hz$^{-1}$, and $L_{46}$ is the bolometric luminosity in 10$^{46}$\,erg\,s$^{-1}$.
The 5\,GHz emission of the RQQ in our sample comes from a radio sphere with a size of $R_{\rm RS} \sim 0.01-0.4$\,pc or $\sim 10-500$ light days.
Such a compact physical size corresponds to an angular scale which is typically below 0.1\,mas (Table~\ref{position}), and is indeed expected to remain unresolved with the VLBA at 5\,GHz.
What is the source of this emission?

\subsection{The compact emission: Jet versus Corona} \label{discussion_jet_corona}

The compact size and the association with the optical position suggest that the emission originates close to the center of the AGN, on a scale comparable in size to the broad-line region, or smaller.
The physical origin is possibly the accretion disk corona or a compact jet.

If the origin of the core emission is a compact jet, the jet needs to be small enough to remain unresolved with the VLBA, that is to be smaller than a few pc.
Since the 5\,GHz core emission is optically thick, the size of the jet is constrained to be below typically $\sim 0.1$\,pc.
A jet which extends beyond $\sim 0.1$\,pc becomes optically thin.
It is therefore not clear why the jet would be confined to $\sim 0.1$\,pc in about half of the objects without evidence of extended emission.

The core emission may be dominated by the jet base, i.e.\ the jet launching region, which may spatially overlap with the accretion disk corona.
VLBA observations of RL AGN indeed often show only compact core emission \citep{Lister2009,Wang2023a}, but such objects do generally show also significant extended 10--100\,kpc jet emission \citep{Baghel2023}.
This is in contrast with RQ PG quasars, where in about half of the objects, the spectral slopes are flat, and there is no significant extended emission.
Similarly, low-luminosity RQ AGN show flat-spectrum compact cores on sub-kpc scales \citep{Baldi2018,Baldi2021,Kharb2021}.

In contrast with the jet interpretation, the accretion disk corona is expected to be extremely compact, on a scale of $\sim$ 10--100 gravitational radii, or $\sim 0.001$\,pc, which naturally explains the compact size of $\lesssim 0.1$\,pc in all objects.
Coronal mass ejections, as observed in coronally active stars, can produce more extended radio emission.
However, the expected drop in the gas synchrotron emission as it expands outward, will prevent the extended optically thin emission from becoming dominant.

Additionally, the rather tight correlation of the VLBA core flux and the X-ray flux (Figure~\ref{Lrx+Mbh}) also points toward a coronal origin as the more likely explanation.
In the near future, we have a VLBA program of observing this sample at higher frequencies and thus on higher resolutions, which will test if the core emission remains flat and unresolved, as expected for the coronal emission origin.

\subsection{The extended emission} \label{discussion_extended}

The extended radio emission detected in some of the RQQ is generally spatially resolved, has a steep slope and a low brightness temperature, and is therefore optically thin synchrotron emission.
The emission is offset from the {\it Gaia} position by $\sim 10-25$\,mas, which corresponds to a physical distance of $\sim 5-100$\,pc from the center.
The extended clumped emission may be produced by an outflow of magnetized plasma from the center, in the form of a weak jet with intermittent activity, or coronal mass ejection events, or it may be produced by an AGN driven wind, which shocks on the clumped ambient gas and leads to particle acceleration and synchrotron emission.

The slope of the sum of the compact and extended emission, i.e.\ $\alpha_{\rm total}$, shows a clear trend with $L/L_{\rm Edd}$ (Figure~\ref{total}).
The flat core emission dominates in the lower $L/L_{\rm Edd}$ ($< 0.3$) objects, and the steep extended emission dominates in about half of the higher $L/L_{\rm Edd}$ ($> 0.3$) objects.
This is consistent with the trend suggested in \citet{Alhosani2022}, for a relation of the emission compactness with $L/L_{\rm Edd}$, though the trend found here is less sharp than that found in \citet{Alhosani2022}.
The sample of eight objects in \citet{Alhosani2022} was selected to have the flattest and steepest VLA scale slopes at 5--8.5\,GHz in \citet{Laor2019}, while the sample observed here was selected to be more representative of the RQ PG quasars.
The conclusion remains that the radio emission in low $L/L_{\rm Edd}$ ($< 0.3$) RQQ is generally core-dominated, while some high $L/L_{\rm Edd}$ ($> 0.3$) RQQ can have significant extended emission. 
However, the core emission is detected also in high $L/L_{\rm Edd}$ sources, and the extended components are also detected in low $L/L_{\rm Edd}$ sources.

The correlation between the $\alpha_{\rm total}$ and the compactness, i.e.\ the VLBA core/total flux ratio and the VLBA/VLA flux ratio (Figure~\ref{fc}), further supports that the flat slope objects are compact on pc scales and are dominated by the core, possibly the coronal emission, and the steep slope objects show extended emission on sub-kpc scales, likely in the form of an AGN wind.
This result is consistent with the earlier study by \citet{Orienti2010} of a few nearby low luminosity Seyfert galaxies, where the steep-spectrum objects show a low VLBA/VLA flux ratio, whereas in flat-spectrum objects the flux ratio is close to unity.
In a jet scenario, the low VLBA/VLA flux ratio may be due to a jet-ISM interaction \citep{Nagar2005,Kharb2021}.

The extended emission is two-sided only in one object, PG0923+129.
In the other three spatially resolved objects, PG0921+525, PG1351+640, and PG1534+580, detected here (Figure~\ref{overlap}), and also in PG0050+124 \citep{Alhosani2022}, the extended emission is one-sided, although a weak counter-jet is detected in an EVN + e-MERLIN 5\,GHz observation of PG0050+124 \citep{Yang2023}.
In a jet scenario, the one-sided emission could be caused by the Doppler boosting and dimming effects.
However, a significant Doppler effect requires a bulk motion at a relativistic speed, which will also produce a strong enhancement to the emission with a high $T_{\rm B}$.
Evidence for mild relativistic speed is observed only in PG1351+640 \citep{Wang2023b}, which may be large enough to explain the lack of counter-jet detection.
In PG0921+525 and PG1534+580, there is no evidence for highly beamed emission ($T_{\rm B} < 10^7$\,K).
In these two objects, the observed asymmetry of the extended emission may reflect the host ISM asymmetry, in the case of a wind interaction with the ambient medium.

An alternative interpretation is that the radio outflow is symmetric, but the missing counterpart is the receding one, which is further away from the nucleus, and is obscured by free-free absorption from intervening ionized gas.
The free-free absorption drops as $\nu^{-2}$ and a free-free absorbing screen will become optically thin at higher frequencies.
If the free-free absorbing gas is photoionized by the AGN, then a screen located at $\sim 10$\,pc is expected to become optically thin at $\nu > 10$\,GHz, as indeed observed in a number of nearby AGN \citep[see][Section 4.1 therein]{Baskin2021}.
Radio interferometry observations at higher frequencies will allow to test the free-free absorbing screen interpretation.

The extended emission can also have a flat slope, if it is produced by optically thin free-free emission, where $\alpha \simeq -0.1$.
Such a component is not observed here, which is not surprising, as optically thin free-free emission on pc scales is characterized by $T_{\rm B} \ll 10^6$\,K \citep{Baskin2021}, while the extended emission detected here has $T_{\rm B} \gtrsim 10^6$\,K (Figure~\ref{distance}).
This lower limit on $T_{\rm B}$ just reflects the sensitivity limit of current VLBA observations.
Lower resolution and higher sensitivity observations may be able to detect extended free-free emission.

Below we discuss the individual objects in more details, and compare the current VLBA results to earlier observations.

\subsection{Core-dominated sources} \label{core_discussion}

The radio emission on the VLBA scales of four objects (PG0026+129, PG1216+069, PG1501+106, and PG1700+518) is dominated by the compact core, and is possibly associated with the accretion disk corona.

PG0026+129 has an inverted slope with $\alpha_{\rm total} > +0.9$ and $\alpha_{\rm core} > +0.8$.
It was observed with the VLA A configuration at 1.4, 4.8, 8.4, and 45\,GHz.
The 1.4\,GHz image shows a slightly extended structure with a flux level of $\sim 3-4$\,mJy \citep{Kukula1998}.
The flux densities at 4.8, 8.4, and 45\,GHz are $\sim 0.2-0.3$\,mJy \citep{Kellermann1989,Leipski2006,Baldi2022}, which is consistent with our VLBA measurement.
This indicates that the core emission on the VLBA scale dominates also on the VLA scale.
The core emission has an inverted slope at 1--5\,GHz and a flat slope at 5--45\,GHz, which suggests that it is synchrotron self-absorbed and remains optically thick at $\nu > 45$\,GHz \citep[e.g.][]{Raginski2016}.
The source therefore is compact and variability may happen, as demonstrated by the VLBA core/total flux ratio $\sim 0.7$ and the VLBA/VLA flux ratio $> 1$.
The consistency with the {\it Gaia} position ($\sim 5.5$\,mas) and the high $T_{\rm B}$ ($> 10^{7.0}$\,K), again indicate a compact optically thick source.

PG1216+069 has a strongly inverted slope with $\alpha_{\rm total} = +2.2$ and $\alpha_{\rm core} = +2.3$, which indicates synchrotron self-absorption, and again a compact optically thick source smaller than $\sim 0.01$\,pc.
The VLBA coordinate is consistent with the {\it Gaia} position ($\sim$ 5.8\,mas) and the $T_{\rm B}$ is the highest one ($> 10^{8.5}$\,K) in our sample.
The object is highly compact as indicated by the VLBA core/total flux ratio $\sim 1$, and is variable, since the VLBA/VLA flux ratio $> 1$.
Large variability is also seen by comparing the flux density at 5\,GHz in our observation ($S_{\rm total} = 6.9$\,mJy) which is a factor of five higher than that in \citet{Wang2023a} ($S_{\rm total} = 1.3$\,mJy).
The two observations were made a few years apart, so the light crossing times imply a size smaller than pc scales, which is consistent with the small size derived from synchrotron self-absorption.
The object was observed with the VLBA at 1.4, 2.3, 5, and 8.4\,GHz \citep{Blundell1998,Ulvestad2005,Wang2023a}. The 5 and 8.4\,GHz maps show a very compact core and a slightly extended structure, which is consistent with our observations.
The VLBA slope at 1--5\,GHz in \citet{Ulvestad2005} is flat ($\alpha_{1.4-2.3} = +0.4$ and $\alpha_{2.3-5} = -0.1$), while it is inverted ($\alpha_{1.6-4.9} = +2.3$) in our new VLBA observations and is consistent with the slope of synchrotron self-absorption.
A possible explanation is that we may be observing a new compact synchrotron component, and the compact radio emission observed in \citet{Ulvestad2005} may have expanded and became optically thin at 1--5\,GHz with weaker and negligible emission by now.
On the VLA scales, this object shows slightly extended emission with the A configuration \citep{Kellermann1989,Kellermann1994}, which may reflect past activity that faded and expanded to larger scales by now.

PG1501+106 has a slightly inverted total slope $\alpha_{\rm total} = +0.1$ and a more inverted core slope $\alpha_{\rm core} = +0.6$.
The VLBA coordinate is consistent with the {\it Gaia} position ($\sim$ 3.6\,mas), though the $T_{\rm B}$ is not too high ($\sim 10^{6.6}$\,K).
There is some extended emission, as the VLBA core/total flux ratio $< 0.5$.
The VLBA/VLA flux ratio $> 1$ again manifests that the source is variable.
This object was observed with the VLA A configuration at 4.8 and 8.4\,GHz \citep{Kellermann1989,Kukula1995}.
It shows a compact core and an inverted VLA slope ($\alpha_{4.8-8.4} = +0.4$), which indicates that the emission detected in our VLBA observation is still dominant on the VLA scales.
This is consistent with the picture that the radio emission in low $L/L_{\rm Edd}$ objects is dominated by the compact optically thick core.

PG1700+518 is a BAL quasar. It has a flat slope with both $\alpha_{\rm total}$ and $\alpha_{\rm core}$ = $-0.3$.
The VLBA coordinate is consistent with the {\it Gaia} position ($\sim$ 1.4\,mas), and the $T_{\rm B}$ is high ($> 10^{7.8}$\,K).
The object is very compact as indicated by the VLBA core/total flux ratio $\sim 0.8$.
The VLBA/VLA flux ratio $\sim 0.5$ suggests that part of the extended emission is resolved out on the VLBA scales but unresolved on the VLA scales.
This object was previously observed with the VLBA at 4.8 and 8.4\,GHz, and both images show a core and a little extended emission \citep{Blundell1998,Wang2023a}.
Moreover, observations with the VLA A configuration at 4.8 and 8.4\,GHz reveal two components \citep{Kellermann1989,Kellermann1994,Kukula1998}.
An EVN observation at 1.6\,GHz of one of the two components reveals a core plus two-sided extended structures \citep{Yang2012}.
These structures may result from an outflow, as the object is a BAL quasar, which has a high velocity outflow \citep{Young2007,Runnoe2018}.
However, the extended emission detected in \citet{Yang2012} is resolved out on the VLBA scales, and we thus only see a compact core, which is probably the base of the outflow.

\subsection{Radio outflows} \label{outflow_discussion}

Four sources (PG0157+001, PG0921+525, PG0923+129, and PG1534+580) are likely to produce an outflow emission on the VLBA scales, though they are distributed at both high and low $L/L_{\rm Edd}$.

PG0157+001 has a steep slope with $\alpha_{\rm total} = -0.9$ and $\alpha_{\rm core} = -0.8$, which indicates that an optically thick compact core is probably missing or below the detection limit.
The steep slope emission is in agreement with earlier VLBA observations, where it was detected at 5\,GHz \citep{Wang2023a}, but not detected at 8.4\,GHz \citep{Blundell1998}.
Observations with the VLA A configuration at 1.4, 4.8, 8.4, and 45\,GHz were carried out. The 4.8 and 8.4\,GHz maps reveal multiple components, a central core plus one/two-sided aligned extended structures \citep{Kellermann1989,Kellermann1994,Kukula1998,Leipski2006}.
The VLA slope at 5--45\,GHz is still steep with $\alpha_{5-45} = -1$ \citep{Baldi2022}, which is comparable to the VLBA slope at 1--5\,GHz.
This suggests that the emission detected with the VLBA and the VLA, though at different scales, may have the same origin, possibly an optically thin outflow, which extends from pc to sub-kpc scales.
Moreover, the large distance from the {\it Gaia} position ($\sim$ 26\,mas), the relatively low $T_{\rm B}$ ($\sim 10^{6.4}$\,K), the VLBA core/total flux ratio ($\sim 0.1$), and the VLBA/VLA flux ratio ($\sim 0.3$), again manifest that the emission is likely associated with an extended outflow.

PG0921+525 has three components.
The slopes of C1 ($\alpha_{\rm total} = -0.1$) and C2 ($\alpha_{\rm total} = +0.4$) are flat/inverted, and that of C3 ($\alpha_{\rm total} < -0.8$) is steep.
The $T_{\rm B}$ is high for C1 ($> 10^{7.7}$\,K), and is lower for C2 ($\sim 10^{6.4}$\,K) and C3 ($\sim 10^{6.8}$\,K).
C1 and C2 are consistent with the {\it Gaia} position ($\sim$ 1.9 and 2.3\,mas), and C3 is farther away ($\sim$ 9.5\,mas).
The above properties suggest that C1 may be the core, C2 may be a new ejection as it is still optically thick and very close to the core, and C3 may be an older ejection as it is already optically thin and farther away from the core.
The radio emission is dominated by the core component, as indicated by the VLBA core/total flux ratio ($> 0.7$) and the VLBA/VLA flux ratio ($< 0.9$).
This object was observed with the VLBA at 1.7 and 5\,GHz. Both images show a compact core with comparable flux densities ($\sim 1.0-1.2$\,mJy) \citep{Doi2013,Wang2023a}, which indicates a flat slope ($\alpha_{1.7-5} = -0.1$) and is consistent with our measurement.
The extended structure was not detected in earlier VLBA observations, which may be due to the detection limit.
On the VLA scales, it was observed with the A configuration at 1.4, 4.8, and 8.4\,GHz.
The 1.4 and 4.8\,GHz images show a core plus extended structure, while the 8.4\,GHz image only shows a core \citep{Kellermann1989,Kellermann1994,Kukula1998}.
The flux density of the core at 1.4\,GHz ($\sim 5.8$\,mJy) is three times higher than that at 4.8 and 8.4\,GHz ($\sim 1.7-1.9$\,mJy), which suggests that a steep slope ($\alpha_{1.4-4.8} = -0.9$) component is dominant at 1--5\,GHz, and a flat slope ($\alpha_{4.8-8.4} = -0.2$) component becomes prominent at 5--8\,GHz.
In addition, radio variability was detected on both the VLBA and the VLA scales \citep{Panessa2022a,Chen2022b}.

PG0923+129 also has three components, the slope of C1 is flat ($\alpha_{\rm total} = -0.5$), and that of C2 ($\alpha_{\rm total} < -1.4$) and C3 ($\alpha_{\rm total} < -2.0$) is steep, which suggests that C1 is likely the core emission, and C2 and C3 are probably the linear extended outflow emission.
The source is indeed extended, as indicated by the VLBA core/total flux ratio and the VLBA/VLA flux ratio both $< 0.5$.
The $T_{\rm B}$ of all three components are not too high ($\sim 10^{6.5} - 10^{6.8}$\,K).
They are also far away from the {\it Gaia} position ($\sim$ 17--25\,mas).
An earlier VLBA observation of this object show a compact core plus a weak extended structure at 1.7\,GHz \citep{Doi2013}.
The two-sided linear extended emission at 1.6\,GHz in our new VLBA observation is not detected in \citet{Doi2013}, though the flux density of the two extended components is higher than their detection limit.
In addition, the core flux density ($\sim 0.9$\,mJy) in our observation is only half of the value ($\sim 1.8$\,mJy) in \citet{Doi2013}.
This may demonstrate that the core emission becomes fainter while the extended emission becomes brighter.
A possible explanation is that the outflow carries energy and propagates from the central core when it launches, which makes the core emission decreases and the extended emission increases.
However, the core component at 4.9\,GHz in our new VLBA observation is not detected in earlier VLBA observations at 4.8 and 8.4\,GHz \citep{Blundell1998,Wang2023a}, which may be due to different detection limits.
A compact core with a flux density of $\sim 2.0$\,mJy at 8.5\,GHz was detected with the VLA A configuration \citep{Schmitt2001a}.

PG1534+580 has two components.
The core component (C1) has a flat slope ($\alpha_{\rm total} = -0.4$), a high $T_{\rm B}$ ($> 10^{6.7}$\,K), and is consistent with the {\it Gaia} position ($\sim$ 6.2\,mas).
The extended component (C2) has a steep slope ($\alpha_{\rm total} < -1.0$), a lower $T_{\rm B}$ ($\sim 10^{6.6}$\,K), and is farther away from the {\it Gaia} position ($\sim$ 10\,mas), which is possibly an outflow emission.
The extended emission is in a scale of being resolved out with the VLBA and unresolved with the VLA, as indicated by the VLBA core/total flux ratio $> 0.6$ and the VLBA/VLA flux ratio $< 0.1$.
This object shows a compact core with a flux density of $\sim 1.8-2.2$\,mJy at 4.8\,GHz with the VLA A configuration \citep{Kellermann1989,Leipski2006}.

\subsection{A mildly relativistic jet?} \label{jet_discussion}

PG1351+640 is an intriguing object exhibiting two components.
The core component C1 has a flat slope with $\alpha_{\rm total} = +1.0$, a high $T_{\rm B}$ ($> 10^{8.0}$\,K), and is consistent with the {\it Gaia} position ($\sim$ 1.9\,mas), which is in agreement with a compact optically thick synchrotron source.
The extended component C2 is optically thin with a steep slope $\alpha_{\rm total} = -0.9$, and is farther away from the {\it Gaia} position ($\sim$ 7.1\,mas).
Its $T_{\rm B}$ is still high ($> 10^{8.0}$\,K), which is 1--2 orders of magnitude higher than the extended emission in the other objects, which makes this extended component resemble the lobe of a radio jet.
Furthermore, the object has a large fraction of extended emission as indicated by the VLBA core/total flux ratio and the VLBA/VLA flux ratio both $< 0.5$ (Figure~\ref{fc}).

PG1351+640 was observed with the VLBA at 1.4, 2.3, and 5\,GHz in 2000 \citep{Ulvestad2005} and at 8.4\,GHz in 1996 \citep{Blundell1998}.
The 5\,GHz map shows a single unresolved component, and the 8.4\,GHz map shows a core and some extended emission $\lesssim 2$\,mas to the northwest, which may be because the extended structure was closer inwards, marginally resolved at 8\,GHz and unresolved at 5\,GHz.
This object also has VLBI observations at 2.3 and 8.4\,GHz in 2011. The maps from the Astrogeo Center \footnote{\url{http://astrogeo.org/index.html}} show only one component at 2.3\,GHz and two components, a core plus an extended component $\sim 5$\,mas to the northwest, at 8.4\,GHz.

The extended emission at 8\,GHz in 2011 was $\sim 5$\,mas away from the core, while it was $\lesssim 2$\,mas from the core in 1996.
This implies that the jet moved $\gtrsim 3$\,mas, corresponding to $\gtrsim 5.7$\,pc, in $\lesssim$ 15 years with an apparent velocity of $\gtrsim 1.2c$ ($c$ is the speed of light).
A recent study by \citet{Wang2023b} who observed this object with the VLBA at 5\,GHz in 2015, also finds that this object has a core + jet morphology which is similar to our result.
A comparison on the separation of the two components in 2015 and 2022 shows that the jet only moved $\sim 0.5$\,mas, corresponding to $\sim 1.0$\,pc, in $\sim$ 7 years, which suggests that the jet moved at a speed of only $\sim 0.5c$ \citep{Wang2023b}.
The above estimates are rough, and more VLBI monitoring are necessary to clarify the proper motion of the jet.
The presence of the proper motion in this object clearly excludes a possible binary BH interpretation, and confirms that it is a core + jet system.

The spectral slope at 1--5\,GHz in \citet{Ulvestad2005} is steep ($\alpha_{1.4-2.3} = -0.8$ and $\alpha_{2.3-5} = -1.0$), while the measurement in our observations is flat ($\alpha_{\rm total} = -0.4$).
A possible explanation is that we may be observing a new compact synchrotron component, and the extended emission observed in \citet{Ulvestad2005} may have expanded and become faint and undetectable by now.

PG1351+640 was also observed with the VLA A configuration at 5\,GHz.
It shows a central core with a flux density of $\sim$ 20\,mJy in \citet{Kellermann1989}.
The flux density decreased to $\sim$ 7--9\,mJy, i.e.\ by a factor of 2--3 in \citet{Leipski2006}, and the source shows two-sided extended emission, in addition to the central core.
The linear extended morphology and the high variability \citep{Barvainis1989} are in agreement with the jet origin of the radio emission.

\section{Summary} \label{summary_section}

In this work, we present the mas-scale radio emission properties of optically selected RQQ.
The sample is selected from the 71 RQ PG quasars at $z < 0.5$, and covers a wide range in luminosity ($-27 < M_{\rm V} < -21$) and H$\beta$ FWHM ($\sim$ 1000--10,000\,km\,s$^{-1}$).
The study is based on our new VLBA observations of ten RQQ at 1.6 and 4.9\,GHz, of which nine were detected, and earlier similar observations of eight more RQQ, of which five were detected \citep{Alhosani2022}, which together yields a sample of 18 objects, of which 14 were detected.
The main results are summarized below.

1. The mas-scale emission in most of the RQQ is composed of a compact core, which is consistent with the {\it Gaia} optical position.
The spectral slope of the compact core emission is flat or inverted in nearly all cases with $T_{\rm B}$ extending to $> 10^8$\,K.
The compact optically thick sources have a physical size of $\sim 0.01-0.4$\,pc, and possibly originate from the accretion disk corona.

2. The extended emission detected in some of the RQQ is generally optically thin synchrotron emission, and is characterized by a steep slope and a lower $T_{\rm B}$ ($< 10^7$\,K).
The extended sources located $\sim 5-100$\,pc from the center, may be formed by a clumped outflow of magnetized plasma, e.g.\ coronal mass ejection or an intermittent jet.
Alternatively, it may originate from a shocked nuclear wind. 

3. The VLBA core luminosity of all objects at $\log M_{\rm BH}/M_{\odot} < 8.5$ is correlated with the X-ray luminosity following $\log L_{\rm R}/L_{\rm X} \simeq -6$, which may indicate a coronal origin of the radio core emission.
Three of the five objects at $\log M_{\rm BH}/M_{\odot} > 8.5$ have $\log L_{\rm R}/L_{\rm X} \gtrsim -4$, of which one potentially shows a mildly relativistic jet/outflow.

4. The larger scale VLA emission ($\sim 300$\,mas) of the 18 objects follows $\log L_{\rm R}/L_{\rm X} \simeq -5$, as found in earlier studies.
The $L_{\rm R}/L_{\rm X}$ shows a gradual rise with $M_{\rm BH}$ above $\sim 10^7 M_{\odot}$, which suggests that the larger the BH mass, the higher the chances to produce extended emission.

5. In one object PG1351+640, a comparison with earlier VLBI observations suggests a mildly relativistic jet in the past $\sim$ 7 years, and potentially a super-luminal jet in the past $\sim$ 15 years.
This may well represent an intermediate RQQ, which is powered by a jet weaker than that in RLQ.

Our recent e-MERLIN and uGMRT observations of this sample (Chen et al.\ in preparation), forthcoming higher frequency VLBA observations, and future VLA observations, will help to establish the dominant radio emission mechanisms on different scales, and the interaction of the AGN with the host galaxy from pc to kpc scales.
In addition, future observations at higher frequencies, especially in mm-wave bands \citep[e.g.][]{Behar2015,Behar2018,Ricci2023}, can help to clarify the nature of the AGN core emission.

Finally, the results above have to be taken with caution given the small sample size.
Further studies of a large sample are necessary to confirm these findings and draw a more reliable picture of the radio emission in RQ AGN.

\section*{Acknowledgements}

We thank Amy Kimball, Justin Linford, and Eric Greisen for the support of the VLBA data reduction.
We thank the anonymous referee for a thorough review and for the many insightful and helpful comments.
A.L. acknowledges support by the Israel Science Foundation (grant no.1008/18).
E.B. acknowledges support by a Center of Excellence of the Israel Science Foundation (grant no.2752/19).
S.C. is supported in part by a Technion fellowship.
The National Radio Astronomy Observatory is a facility of the National Science Foundation operated under cooperative agreement by Associated Universities, Inc.
This work has made use of data from the European Space Agency (ESA) mission {\it Gaia} (\url{https://www.cosmos.esa.int/gaia}), processed by the {\it Gaia} Data Processing and Analysis Consortium (DPAC, \url{https://www.cosmos.esa.int/web/gaia/dpac/consortium}). Funding for the DPAC has been provided by national institutions, in particular the institutions participating in the {\it Gaia} Multilateral Agreement.
This research has made use of data obtained from the 4XMM XMM-Newton serendipitous source catalogue compiled by the XMM-Newton Survey Science Centre consortium.

\section*{Data availability}

The radio data underlying this article are available in the NRAO Science Data Archive at \url{https://archive.nrao.edu/archive/advquery.jsp}, and can be accessed with the project code of BC273.

\bibliographystyle{mnras}
\bibliography{VLBA.bbl}


\begin{table*}
\begin{center}
\caption{The VLBA positions and their distances from the {\it Gaia} positions of the 10 RQ PG quasars observed with VLBA. Columns: 
(1) name,
(2) redshift,
(3) physical scale,
(4) observation date,
(5) frequency,
(6) right ascension of the centroid of VLBA emission determined using IMFIT,
(7) declination of the centroid of VLBA emission determined using IMFIT,
(8) separation between the VLBA and the {\it Gaia} positions.
The positions in C band are reported if the emission is detected, otherwise the positions in L band are listed.}
\label{position}
\begin{tabular}{cccccccc}
\hline
\hline
\multirow{2}{*}{Name} & \multirow{2}{*}{$z$} & Scale & \multirow{2}{*}{Obs. Date} & $\nu$ & R.A. & Dec. & $\Delta$ \\
& & (pc\,mas$^{-1}$) & & (GHz) & (hh:mm:ss) & (dd:mm:ss) & (mas) \\
(1) & (2) & (3) & (4) & (5) & (6) & (7) & (8) \\
\hline
PG0026+129 & 0.142 & 3.26 & 06 Mar 2022 & 4.9 & 00:29:13.7014 & $+$13:16:03.9443 & 5.5 \\
\hline
PG0157+001 & 0.164 & 3.81 & 21 Mar 2022 & 4.9 & 01:59:50.2543 & $+$00:23:40.8650 & 25.5 \\
\hline
\multirow{3}{*}{PG0921+525} & \multirow{3}{*}{0.035} & \multirow{3}{*}{0.75} & \multirow{3}{*}{18 Mar 2022} & 4.9 C1 & 09:25:12.8480 & $+$52:17:10.3871 & 1.9 \\
& & & & 4.9 C2 & 09:25:12.8476 & $+$52:17:10.3879 & 2.3 \\
& & & & 1.6 C3 & 09:25:12.8469 & $+$52:17:10.3908 & 9.5 \\
\hline
\multirow{3}{*}{PG0923+129} & \multirow{3}{*}{0.029} & \multirow{3}{*}{0.62} & \multirow{3}{*}{27 Feb 2022} & 4.9 C1 & 09:26:03.2697 & $+$12:44:03.7480 & 17.3 \\
& & & & 1.6 C2 & 09:26:03.2709 & $+$12:44:03.7466 & 25.2 \\
& & & & 1.6 C3 & 09:26:03.2685 & $+$12:44:03.7424 & 19.5 \\
\hline
PG1216+069 & 0.334 & 8.52 & 28 Feb 2022 & 4.9 & 12:19:20.9314 & $+$06:38:38.4698 & 5.8 \\
\hline
PG1351+236 & 0.055 & 1.19 & 14 Mar 2022 & - & - & - & - \\
\hline
\multirow{2}{*}{PG1351+640} & \multirow{2}{*}{0.087} & \multirow{2}{*}{1.92} & \multirow{2}{*}{05 Mar 2022} & 4.9 C1 & 13:53:15.8310 & $+$63:45:45.6844 & 1.9 \\
& & & & 4.9 C2 & 13:53:15.8304 & $+$63:45:45.6876 & 7.1 \\
\hline
PG1501+106 & 0.036 & 0.77 & 24 Mar 2022 & 4.9 & 15:04:01.1937 & $+$10:26:15.7833 & 3.6 \\
\hline
\multirow{2}{*}{PG1534+580} & \multirow{2}{*}{0.030} & \multirow{2}{*}{0.64} & \multirow{2}{*}{19 Mar 2022} & 4.9 C1 & 15:35:52.4035 & $+$57:54:09.5187 & 6.2 \\
& & & & 1.6 C2 & 15:35:52.4029 & $+$57:54:09.5238 & 10.0 \\
\hline
PG1700+518 & 0.292 & 7.30 & 13 Apr 2022 & 4.9 & 17:01:24.8266 & $+$51:49:20.4492 & 1.4 \\
\hline
\hline
\multirow{2}{*}{PG0050+124} & \multirow{2}{*}{0.060} & \multirow{2}{*}{1.17} & \multirow{2}{*}{05 May 2020} & 4.8 C2 & 00:53:34.9337 & $+$12:41:35.9289 & 5.7 \\
& & & & 4.8 C1 & 00:53:34.9342 & $+$12:41:35.9306 & 12.6 \\
\hline
PG0052+251 & 0.155 & 2.71 & 16 May 2020 & 4.8 & 00:54:52.1182 & $+$25:25:38.9859 & 1.9 \\
\hline
PG1149$-$110 & 0.050 & 0.98 & 31 May 2020 & 4.8 & 11:52:03.5505 & $-$11:22:24.0932 & 1.3 \\
\hline
PG1612+261 & 0.131 & 2.35 & 12 May 2020 & 1.4 & 16:14:13.2058 & $+$26:04:16.2230 & 10.2 \\
\hline
PG2304+042 & 0.042 & 0.83 & 07 May 2020 & 4.8 & 23:07:02.9147 & $+$04:32:57.1019 & 0.2 \\
\hline
\hline
\end{tabular}
\end{center}
\end{table*}

\begin{table*}
\begin{center}
\caption{The beam and source size of the 10 RQ PG quasars observed with VLBA. Columns: 
(1) name, 
(2) frequency, 
(3) major axis of the beam in unit of mas, 
(4) minor axis of the beam in unit of mas, 
(5) position angle of the beam in unit of degree, 
(6) major axis of the source in unit of mas, 
(7) minor axis of the source in unit of mas, 
(8) position angle of the source in unit of degree,
(9) deconvolved major axis of the source in unit of mas, 
(10) deconvolved minor axis of the source in unit of mas, 
(11) deconvolved position angle of the source in unit of degree.}
\label{size}
\begin{tabular}{ccccccccccc}
\hline
\hline
\multirow{3}{*}{Name} & Frequency & \multicolumn{3}{c}{Beam size} & \multicolumn{3}{c}{Source size} & \multicolumn{3}{c}{Deconvolved source size} \\
& $\nu$ & $\theta_{\rm maj}$ & $\theta_{\rm min}$ & PA & $\theta_{\rm maj}$ & $\theta_{\rm min}$ & PA & $\theta_{\rm maj}$ & $\theta_{\rm min}$ & PA \\
& (GHz) & (mas) & (mas) & (degree) & (mas) & (mas) & (degree) & (mas) & (mas) & (degree) \\
(1) & (2) & (3) & (4) & (5) & (6) & (7) & (8) & (9) & (10) & (11) \\
\hline
PG0026+129 & 4.9 & 4.06 & 1.83 & 1.0 & 4.32 & 2.57 & 1.8 & $<$2.03 & 1.45 & - \\
\hline
\multirow{2}{*}{PG0157+001} & 4.9 & 3.69 & 1.53 & 6.0 & 9.01 & 7.50 & 8.2 & 8.22 & 7.34 & 10.1 \\
& 1.6 & 11.26 & 4.49 & 0.7 & 22.16 & 16.97 & 126.2 & 21.10 & 13.67 & 114.7 \\
\hline
\multirow{5}{*}{PG0921+525} & 4.9 C1 & \multirow{2}{*}{4.12} & \multirow{2}{*}{1.86} & \multirow{2}{*}{32.9} & 4.29 & 1.98 & 32.8 & $<$2.06 & $<$0.93 & - \\
& 4.9 C2 & & & & 5.42 & 3.71 & 41.3 & 3.71 & 3.00 & 68.9 \\
& 1.6 C1 & \multirow{3}{*}{11.73} & \multirow{3}{*}{4.26} & \multirow{3}{*}{36.6} & 12.82 & 5.54 & 34.2 & $<$5.87 & 3.29 & - \\
& 1.6 C2 & & & & 15.06 & 9.16 & 62.5 & 11.65 & 4.38 & 89.3 \\
& 1.6 C3 & & & & 21.48 & 6.68 & 56.8 & 18.51 & 2.77 & 63.5 \\
\hline
\multirow{4}{*}{PG0923+129} & 4.9 C1 & 4.14 & 1.80 & 178.2 & 4.64 & 2.90 & 174.5 & 2.41 & 1.97 & 120.0 \\
& 1.6 C1 & \multirow{3}{*}{10.30} & \multirow{3}{*}{3.59} & \multirow{3}{*}{13.6} & 13.54 & 10.39 & 23.2 & 10.13 & 8.33 & 79.4 \\
& 1.6 C2 & & & & 14.10 & 13.02 & 25.4 & 12.58 & 9.54 & 98.5 \\
& 1.6 C3 & & & & 21.85 & 16.70 & 42.9 & 20.09 & 15.29 & 56.9 \\
\hline
\multirow{2}{*}{PG1216+069} & 4.9 & 4.43 & 1.88 & 173.0 & 4.51 & 1.88 & 172.2 & $<$2.22 & $<$0.94 & - \\
& 1.6 & 29.56 & 6.87 & 161.6 & 28.00 & 7.17 & 162.4 & $<$14.78 & $<$3.44 & - \\
\hline
\multirow{4}{*}{PG1351+640} & 4.9 C1 & \multirow{2}{*}{4.16} & \multirow{2}{*}{1.91} & \multirow{2}{*}{30.0} & 4.20 & 2.01 & 27.6 & $<$2.08 & $<$0.96 & - \\
& 4.9 C2 & & & & 4.25 & 2.53 & 31.1 & $<$2.13 & $<$1.27 & - \\
& 1.6 C1 & \multirow{2}{*}{10.14} & \multirow{2}{*}{3.26} & \multirow{2}{*}{29.2} & 19.18 & 6.54 & 72.4 & 17.74 & $<$1.63 & - \\
& 1.6 C2 & & & & 11.30 & 7.00 & 27.8 & 6.22 & 4.96 & 127.1 \\
\hline
\multirow{2}{*}{PG1501+106} & 4.9 & 4.07 & 1.77 & 4.4 & 5.99 & 3.55 & 18.1 & 4.56 & 2.82 & 32.5 \\
& 1.6 & 10.43 & 3.41 & 6.8 & 14.20 & 6.04 & 175.1 & 10.07 & 4.05 & 161.7 \\
\hline
\multirow{3}{*}{PG1534+580} & 4.9 C1 & 4.58 & 1.89 & 37.0 & 3.06 & 1.59 & 34.5 & $<$2.29 & $<$0.95 & - \\
& 1.6 C1 & \multirow{2}{*}{11.74} & \multirow{2}{*}{3.53} & \multirow{2}{*}{40.2} & 15.75 & 12.74 & 35.4 & 12.31 & 10.42 & 139.9 \\
& 1.6 C2 & & & & 15.94 & 14.15 & 77.1 & 14.59 & 9.54 & 117.6 \\
\hline
\multirow{2}{*}{PG1700+518} & 4.9 & 3.87 & 1.56 & 15.0 & 4.20 & 1.77 & 13.2 & $<$1.94 & $<$0.78 & - \\
& 1.6 & 12.65 & 7.04 & 170.5 & 15.05 & 8.04 & 158.7 & 8.85 & $<$3.52 & - \\
\hline
\hline
\end{tabular}
\end{center}
\end{table*}

\begin{table*}
\begin{center}
\caption{The flux densities of the 10 RQ PG quasars observed with VLBA at 4.9 and 1.6\,GHz. Columns: 
(1) name, 
(2) frequency, 
(3) core flux density of the full resolution map, 
(4) total flux density of the full resolution map, 
(5) background noise of the full resolution map, 
(6) core flux density of the tapered map, 
(7) total flux density of the tapered map, 
(8) background noise of the tapered map. 
The tapered maps have a UV range of 3000--50000\,$k\lambda$ in both bands.}
\label{flux}
\begin{tabular}{cccccccc}
\hline
\hline
\multirow{3}{*}{Name} & Frequency & \multicolumn{3}{c}{Full resolution maps} & \multicolumn{3}{c}{Tapered (3000--50000 $k\lambda$) maps} \\
& $\nu$ & $S_{\rm core}$ & $S_{\rm total}$ & RMS & $S_{\rm core}$ & $S_{\rm total}$ & RMS \\
& (GHz) & (mJy\,beam$^{-1}$) & (mJy) & (mJy\,beam$^{-1}$) & (mJy\,beam$^{-1}$) & (mJy) & (mJy\,beam$^{-1}$) \\
(1) & (2) & (3) & (4) & (5) & (6) & (7) & (8) \\
\hline
\multirow{2}{*}{PG0026+129} & 4.9 & 0.21 $\pm$ 0.02 & 0.31 $\pm$ 0.04 & 0.019 & 0.25 $\pm$ 0.02 & 0.28 $\pm$ 0.04 & 0.024 \\
& 1.6 & - & - & 0.020 & - & - & 0.021 \\
\hline
\multirow{2}{*}{PG0157+001} & 4.9 & 0.15 $\pm$ 0.02 & 1.79 $\pm$ 0.30 & 0.028 & 0.27 $\pm$ 0.03 & 0.95 $\pm$ 0.16 & 0.038 \\
& 1.6 & 0.82 $\pm$ 0.03 & 6.10 $\pm$ 0.28 & 0.037 & 0.62 $\pm$ 0.03 & 2.60 $\pm$ 0.17 & 0.037 \\
\hline
\multirow{6}{*}{PG0921+525} & 4.9 C1 & 1.10 $\pm$ 0.02 & 1.21 $\pm$ 0.04 & \multirow{3}{*}{0.022} & 1.02 $\pm$ 0.03 & 1.02 $\pm$ 0.03 & \multirow{3}{*}{0.027} \\
& 4.9 C2 & - & 0.35 $\pm$ 0.07 & & - & 0.43 $\pm$ 0.06 & \\
& 4.9 C3 & - & $<$ 0.11 & & - & $<$ 0.14 & \\
& 1.6 C1 & 0.87 $\pm$ 0.02 & 1.24 $\pm$ 0.05 & \multirow{3}{*}{0.025} & 0.84 $\pm$ 0.03 & 1.19 $\pm$ 0.06 & \multirow{3}{*}{0.026} \\
& 1.6 C2 & - & 0.36 $\pm$ 0.09 & & - & 0.28 $\pm$ 0.08 & \\
& 1.6 C3 & - & 0.44 $\pm$ 0.09 & & - & 0.33 $\pm$ 0.08 & \\
\hline
\multirow{6}{*}{PG0923+129} & 4.9 C1 & 0.11 $\pm$ 0.02 & 0.21 $\pm$ 0.05 & \multirow{3}{*}{0.019} & 0.17 $\pm$ 0.02 & 0.17 $\pm$ 0.02 & \multirow{3}{*}{0.025} \\
& 4.9 C2 & - & $<$ 0.10 & & - & $<$ 0.13 & \\
& 4.9 C3 & - & $<$ 0.10 & & - & $<$ 0.13 & \\
& 1.6 C1 & 0.24 $\pm$ 0.02 & 0.91 $\pm$ 0.09 & \multirow{3}{*}{0.021} & 0.20 $\pm$ 0.02 & 0.29 $\pm$ 0.04 & \multirow{3}{*}{0.021} \\
& 1.6 C2 & - & 1.15 $\pm$ 0.11 & & - & 0.58 $\pm$ 0.10 & \\
& 1.6 C3 & - & 2.39 $\pm$ 0.20 & & - & 1.13 $\pm$ 0.14 & \\
\hline
\multirow{2}{*}{PG1216+069} & 4.9 & 6.77 $\pm$ 0.02 & 6.91 $\pm$ 0.04 & 0.021 & 6.74 $\pm$ 0.03 & 6.89 $\pm$ 0.05 & 0.027 \\
& 1.6 & 0.47 $\pm$ 0.03 & 0.47 $\pm$ 0.03 & 0.030 & 0.50 $\pm$ 0.03 & 0.60 $\pm$ 0.07 & 0.035 \\
\hline
\multirow{2}{*}{PG1351+236} & 4.9 & - & - & 0.024 & - & - & 0.025 \\
& 1.6 & - & - & 0.030 & - & - & 0.034 \\
\hline
\multirow{4}{*}{PG1351+640} & 4.9 C1 & 2.16 $\pm$ 0.02 & 2.30 $\pm$ 0.04 & \multirow{2}{*}{0.021} & 2.26 $\pm$ 0.03 & 2.42 $\pm$ 0.05 & \multirow{2}{*}{0.026} \\
& 4.9 C2 & - & 2.67 $\pm$ 0.04 & & - & 2.58 $\pm$ 0.05 & \\
& 1.6 C1 & 0.16 $\pm$ 0.01 & 0.59 $\pm$ 0.07 & \multirow{2}{*}{0.024} & 0.59 $\pm$ 0.01 & 0.78 $\pm$ 0.03 & \multirow{2}{*}{0.021} \\
& 1.6 C2 & - & 8.13 $\pm$ 0.05 & & - & 7.29 $\pm$ 0.04 & \\
\hline
\multirow{2}{*}{PG1501+106} & 4.9 & 0.21 $\pm$ 0.02 & 0.62 $\pm$ 0.08 & 0.022 & 0.33 $\pm$ 0.03 & 0.40 $\pm$ 0.05 & 0.028 \\
& 1.6 & 0.15 $\pm$ 0.02 & 0.37 $\pm$ 0.07 & 0.022 & 0.16 $\pm$ 0.02 & 0.37 $\pm$ 0.07 & 0.024 \\
\hline
\multirow{4}{*}{PG1534+580} & 4.9 C1 & 0.14 $\pm$ 0.02 & 0.14 $\pm$ 0.02 & \multirow{2}{*}{0.022} & 0.14 $\pm$ 0.02 & 0.14 $\pm$ 0.02 & \multirow{2}{*}{0.024} \\
& 4.9 C2 & - & $<$ 0.11 & & - & $<$ 0.12 & \\
& 1.6 C1 & 0.13 $\pm$ 0.02 & 0.63 $\pm$ 0.09 & \multirow{2}{*}{0.018} & 0.08 $\pm$ 0.02 & 0.20 $\pm$ 0.06 & \multirow{2}{*}{0.018} \\
& 1.6 C2 & - & 0.84 $\pm$ 0.10 & & - & 0.35 $\pm$ 0.07 & \\
\hline
\multirow{2}{*}{PG1700+518} & 4.9 & 0.75 $\pm$ 0.02 & 0.92 $\pm$ 0.04 & 0.020 & 0.77 $\pm$ 0.03 & 1.00 $\pm$ 0.05 & 0.026 \\
& 1.6 & 1.07 $\pm$ 0.06 & 1.46 $\pm$ 0.13 & 0.062 & 1.06 $\pm$ 0.06 & 1.36 $\pm$ 0.13 & 0.065 \\
\hline
\hline
\end{tabular}
\end{center}
\end{table*}

\begin{table*}
\begin{center}
\caption{The BH mass, the Eddington ratio, the unresolved flux density at 5\,GHz with VLA A configuration, the X-ray flux at 0.2--12.0\,keV, and the ratio of radio to X-ray luminosity of the 18 RQ PG quasars observed with VLBA, including 10 objects from our new observations and 8 objects from \citet{Alhosani2022}. Columns: 
(1) name,
(2) logarithm of BH mass,
(3) logarithm of bolometric luminosity in unit of erg\,s$^{-1}$,
(4) logarithm of Eddington ratio,
(5) flux density at 5\,GHz with VLA A configuration from literature,
(6) reference for the VLA 5\,GHz flux density,
(7) X-ray flux at 0.2--12.0\,keV in unit of $10^{-12}$\,erg\,s$^{-1}$\,cm$^{-2}$ from the {\it XMM-Newton} DR12 catalog,
(8) radio to X-ray luminosity ratio using the VLBA core flux,
(9) radio to X-ray luminosity ratio using the unresolved VLA A configuration flux.
The BH mass and bolometric luminosity are from \citet{Davis2011} or \citet{Laor2019}. Reference: (a) \citet{Kellermann1989}, (b) \citet{Berton2018}, (c) \citet{Alhosani2022}.}
\label{literature}
\begin{tabular}{ccccccccc}
\hline
\hline
\multirow{2}{*}{Name} & \multirow{2}{*}{$\log M_{\rm BH}/M_{\odot}$} & \multirow{2}{*}{$\log L_{\rm bol}$} & \multirow{2}{*}{$\log L/L_{\rm Edd}$} & $S^{\rm VLA}_{\rm 5GHz}$ & \multirow{2}{*}{Ref.} & $f_{\rm 0.2-12.0keV}$ & VLBA & VLA \\
& & & & (mJy) & & ($10^{-12}$\,erg\,s$^{-1}$\,cm$^{-2}$) & $\log L_{\rm R}/L_{\rm X}$ & $\log L_{\rm R}/L_{\rm X}$ \\
(1) & (2) & (3) & (4) & (5) & (6) & (7) & (8) & (9) \\
\hline
PG0026+129 & 7.74 & 46.15 & $+0.30$ & 0.20 $\pm$ 0.06 & a & 9.25 $\pm$ 0.10 & $-5.88 \pm 0.04$ & $-5.97 \pm 0.13$ \\
PG0157+001 & 8.00 & 45.93 & $-0.18$ & 5.58 $\pm$ 0.06 & a & 2.82 $\pm$ 0.05 & $-5.34 \pm 0.06$ & $-4.00 \pm 0.01$ \\
PG0921+525 & 6.87 & 44.47 & $-0.51$ & 1.87 $\pm$ 0.06 & a & 55.80 $\pm$ 0.10 & $-6.05 \pm 0.01$ & $-5.78 \pm 0.01$ \\
PG0923+129 & 6.82 & 44.53 & $-0.40$ & 2.82 $\pm$ 0.02 & b & 34.10 $\pm$ 0.13 & $-6.61 \pm 0.06$ & $-5.38 \pm 0.01$ \\
PG1216+069 & 9.06 & 46.61 & $-0.56$ & 4.95 $\pm$ 0.06 & a & 3.65 $\pm$ 0.03 & $-4.04 \pm 0.01$ & $-4.17 \pm 0.01$ \\
PG1351+236 & 8.10 & 44.57 & $-1.64$ & $\lesssim$ 0.25 & a & - & - & - \\
PG1351+640 & 8.49 & 45.31 & $-1.29$ & 20.0 $\pm$ 0.06 & a & 0.95 $\pm$ 0.02 & $-3.93 \pm 0.01$ & $-2.98 \pm 0.01$ \\
PG1501+106 & 8.11 & 44.90 & $-1.32$ & 0.50 $\pm$ 0.06 & a & 32.90 $\pm$ 0.13 & $-6.31 \pm 0.04$ & $-6.12 \pm 0.05$ \\
PG1534+580 & 7.71 & 44.49 & $-1.33$ & 1.80 $\pm$ 0.06 & a & 17.10 $\pm$ 0.10 & $-6.41 \pm 0.08$ & $-5.28 \pm 0.01$ \\
PG1700+518 & 8.59 & 46.63 & $-0.07$ & 2.05 $\pm$ 0.06 & a & 0.02 $\pm$ 0.01 & $-2.70 \pm 0.13$ & $-2.27 \pm 0.13$ \\
\hline
PG0050+124 & 6.99 & 45.12 & $+0.03$ & $2.41 \pm 0.12$ & c & $14.90 \pm 0.04$ & $-6.10 \pm 0.09$ & $-5.10 \pm 0.02$ \\
PG0052+251 & 8.64 & 46.06 & $-0.68$ & $0.68 \pm 0.04$ & c & $13.70 \pm 0.06$ & $-6.06 \pm 0.03$ & $-5.61 \pm 0.03$ \\
PG1149$-$110 & 7.34 & 44.75 & $-0.70$ & $2.27 \pm 0.05$ & c & $9.12 \pm 0.08$ & $-5.64 \pm 0.02$ & $-4.91 \pm 0.01$ \\
PG1612+261 & 7.69 & 45.38 & $-0.41$ & $5.58 \pm 0.08$ & c & $8.07 \pm 0.09$ & $< -6.31 \pm 0.01$ & $-4.47 \pm 0.01$ \\
PG2304+042 & 7.91 & 44.49 & $-1.52$ & $0.77 \pm 0.07$ & c & $30.00 \pm 0.13$ & $-6.35 \pm 0.03$ & $-5.90 \pm 0.04$ \\
PG1440+356 & 7.09 & 45.62 & $+0.43$ & $1.24 \pm 0.07$ & c & $10.90 \pm 0.05$ & $< -6.44 \pm 0.01$ & $-5.25 \pm 0.02$ \\
PG1613+658 & 8.89 & 45.89 & $-1.10$ & $3.03 \pm 0.07$ & c & $10.70 \pm 0.19$ & $< -6.44 \pm 0.01$ & $-4.86 \pm 0.01$ \\
PG2130+099 & 7.49 & 45.52 & $-0.07$ & $2.18 \pm 0.07$ & c & $10.00 \pm 0.06$ & $< -6.46 \pm 0.01$ & $-4.97 \pm 0.01$ \\
\hline
\hline
\end{tabular}
\end{center}
\end{table*}

\begin{table*}
\begin{center}
\caption{The spectral slopes, brightness temperature, and compactness of the 14 RQ PG quasars observed with VLBA, including 9 objects detected in our new observations and 5 objects detected in \citet{Alhosani2022}. Columns: 
(1) name, 
(2) component, 
(3) spectral slope of the core flux density at 1.6--4.9\,GHz measured in the tapered maps, 
(4) spectral slope of the total flux density at 1.6--4.9\,GHz measured in the tapered maps, 
(5) spectral slope of the extended flux density at 1.6--4.9\,GHz measured in the tapered maps, 
(6) frequency at which the brightness temperature calculated, 
(7) logarithm of surface brightness temperature at 4.9 or 1.6\,GHz in unit of K, 
(8) ratio of the VLBA core/total flux at 5\,GHz, 
(9) ratio of the VLBA/VLA flux at 5\,GHz.}
\label{slope}
\begin{tabular}{ccccccccc}
\hline
\hline
\multirow{2}{*}{Name} & \multirow{2}{*}{Comp.} & \multirow{2}{*}{$\alpha_{\rm core}$} & \multirow{2}{*}{$\alpha_{\rm total}$} & \multirow{2}{*}{$\alpha_{\rm extended}$} & $\nu$ & \multirow{2}{*}{$\log T_{\rm B}$} & \multirow{2}{*}{$f(\frac{S_{\rm core}}{S_{\rm total}})$} & \multirow{2}{*}{$f(\frac{S_{\rm VLBA}}{S_{\rm VLA}})$} \\
& & & & & (GHz) & & & \\
(1) & (2) & (3) & (4) & (5) & (6) & (7) & (8) & (9) \\
\hline
PG0026+129 & - & $> +0.79 \pm 0.09$ & $> +0.89 \pm 0.14$ & - & 4.9 & $>$ 7.0 & $0.67 \pm 0.11$ & $1.55 \pm 0.51$ \\
\hline
PG0157+001 & - & $-0.75 \pm 0.13$ & $-0.90 \pm 0.16$ & $-0.95 \pm 0.22$ & 4.9 & 6.4 & $0.08 \pm 0.02$ & $0.32 \pm 0.05$ \\
\hline
\multirow{4}{*}{PG0921+525} & All & $+0.18 \pm 0.04$ & $< -0.12 \pm 0.07$ & $< -0.48 \pm 0.16$ & - & - & $> 0.65 \pm 0.03$ & $< 0.90 \pm 0.05$ \\
& C1 & - & $-0.14 \pm 0.05$ & - & 4.9 & $>$ 7.7 & - & - \\
& C2 & - & $+0.37 \pm 0.28$ & - & 4.9 & 6.4 & - & - \\
& C3 & - & $< -0.80 \pm 0.21$ & - & 1.6 & 6.8 & - & - \\
\hline
\multirow{4}{*}{PG0923+129} & All & $-0.16 \pm 0.16$ & $< -1.40 \pm 0.09$ & $< -1.78 \pm 0.15$ & - & - & $> 0.28 \pm 0.06$ & $< 0.14 \pm 0.02$ \\
& C1 & - & $-0.48 \pm 0.19$ & - & 4.9 & 6.5 & - & - \\
& C2 & - & $< -1.39 \pm 0.15$ & - & 1.6 & 6.8 & - & - \\
& C3 & - & $< -1.98 \pm 0.11$ & - & 1.6 & 6.8 & - & - \\
\hline
PG1216+069 & - & $+2.33 \pm 0.06$ & $+2.18 \pm 0.10$ & - & 4.9 & $>$ 8.5 & $0.98 \pm 0.01$ & $1.40 \pm 0.02$ \\
\hline
\multirow{3}{*}{PG1351+640} & All & $+1.20 \pm 0.02$ & $-0.43 \pm 0.01$ & $-0.90 \pm 0.02$ & - & - & $0.44 \pm 0.01$ & $0.25 \pm 0.01$ \\
& C1 & - & $+1.01 \pm 0.04$ & - & 4.9 & $>$ 8.0 & - & - \\
& C2 & - & $-0.93 \pm 0.02$ & - & 4.9 & $>$ 8.0 & - & - \\
\hline
PG1501+106 & - & $+0.63 \pm 0.14$ & $+0.08 \pm 0.21$ & - & 4.9 & 6.6 & $0.34 \pm 0.05$ & $1.25 \pm 0.22$ \\
\hline
\multirow{3}{*}{PG1534+580} & All & $+0.45 \pm 0.24$ & $< -0.70 \pm 0.17$ & $< -1.24 \pm 0.30$ & - & - & $> 0.57 \pm 0.10$ & $< 0.14 \pm 0.01$ \\
& C1 & - & $-0.36 \pm 0.29$ & - & 4.9 & $>$ 6.7 & - & - \\
& C2 & - & $< -0.98 \pm 0.17$ & - & 1.6 & 6.6 & - & - \\
\hline
PG1700+518 & - & $-0.29 \pm 0.06$ & $-0.28 \pm 0.10$ & $> -0.59 \pm 0.23$ & 4.9 & $>$ 7.8 & $0.81 \pm 0.04$ & $0.45 \pm 0.02$ \\
\hline
\hline
\multirow{3}{*}{PG0050+124} & All & - & $-0.90 \pm 0.15$ & - & - & - & $0.44 \pm 0.12$ & $0.23 \pm 0.04$ \\
& C2 & - & $0.00 \pm 0.29$ & - & 4.8 & 6.1 & - & - \\
& C1 & - & $-1.21 \pm 0.19$ & - & 4.8 & 6.0 & - & - \\
\hline
PG0052+251 & - & - & $-0.13 \pm 0.16$ & - & 4.8 & 7.6 & $0.81 \pm 0.10$ & $0.44 \pm 0.05$ \\
\hline
PG1149$-$110 & - & - & $-0.31 \pm 0.12$ & - & 4.8 & 6.8 & $0.72 \pm 0.06$ & $0.26 \pm 0.02$ \\
\hline
PG1612+261 & - & - & $< -1.61 \pm 0.11$ & - & 1.4 & 7.1 & - & $< 0.01 \pm 0.01$ \\
\hline
PG2304+042 & - & - & $-0.09 \pm 0.12$ & - & 4.8 & 6.6 & $0.56 \pm 0.06$ & $0.64 \pm 0.08$ \\
\hline
\hline
\end{tabular}
\end{center}
\end{table*}

\label{lastpage}
\end{document}